\documentclass[12pt]{elsart}
\usepackage{epsfig}
\usepackage{colordvi}
\begin{document}

\begin{frontmatter}

\title{
Systematics of azimuthal asymmetries in heavy ion collisions in the 
$1A$ GeV regime}

\author[GSI]{W.~Reisdorf, \thanksref{info}},
\author[GSI]{Y.~Leifels},
\author[GSI]{A.~Andronic},
\author[GSI]{R.~Averbeck},
\author[CLER]{V.~Barret},
\author[ZAG]{Z.~Basrak},
\author[CLER]{N.~Bastid},
\author[HEID]{M.L.~Benabderrahmane},
\author[ZAG]{R.~\v{C}aplar},
\author[CLER]{P.~Crochet},
\author[CLER]{P.~Dupieux},
\author[ZAG]{M.~D\v{z}elalija},
\author[BUD]{Z.~Fodor},
\author[WAR]{P.~Gasik},
\author[ITEP]{Y.~Grishkin},
\author[GSI]{O.N.~Hartmann},
\author[HEID]{N.~Herrmann},
\author[GSI]{K.D.~Hildenbrand},
\author[KOR]{B.~Hong},
\author[GSI,KOR]{T.I.~Kang},
\author[BUD]{J.~Kecskemeti},
\author[GSI]{Y.J.~Kim},
\author[WAR]{M.~Kirejczyk},
\author[GSI,ZAG]{M.~Ki\v{s}},
\author[GSI]{P.~Koczo\'{n}},
\author[ZAG]{M.~Korolija},
\author[ROSS]{R.~Kotte},
\author[GSI]{T.~Kress},
\author[ITEP]{A.~Lebedev},
\author[CLER]{X.~Lopez},
\author[WAR]{T.~Matulewicz},
\author[HEID]{M.~Merschmeyer},
\author[ROSS]{W.~Neubert},
\author[BUC]{M.~Petrovici},
\author[HEID,WAR]{K.~Piasecki},
\author[STRAS]{F.~Rami},
\author[KOR]{M.S.~Ryu},
\author[GSI]{A.~Sch\"{u}ttauf},
\author[BUD]{Z.~Seres},
\author[WAR]{B.~Sikora},
\author[KOR]{K.S.~Sim},
\author[BUC]{V.~Simion},
\author[WAR]{K.~Siwek-Wilczy\'nska},
\author[ITEP]{V.~Smolyankin},
\author[HEID]{M.~Stockmeier},
\author[BUC]{G.~Stoicea},
\author[WAR]{Z.~Tymi\'{n}ski},
\author[WAR]{K.~Wi\'{s}niewski},
\author[ROSS]{D.~Wohlfarth},
\author[GSI,IMP]{Z.G.~Xiao},
\author[IMP]{H.S.~Xu},
\author[KUR]{I.~Yushmanov},
\author[ITEP]{A.~Zhilin}

(FOPI Collaboration)
\address[GSI]{GSI Helmholtzzentrum f\"ur Schwerionenforschung GmbH,
 Darmstadt, Germany}
\address[COLL]{FOPI}
\address[BUC]{National Institute for Nuclear Physics and Engineering,
Bucharest,Romania}
\address[BUD]{Central Research Institute for Physics, Budapest, Hungary}
\address[CLER]{Clermont Universit\'e, Universit\'e Blaise Pascal, CNRS/IN2P3,
 Laboratoire de Physique Corpusculaire, Clermont-Ferrand, France}
\address[ROSS]{ Institut f\"ur Strahlenphysik,
 Helmholtz-Zentrum Dresden-Rossendorf,
Dresden, Germany}
\address[HEID]{Physikalisches Institut der Universit\"at Heidelberg,Heidelberg,
Germany}
\address[ITEP]{Institute for Theoretical and Experimental Physics,
Moscow,Russia}
\address[KUR]{Kurchatov Institute, Moscow, Russia}
\address[IMP]{Institute of Modern Physics, Chinese Academy of Sciences,
Lanzhou, China}
\address[KOR]{Korea University, Seoul, South Korea}
\address[STRAS] {Institut Pluridisciplinaire Hubert Curien, IN2P3-CNRS,
Universit\'e de Strasbourg, Strasbourg, France}
\address[WAR]{Institute of Experimental Physics, University of Warsaw, Poland}
\address[ZAG]{Rudjer Boskovic Institute, Zagreb, Croatia}

\thanks[info]{Email:~W.Reisdorf@gsi.de}

\begin{abstract}
Using the large acceptance apparatus FOPI, we study central 
and semi-central collisions in
the reactions (energies in $A$ GeV are given in parentheses):
$^{40}$Ca+$^{40}$Ca (0.4, 0.6, 0.8, 1.0, 1.5, 1.93),
$^{58}$Ni+$^{58}$Ni (0.15, 0.25, 0.4),
$^{96}$Ru+$^{96}$Ru (0.4, 1.0, 1.5),
$^{96}$Zr+$^{96}$Zr (0.4, 1.0, 1.5),
$^{129}$Xe+CsI (0.15, 0.25, 0.4),
$^{197}$Au+$^{197}$Au (0.09, 0.12, 0.15, 0.25, 0.4, 0.6, 0.8, 1.0, 1.2, 1.5).
The observables include directed and elliptic flow.
The data are compared to earlier data where possible and to transport model
simulations.
A stiff nuclear equation of state is found to be incompatible with the data.
Evidence for extra-repulsion of neutrons in compressed asymmetric matter is
found.

\end{abstract}

\end{frontmatter}

\section{Introduction}\label{intro}
Over the past nearly three decades the phenomenon of azimuthally asymmetric
particle emission in high energy heavy ion reactions,  called 'flow' in most
of the relevant literature,
has emerged as a general phenomenon observed over a very wide range of 
energies and  systems.
Ever since flow in heavy ion collisions was discovered
 \cite{gustafsson84,renfordt84} it was believed that basic information
on the nuclear equation of state, EoS, and other hot and compressed nuclear
matter properties, such as viscosity, could be inferred from the
data with use of hydrodynamical approaches \cite{scheid74,stoecker86,clare86}.
This approach continues to be important in the highest energy regimes
available today, see  recent examples in \cite{hsong11,niemi11}.
However, special measures have to be taken to handle both the 
initial and the final ('freeze-out') phases of the collision which by
definition are far from the close-to-local equilibrium situation assumed in
hydrodynamical theories. 

In the SIS energy range ($0.1A-2.0A$ GeV) which we cover here and which is
expected to access nuclear matter up to 2.5 times saturation density, it was
recognized early that microscopic transport theory that relaxes the local
equilibrium assumption was necessary.
Such dynamical theories suitable for heavy ion reactions were developed in
the eighties \cite{bertsch88} and continue to be improved
\cite{blaettel93,cassing00,aichelin91,hartnack98,yzhang05,bass98,qfli11,colonna04,santini05,baran05,gaitanos05,lwchen04,bali04,danielewicz92,danielewicz00,buss11}.
Since then many works confronting flow data with transport code simulations
have been done.
The situation towards the end of the nineties was reviewed in 
\cite{reisdorf97r,herrmann99}.
More recently a courageous attempt to present an EoS constrained by
heavy ion
flow data has found wide-spread attention \cite{danielewicz02}.
However the authors state that they 'did not find a unique formulation of
the EoS that reproduces all the data'.
Other published experimental works on flow comparing data with various
transport codes come to similar conclusions 
\cite{partlan95,andronic05}:
not only does one fail to reproduce all the data with a given code,
there is unfortunately so far no really satisfactory agreement between
the various theoretical approaches.
Observables other than conventional flow parameterizations have been
proposed with some success.
In \cite{swang96,stoicea04} the azimuthal modulations of radial flow
were confronted with simulations using different codes. 
While no definite conclusions were forwarded in \cite{swang96} using
a quantum molecular dynamics code \cite{peilert89}, 
a clear preference for a 'soft' EoS was found \cite{stoicea04} using a 
Boltzmann-Uehling-Uhlenbeck (BUU) transport code 
\cite{danielewicz92,danielewicz00}.
A drawback of \cite{stoicea04} was that conclusions were drawn from one code
using
one observable at one incident energy.
A similar limitation holds for subthreshold kaon production 
\cite{sturm01,foerster07} which shows sensitivity
\cite{fuchs06,hartnack06} only in a narrow beam energy interval
just below threshold and above sufficient detection probability.

It is clear  that a large and rather complete data base on all the observable
aspects of heavy ion reactions spanning large energy ranges in sufficiently
dense steps is needed in order to be able to perform a stringent testing
of the quality of a given transport code.
Only then can reliable and convincing conclusions on basic nuclear properties
be derived.
Recently our Collaboration, using the large acceptance apparatus FOPI
 \cite{gobbi93,ritman95} at the SIS accelerator in Darmstadt, Germany,
has started a large effort to complement earlier, more specfic, works
 (see some references
given in subsequent sections) by a  systematic investigation
encompassing 25 system-energies.

A first paper written in this spirit \cite{reisdorf07} has offered a detailed
overview of pion production. 
Pions are in this energy regime by far the most copiously
created particle.
This study contained also information on the pion's longitudinal and transverse
rapidity ditributions, as well as the two forms of azimuthally asymmetric
pionic flow: directed and elliptic
 (the definition will be recalled in the next section).
In a second publication \cite{reisdorf10} we have established a systematics
for identified light charged particles in the most central collisions,
studying the freeze-out chemistry, the degree of stopping and radial flow.
The present work, finally, will present a systematics of directed 
and elliptic flow of these identified light particles in the same, SIS,
 energy regime. 

In this energy regime the passing time and the expansion time are comparable.
While this offers a possibility to obtain a handle on such basic properties as
the sound propagation velocity, it also has the consequence that rather complex
collision geometries result, as shown in early hydrodynamic simulations
\cite{clare86}.
The presence of relatively cold 'spectator' matter at the time where the
fireball is already expanding leads to cross talk between the various zones of
very different local temperature.
Also, quantum effects such as Fermi motion cannot be ignored as the initial
rapidity gaps between projectile and target are not extremely large.
The resulting flow patterns are rather different from those observed
\cite{RHIC05} for instance at the relativistic heavy ion collider RHIC,
although, formally, one is using the same parameterizations in terms of 
Fourier expansion coefficients \cite{voloshin96,poskanzer98},
 such as $v_1$ and $v_2$ to be defined in section \ref{apparatus}.
To cope with the complexity of the flow in heavy ion collisions at SIS we will
present an overview covering dependences on incident energy, system size and
centrality.
As some of the observables are strongly dependent on centrality, we
make an effort to define centrality in a  way
that allows a good match to centralities in simulations.
It will become clear furthermore that a separation of flow contributions by the 
mass and charge of the ejectiles covering differentially 
a large part of a sharply defined  phase space section  both
in transverse and longitudinal dimensions is needed to be able to draw unique
conclusions from comparison with transport model simulations.

We also devote some effort to a more recent extension of this physics to
include the isospin degree of freedom \cite{baran05,bali01} by varying the 
system's isospin
on one hand, and by measuring the flow differences of the isospin pair
$^3$H and $^3$He on the other hand
 (the $\pi^-/\pi^+$ pair was studied in \cite{reisdorf07}).

In the sequel we start by briefly describing the FOPI apparatus used in
this work, the
centrality selection and by defining some of the most used terms in section
\ref{apparatus}.
An overview of the experimental data on directed flow, section \ref{v1},
and elliptic flow, section \ref{v2}, follows.
An orientational confrontation of our data with transport model calculations
will be given in section \ref{iqmd} and we will terminate with a summary of
the main conclusions.


\section{Apparatus and data analysis}\label{apparatus}
The  experiments were performed at the heavy ion accelerator SIS of
GSI/Darmstadt using the large acceptance FOPI detector \cite{gobbi93,ritman95}.
The equipment and analysis methods in the present work largely correspond to
those used in two of our earlier publications 
on pion emission \cite{reisdorf07} and central collision observables
\cite{reisdorf10} in the same energy regime.
We therefore confine ourselves to mentioning a few points that should be useful
in making the present work reasonably self-contained.
A total of 25 system-energies are analysed for this work (energies 
per nucleon, $E/u$, in  GeV
are given in parentheses):
$^{40}$Ca+$^{40}$Ca (0.4, 0.,6 0.8, 1.0, 1.5, 1.93),
$^{58}$Ni+$^{58}$Ni (0.15, 0.25),
$^{129}$Xe+CsI (0.15, 0.25),
$^{96}$Ru+$^{96}$Ru (0.4, 1.0, 1.5),
$^{96}$Zr+$^{96}$Zr (0.4, 1.5),
$^{197}$Au+$^{197}$Au (0.09, 0.12, 0.15, 0.25, 0.4, 0.6, 0.8, 1.0, 1.2, 1.5).

Two  setups were used for the low energy data
($E/u < 0.4$ GeV) and the high energy data.
For the latter,
particle tracking  and energy loss determination were
done using two drift chambers, the CDC (covering polar angles
between $35^\circ$ and $135^\circ$) and
the Helitron ($9^\circ-26^\circ$), both
located inside a superconducting solenoid operated at a magnetic field of
0.6\,T.
A set of scintillator arrays,
Zero Degree Detector $(1.2^\circ-7^\circ)$,
 Plastic Wall $(7^\circ-30^\circ)$,
 and Barrel $(42^\circ-120^\circ)$,
 allowed us to measure the time of flight
and also the energy loss.

In the low energy run  the Helitron drift chamber
 was replaced by
 a set of gas ionization chambers (Parabola) installed between 
the solenoid magnet (enclosing the CDC) and the Plastic Wall.
More technical details can be found in Refs. \cite{gobbi93,andronic01}. 
 The combination of the two setups allowed us to identify pions and 
light charged particles (LCP, i.e. isotopes of hydrogen and helium)
 over a broad sector of phase space and to cover
a  large range of incident $E/u$ since there is approximately a factor of
twenty between the lowest and the highest energy. 

As in our earlier work,
collision centrality selection was obtained by binning 
distributions of either the detected
charge particle multiplicity, {\it MUL}, or the ratio of total transverse to
longitudinal kinetic energies in the center-of-mass (c.m.) system, {\it ERAT}.
 To avoid autocorrelations we have always excluded the particle
of interest, for which we build up spectra, from the definition of
{\it ERAT}.
We estimate  the impact parameter $b$
from the measured differential cross sections for the {\it ERAT} or the
multiplicity distributions, using a geometrical sharp-cut approximation.
More detailed discussions of the centrality selection methods used here
can be found in Refs. \cite{andronic06,reisdorf97}.
We characterize the 
centrality by the interval of
the scaled impact parameter $b_0$ defined by $b_0=b/b_{max}$,
 taking $b_{max} = 1.15 (A_{P}^{1/3} + A_{T}^{1/3})$~fm. 
This scaling is useful when comparing systems of different size.
In this work we present data for four centrality bins:
$b_0<0.15$, $b_0<0.25$, $0.25<b_0<0.45$ and $0.45<b_0<0.55$.
The total number of registered events was typically $(0.3-1.5)\times 10^6$
and was triggered by a multiplicity filter.
Some minimum bias events were registered at a lower rate.

\begin{figure}
\epsfig{file=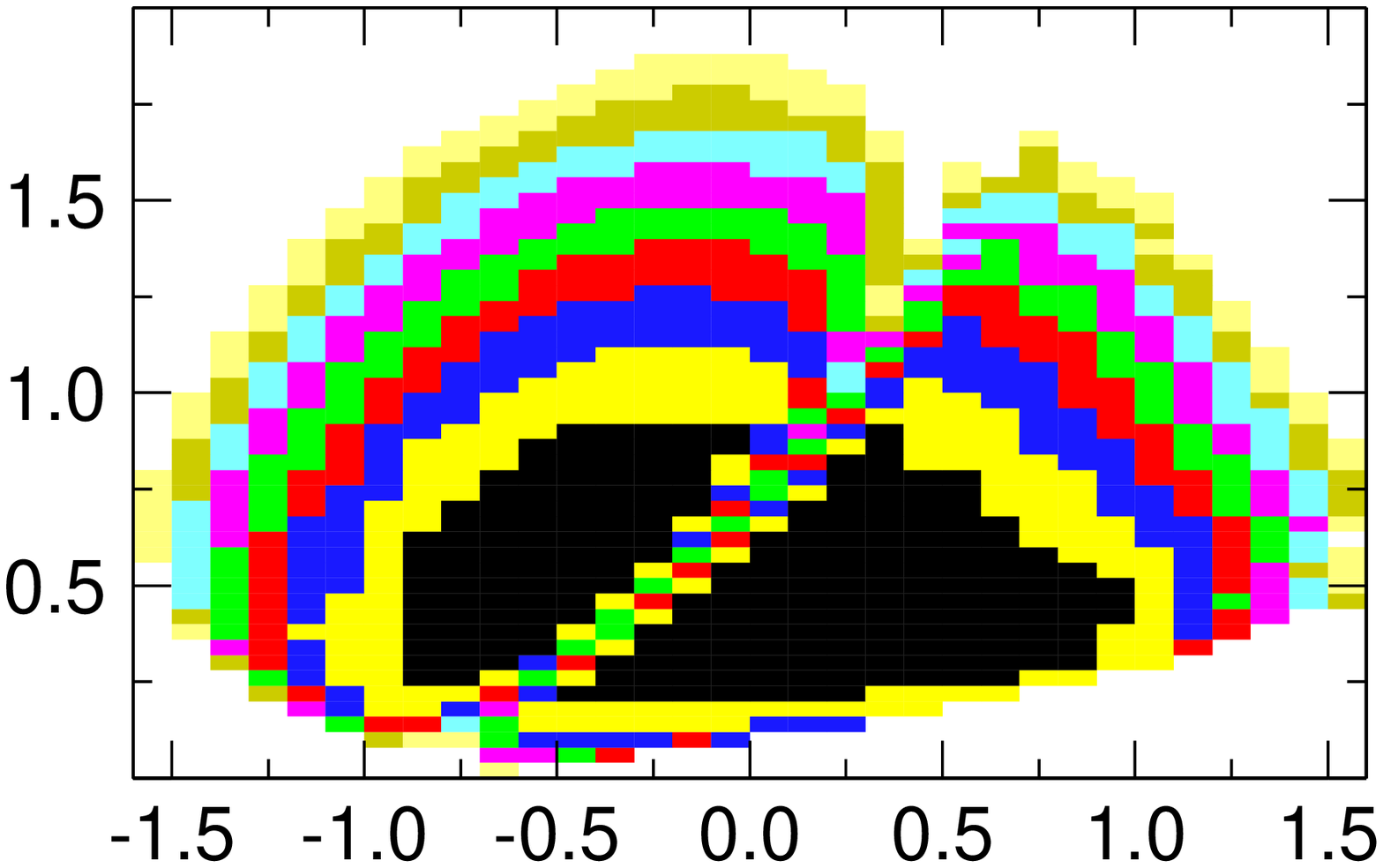,width=75mm}
\epsfig{file=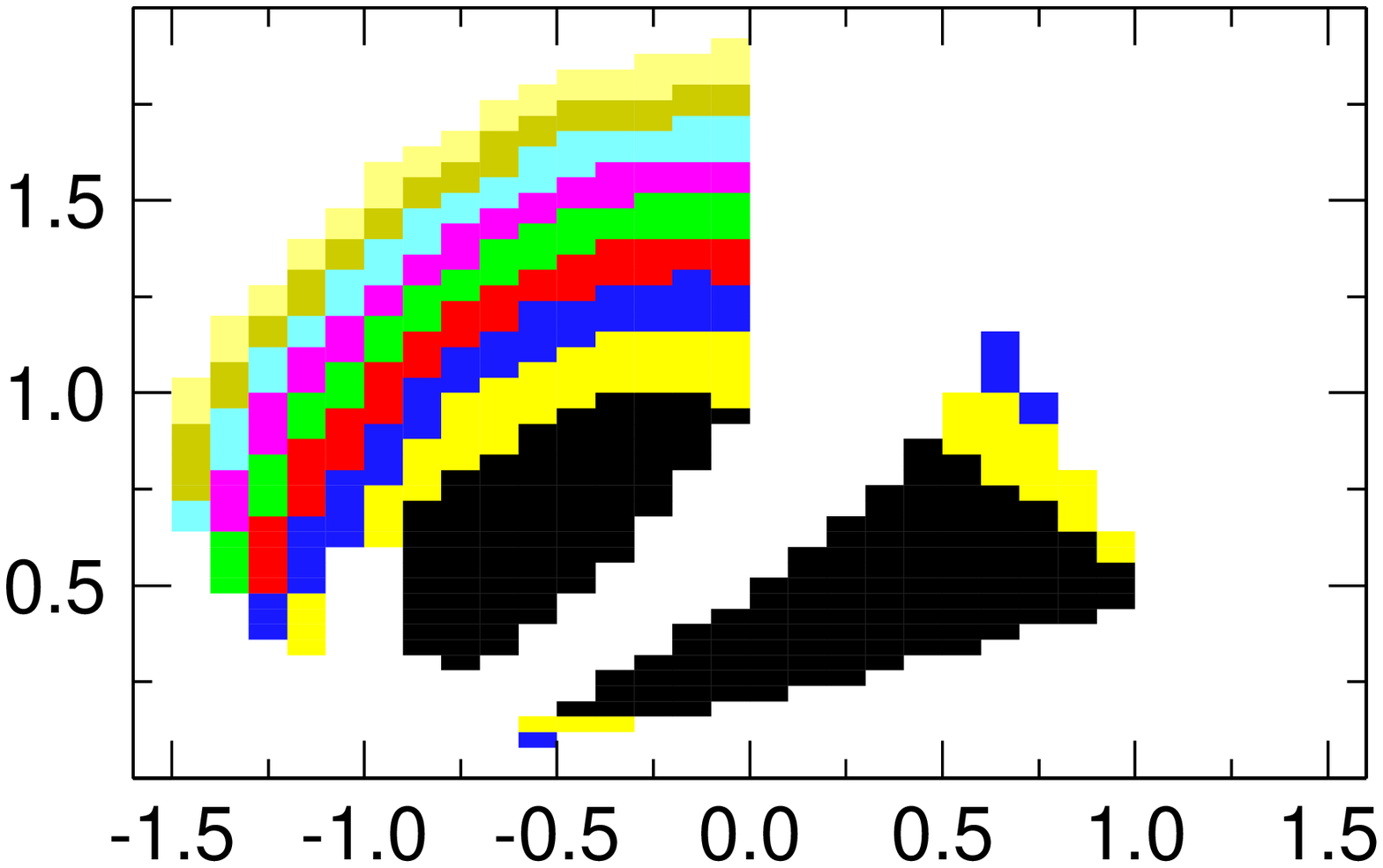,width=75mm}

\epsfig{file=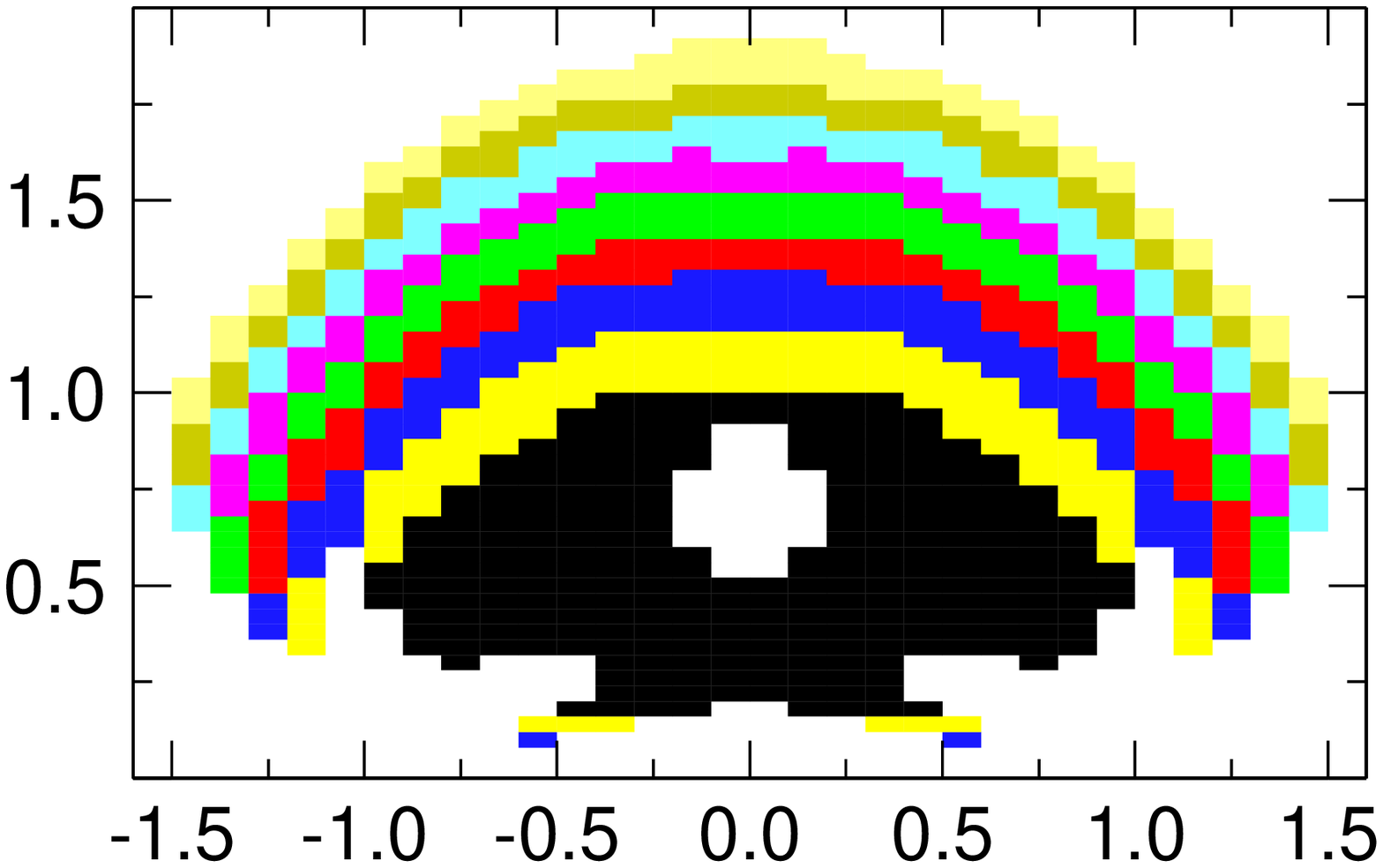,width=75mm}
\epsfig{file=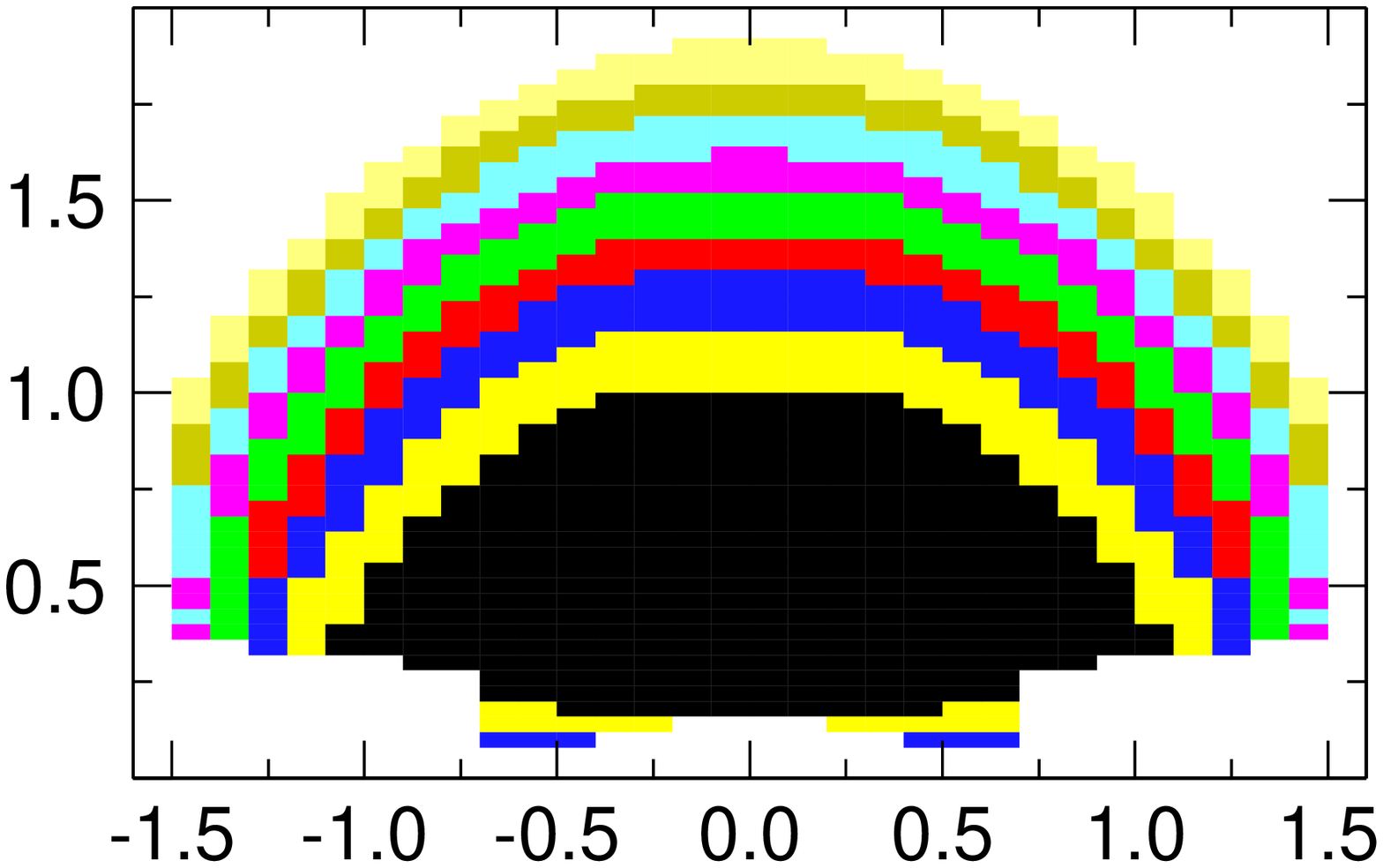,width=75mm}
\caption{
%
Two-dimensional yield distributions (contours differ by a factor 1.5) of
protons in collisions of Au+Au with $1A$ GeV beam energy and centrality
$0.25<b_0<0.45$. The ordinate is the scaled momentum $u_{t0}$, the abscissa 
is the scaled rapidity $y_0$.
Upper left panel: data after matching of various detectors and subtraction
of pionic contributions (in the forward part).
Upper right panel: data after applying sharp cuts to remove problem areas
and edge effects.
Lower left panel: data after symmetrisation of the forward/backward
hemispheres. Lower right panel: data after bi-dimensional smoothening and
closing of some gaps by interpolation.
}
\label{uty-au1000c2Z1A1}
\end{figure}
\begin{figure}
\epsfig{file=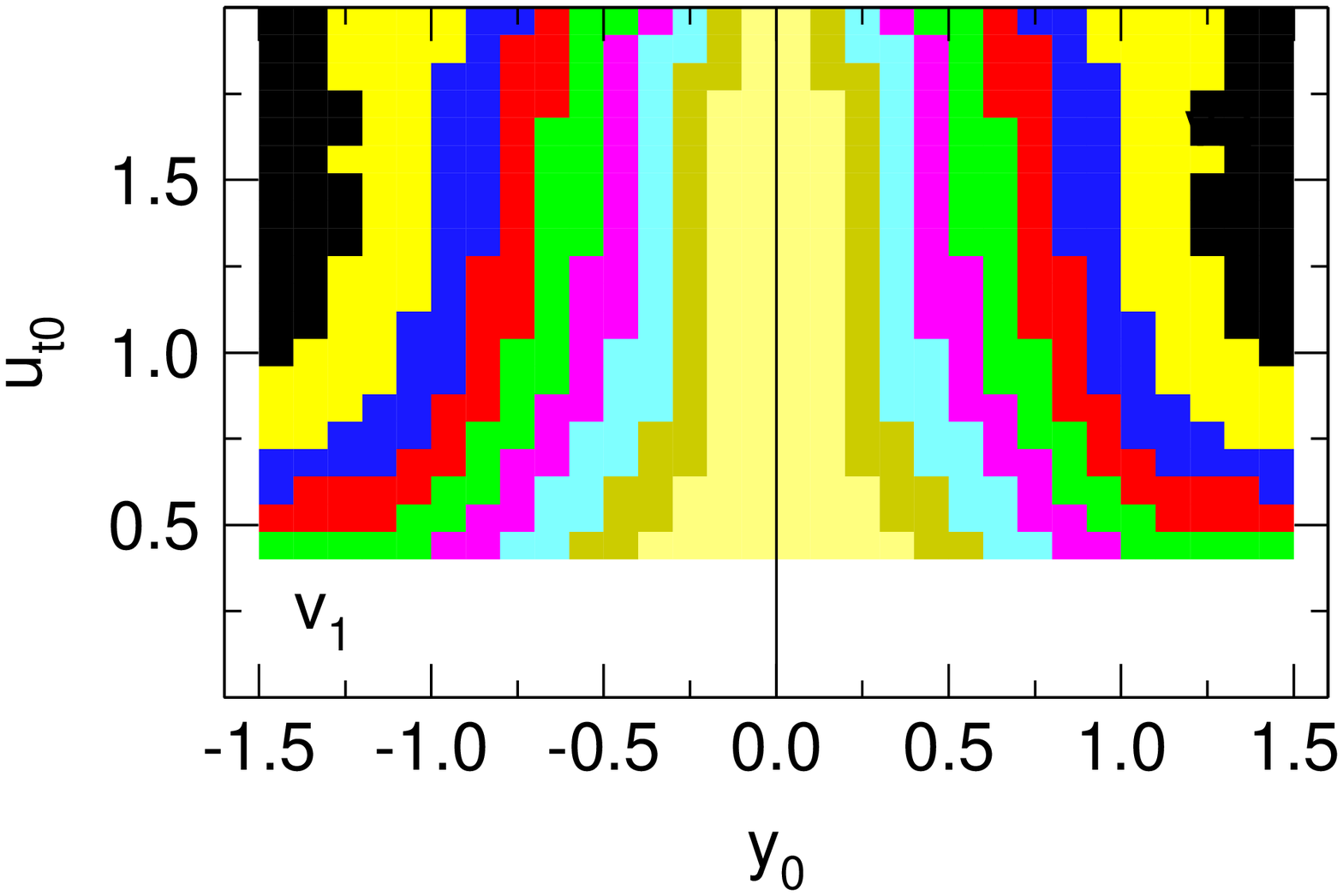,width=80mm}
\epsfig{file=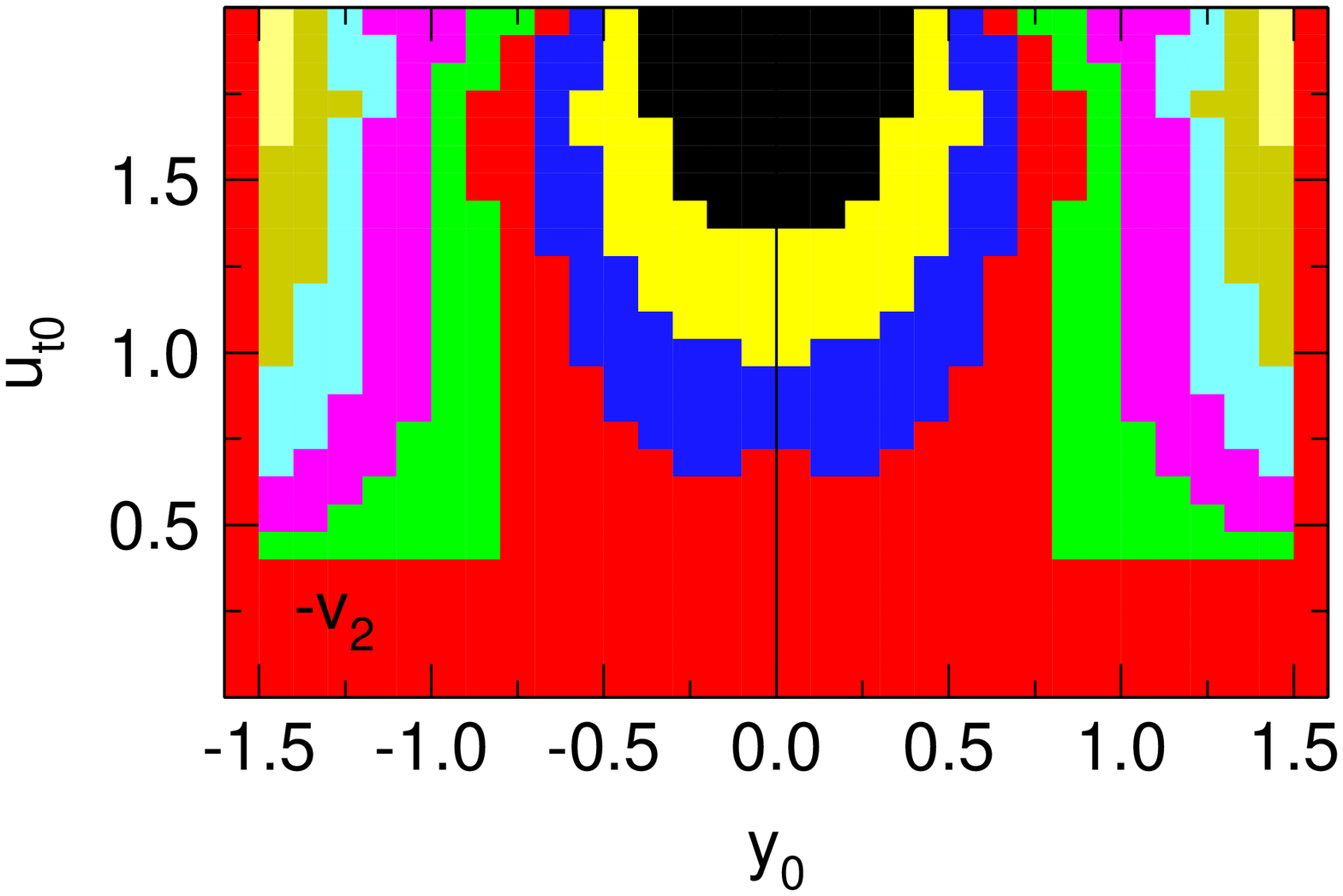,width=80mm}
\caption{
%
Two-dimensional distributions  of directed flow (left) and
elliptic flow (right) for
protons in collisions of Au+Au with $1A$ GeV beam energy and centrality
$0.25<b_0<0.45$. The ordinate is the scaled momentum $u_{t0}$, the abscissa 
is the scaled rapidity $y_0$.
The data are limited to $u_{t0}>0.4$.
In the left panel the colours delimit nine linear cuts varying from a
maximum $v_1=1.3$ to zero (at mid-rapidity). For technical reasons
the values in the backward hemisphere have been sign-inverted leading
to a reflection symmetric (on the $y_0=0$ axis) topology.
In the right panel the colours for $-v_2(y_0,u_{t0})$ again delimit nine
cuts from a maximum around 0.31 reached for high momenta at mid-rapidity 
to a minimum -0.46 at
$|y_0|=1.5$. The red colour marks values close to zero separating
areas of differing sign.    
}
\label{vuty-au1000c2Z1A1}
\end{figure}

Figures \ref{uty-au1000c2Z1A1} and \ref{vuty-au1000c2Z1A1} give an idea of
 the yield and flow data covered by our
apparatus for mass and charge identified LCP in the high energy run.
In these plots the abscissa is the longitudinal (beam axis) rapidity $y$
in the $c.o.m.$ reference system and the ordinate is the transverse
(spatial) component $t$ of the four-velocity $u$, given by $u_t=\beta_t\gamma$.
The 3-vector $\vec{\beta}$ is the velocity in units of the light
velocity and $\gamma=1/\sqrt{1-\beta^2}$.
Throughout we use scaled units $y_0=y/y_p$ and $u_{t0}=u_t/u_p$,
with $u_p=\beta_p \gamma_p$, the index p referring to the incident projectile
in the $c.o.m.$.
In these units the initial target-projectile rapidity gap always extends from
$y_0=-1$ to $y_0=1$.

In Fig. \ref{uty-au1000c2Z1A1}
the upper left panel shows for $1A$ GeV Au on Au collisions the proton yield
distribution in the $(y_0$ vs $u_{t0})$ plane.
The centrality is $0.25<b_0<0.45$.
Two separate parts are visible: a predominantly forward $(y_0>0)$ part
obtained with use of the carefully matched Helitron and Plastic Wall
and a mainly backward part using the matched CDC and Barrel detector systems.
For particle types separated by charge only there is an additional forward
low-momentum sector available from the Zero Degree detector that is not
plotted here, see Ref. \cite{reisdorf10}.
While for the centrality selection the complete detector system is used,
the flow analysis proceeds as follows (as illustrated in the various panels of
Fig. \ref{uty-au1000c2Z1A1}): To remove problem areas and edge effects visible in the
upper left panel we apply rather restrictive sharp cuts as shown in
the upper right panel. 
Taking advantage of the symmetry of the system we then apply reflection
symmetry on the $y_0=0$ axis, lower left panel and, finally, after
some smoothing \cite{reisdorf04b,reisdorf10} we fill some of the small gaps
that remained by two-dimensional interpolation.
The resulting two-dimensional distribution covers most of the phase space
except for low transverse momenta.
It is in this phase space that we are able then to determine the flow
fields describing the azimuthal emission dependences shown in 
Fig. \ref{vuty-au1000c2Z1A1}. 

\vspace{4mm}

Owing to collective flow phenomena, discovered experimentally in 
1984~\cite{gustafsson84,renfordt84},
it is possible to reconstruct the reaction plane 
event-by-event and hence to study
azimuthal correlations relative to that plane.
We have used the transverse momentum method~\cite{danielewicz85}
including all particles
identified outside the midrapidity interval $|y_0|<0.3$ and excluding
identified pions.
Pions, in this context, were only found to be important at the highest
SIS energy, but then contributed primarily to the reaction plane
fluctuation.

We use the well established parameterization \cite{voloshin96,poskanzer98}

\begin{equation}
 u = (\gamma , \vec{\beta}\gamma) \ ; \ \  u_t = \beta_t\gamma 
\end{equation}
\begin{equation}
 \frac{dN}{u_t du_t dy d\phi} = v_0 [1 + 2v_1 \cos(\phi) + 2v_2
\cos(2\phi)] 
\end{equation}
\begin{equation}
 v_0 = v_0(y,u_t) \  ; \ \  v_1 = v_1(y,u_t) \ ; \ \  v_2 = v_2(y,u_t) 
\end{equation}
\begin{equation}
 v_1 = \left<\frac{p_x}{p_t}\right> = \left <cos(\phi) \right >\ ;
 \ \  v_2 = \left <\left( {\frac{p_x}{p_t}}\right)^2 -
  {\left(\frac{p_y}{p_t}\right)}^2 \right > = \left <cos(2\phi)\right >
\end{equation}
\textBlack
where $\phi$ is the azimuth with respect to the reaction plane
and where angle brackets indicate averaging over events (of a specific class).
The Fourier expansion is truncated, so that only three
parameters, $v_0,v_1$ and $v_2$, are used to describe the 
'third dimension' for  fixed intervals of rapidity and transverse
momentum. 
The adequacy of this truncation for the $1A$ GeV beam energy regime
was already noted in Ref. \cite{gutbrod90f} where this kind of
Fourier analysis of azimuhal distributions was probably first used.
Due to finite-number fluctuations the apparent reaction plane determined
experimentally does not coincide event-wise with the true reaction plane,
causing an underestimation of the deduced coefficients $v_1$ and $v_2$
which, however, can be corrected by dividing events into randomly chosen
 sub-events \cite{danielewicz85,barrette97}:
 as explained in more detail in \cite{andronic01}, we have used the 
method of Ollitrault \cite{ollitrault98} to achieve this.
The correction factors (for the {\it ERAT} selection) are listed
in a table that can be found in the Appendix.
The finite resolution of the azimuth determination is also the prime
reason why the measured higher Fourier components turn out to be rather small.

Alternatively to the three Fourier coefficients, one can introduce the
yields $Q_1$, $Q_2$, $Q_3$, $Q_4$ in the four azimuthal quadrants,
of which only three are independent (on the average over many events)
due to symmetry requirements

\begin{eqnarray}
 Q_2 & = &  Q_4 \\
 Q_{0} &= & Q_1 + Q_2 + Q_3 + Q_4 \\
 Q_{24} & = & Q_2 + Q_4 
\end{eqnarray}
More precisely, the flow axis is defined to be the median axis of
quadrant $Q_1$ which extends from $\phi = -45^{\circ}$ to $45^{\circ}$
relative to the reaction plane located by definition at $\phi=0$.
Thus, $Q_3$ is the 'antiflow' quadrant, while $Q_{2,4}$ are the out-of-plane
quadrants.

\Black{The two equivalent triplets}
\begin{center}
$v_0$, $v_1$, $v_2$ $\longleftrightarrow$   $Q_{0}$, $Q_1$, $Q_{24}$
\end{center}
\Black{are related by}
\begin{eqnarray}
Q_0 & = 2\pi\  v_0 & \approx 6.28\  v_0 \\
\frac{Q_1-Q_3}{Q_0} &= \frac{2 \sqrt {2}}{\pi}\  v_1 & \approx  0.900\  v_1  \\
 \frac{Q_{24}}{Q_0} - \frac{1}{2} &= -\frac{2}{\pi}\  v_2 & \approx
  -0.637\  v_2 
\end{eqnarray}
\textBlack
These relations show that $v_1$ is a dipole, while $v_2$ is a quadrupole
strength. Statistical (count rate)
errors can be deduced with use of elementary algebra.
The fact that $v_1$, as well as $v_2$, are found to be non-zero,
is generally called 'flow' in the literature and in particular
the first Fourier coefficient is  taken to be a measure of 'directed flow',
while the second Fourier coefficient has been dubbed 'elliptic flow'.
In the sequel we will occasionally also use instead of $v_1(y_0,u_{t0})$
(number weighted flow) the scaled transverse momentum ($u_{t0}$) weighted flow
$u_{x0}(y_0,u_{t0})$.
Most of the time we will plot elliptic flow with an inverted sign 
(i.e. $-v_2$) for reasons that will become clear later (see also the
interpretation in terms of 'quadrupole' strength given above).

In the language of particle-particle correlations, which avoids the
explicit use of a 'reaction plane' we are actually measuring the correlation
of one chosen identified particle with all the other (measured) ejectiles.
 We are not trying here to separate so called 'non-flow' from 'flow'
contributions or to establish two-, four- or more particle-particle
correlations.
We take the point of view that 'non-flow' correlations (
such as resonance and cluster decays, which are not
obviously disconnected from flow) should be properly taken into account in
a realistic microscopic transport simulation.
On the other hand multi-particle correlations are to a large degree
revealed in our flow data for multi-nucleon clusters.

\vspace{4mm}

In Fig. \ref{vuty-au1000c2Z1A1} we show the two-dimensional data deduced for the same reaction
as in Fig. \ref{uty-au1000c2Z1A1}  for directed flow (left) and elliptic flow (right).
The flat acceptance cut $u_{t0}>0.4$ preserves the backward/forward
(a)symmetry of the apparent ($v_1$) $v_2$.
In the low energy data we can use such a flat cut only if we
require $u_{t0}>0.8$ for mass and charge identified particles since
the Helitron was not available yet for this part of our experimental
campaigns. 
Small symmetry violations present in the original data have been corrected
out. 
The quadrant method is well suited  for this \cite{reisdorf07}.
The corrections turn out to be critical for small isospin differences
such as observed between $^3$H and $^3$He.
 We shall therefore come back to this later
in sections \ref{v1iso} and \ref{v2iso}.

A few general remarks on accuracy will be made here.
In the sequel, all our plotted data are given with errors that consist
of a quadratic addition of systematic uncertainties and statistical
fluctuations on the standard confidence level. 
If these errors are not visible they do not exceed the symbol sizes.
Generallly the {\it systematic} errors of $v_1$ and $v_2$ are most important,
 but there are exceptions that can be seen when the errors tend to blow up
for instance for high momenta, and for some of the Ca+Ca data which were
obtained with lesser statistics.
The estimates of systematic errors are based on observed forward-backward
symmetry violations (a specialty of FOPI is that there is some overlap
of measurements in the forward, resp. the backward hemisphers in the c.m.),
violations of the zero crossing of $v_1$ at mid-rapidity, discontinuities
when switching sub-detectors, deviations from the condition that the data for 
$Z=1$ should be equal to the yield weighted sum of the separated isotopes
(which require for identification additional sub-detectors) and incomplete
isotopic separation
 (primarily $^3$H /$^3$He, see also the discussion in section 3.5).
These errors may vary somewhat with the system-energy, typically we
have for $v_1$ errors of 0.01 (protons and deuterons), 0.015 (mass 3) and
0.02 (alpha's). For $v_2$ typical systematic errors are 0.005 (protons), 0.007
(deuterons) and 0.1 (mass $>2$).

\section{Directed flow}\label{v1}
Even under the contraints of symmetric heavy ion systems,
the flow fields $v_1$ and $v_2$ have  complex multidimensional dependences:
\begin{equation}
v_1=v_1(E/u, A_{sys}, Z_{sys}, b_0, A, Z, y_0, u_{t0})
\end{equation}
\begin{equation}
v_2=v_2(E/u, A_{sys}, Z_{sys}, b_0, A, Z, y_0, u_{t0}) 
\end{equation}
where $E/u$ is the incident beam energy per mass unit, $A_{sys}, Z_{sys}$
 are the system mass and charge, $A, Z$ is the ejectile composition.
As a consequence a complete systematics encompasses an enormous amount
of information.
It is out of question to present all this information in one readable paper:
the chosen one-dimensional cuts through the flow topology are necessarily
restrictive and perhaps even somewhat arbitrary. 
Nevertheless we believe that  theoretical simulations that
reproduce all of the presented results  are likely to reproduce also
the projections not shown here.
 Additional information can be found in some of our earlier
publications on nucleonic flow at SIS energies 
\cite{andronic05,stoicea04,andronic01,andronic06,ramillien95,bastid97,rami99,andronic01a,andronic03,bastid04,bastid05}.

In the sequel we will first treat the centrality $(b_0)$ dependence, then
the ejectile mass $(A)$ and the system size $(A_{sys})$ dependence,
the variation with incident energy $(E/u)$ and, finally, the isospin
$(Z/A)$, $(Z_{sys}/A_{sys})$ effects.
Although direct comparisons with data obtained with different apparatus are
exceedingly difficult owing to different apparatus biases and phase space
cuts we will also present a few such comparisons.
The next main section, on elliptic $(v_2)$ flow, will follow the same pattern.

\subsection{Centrality dependences}\label{v1c}

\begin{figure}
\hspace{\fill}
\epsfig{file=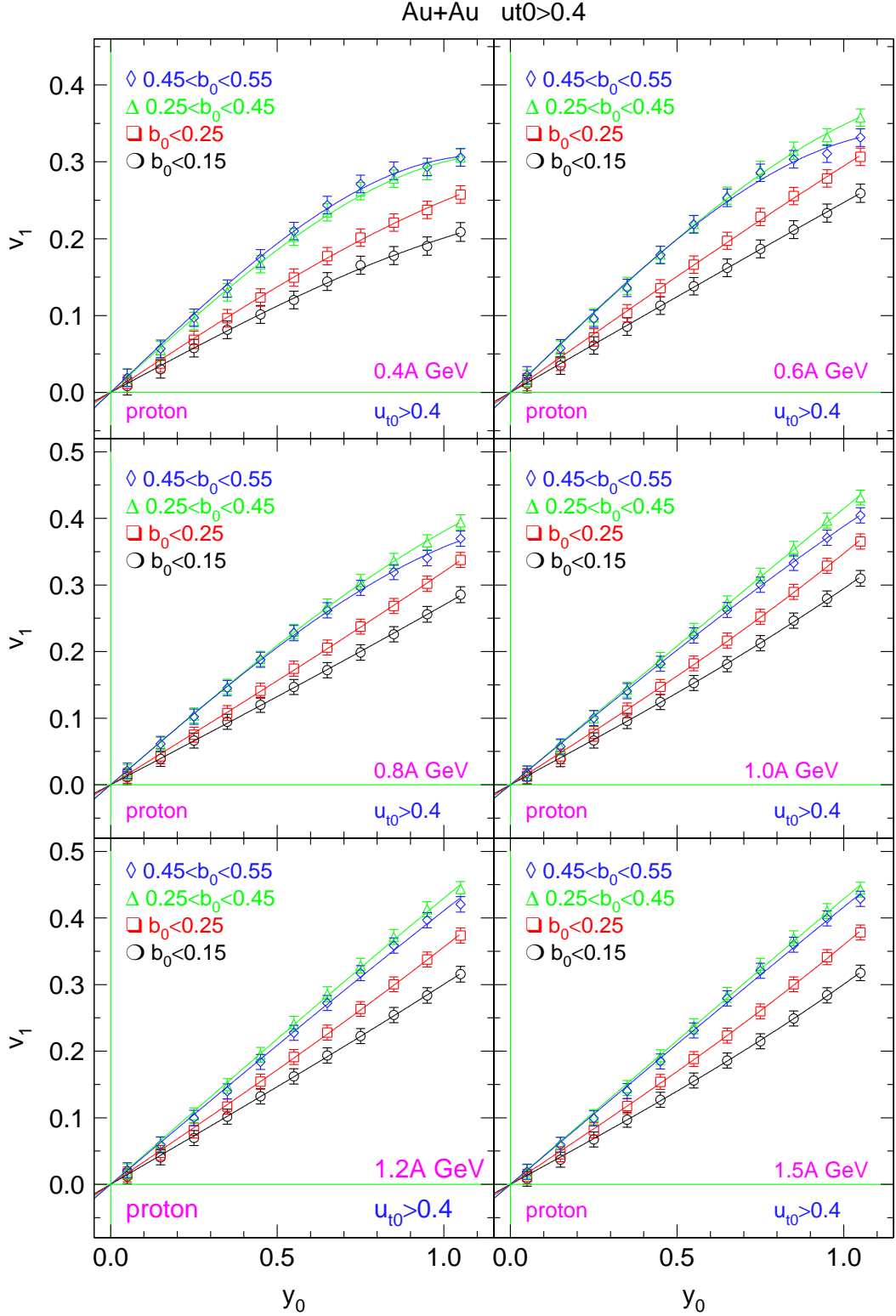,width=150mm}
\hspace{\fill}
\caption{
%
Directed flow $v_1$ of protons in Au+Au collisions for different
indicated centrality ranges and incident energies.
The smooth curves are least squares fits to
 $v_1=v_{11}\cdot y_0 + v_{13}\cdot y_0^3$.
Transverse 4-velocities $u_{t0}$ below 0.4 are cut off.
}
\label{v1-au400Z1A1-grqcv-6}
\end{figure}

In Fig. \ref{v1-au400Z1A1-grqcv-6} we show a sample of our $v_1(y_0)$ data 
integrated over $u_{t0}$, but with data for $u_{t0}<0.4$ cut off (see section
\ref{apparatus}). 
Only the forward hemisphere data are shown, the backward data are
antisymmetric.
It is known experimentally since some time \cite{doss86} that directed flow is
maximal at some intermediate centrality: for $b_0=0$ it must however converge
to zero for symmetry reasons, but does not do so in experiment because of
finite $b_0$ resolution.
Nevertheless, below $b_0 \approx 0.5$ the figure shows that the variation with
centrality is relatively modest as even the most central collisions
($b_0<0.15$ or $2.2\%$ of the sharp radius cross section) still are
associated with a significant signal.
The finite resolution effect is only part of the explanation, since the
other aspects of flow, $v_2$, see section \ref{v2}, and stopping
\cite{reisdorf10} are found to vary rather strongly with $b_0$.
One may conjecture that the geometrical decrease of asymmetry with decreasing
$b_0$ is to some degree compensated by the increase of pressure expected 
from a more efficient stopping in more central collisions \cite{reisdorf10}.
As we shall see in section \ref{iqmd}, sensitivity to the stiffness of the
EoS is highest for the most central $b_0$ interval.

Looking at the various panels of
 Fig. \ref{v1-au400Z1A1-grqcv-6} a striking feature is the large similarity of
the data in the shown energy interval ($0.4$ to $1.5A$ GeV) which covers the
high energy part of the present study.
This is in strong contrast to the behaviour at lower energy (see later).
The use of scaled rapidities is essential to see this.
In general, for proton flow, which is strongly 'smoothened' by superimposed 
random motion, the data can be well described by two-parameter fits
$v_1(y_0)=v_{11}\cdot y_0 + v_{13}\cdot y_0^3$ which are also plotted in the
figure.
These fits take care of smaller, more subtle, changes with incident energy
that are visible on closer inspection of the $v_1(y_0)$ data.

\begin{figure}
\hspace{\fill}
\epsfig{file=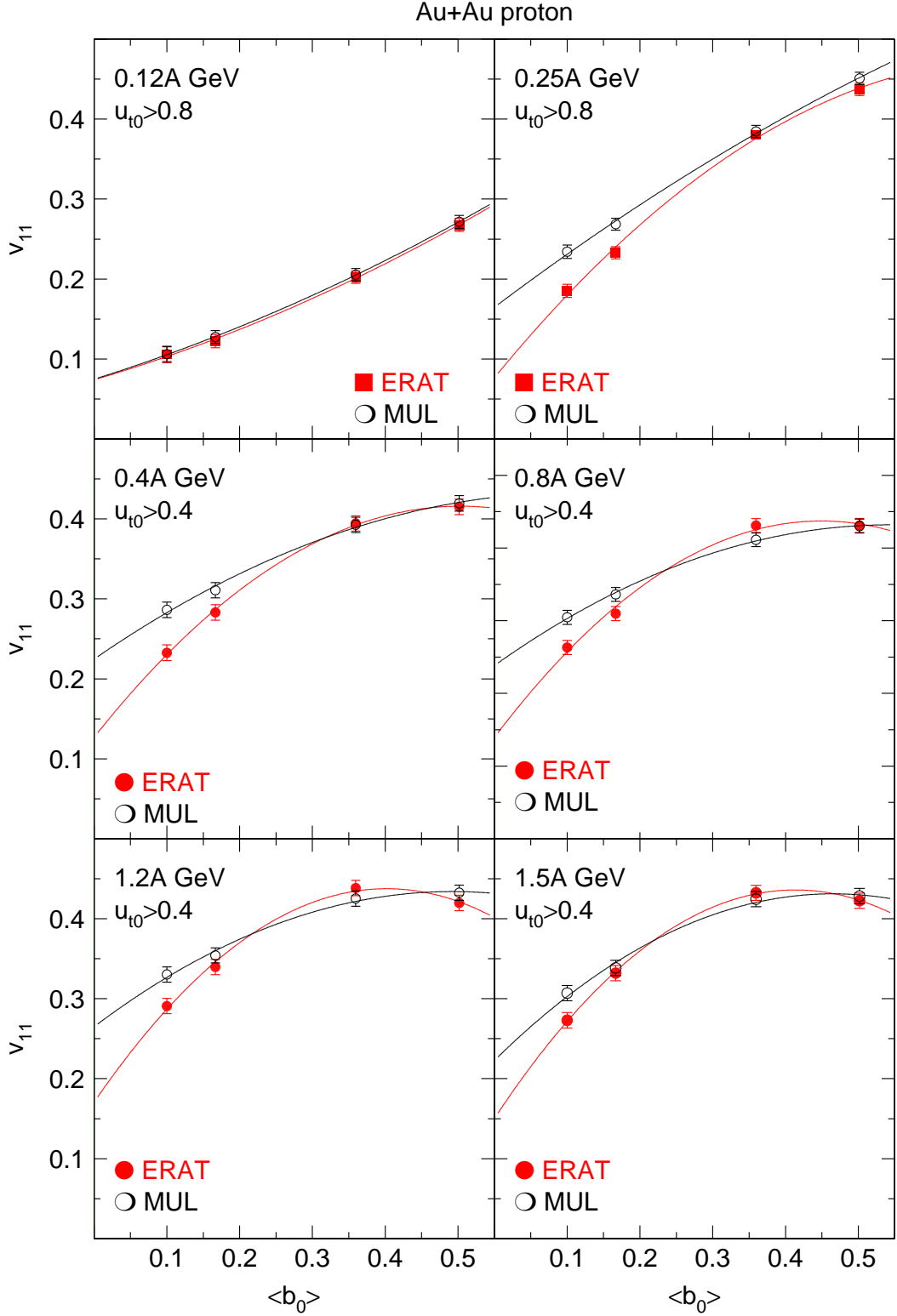,width=150mm}
\hspace{\fill}
\caption{
%
Mid-rapidity slopes $v_{11}$ of directed flow  of protons in Au+Au collisions
 for different indicated incident energies as function of average centrality.
Centrality selection is done using either {\it ERAT} (closed red circles) 
or charged particle multiplicity {\it MUL} (open black circles).
The smooth curves are least squares polynomial fits.
Transverse 4-velocities $u_{t0}$ below 0.4 are cut off, except for the two
lowest energies (uppermost panels) where the cutoff is 0.8 (as indicated).
}
\label{v1-au1500Z1A1-grqcv-b}
\end{figure}

Limiting ourselves to the information contained in the mid-rapidity slope
parameter $v_{11}$ 
\footnote{Such slope parameters for the transverse momentum weighted flow
were first introduced in Ref. \cite{doss86}}
we obtain in a compact form data comparing the centrality $(b_0)$ dependence
of multiplicity selected ({\it MUL}) with {\it ERAT} selected data,
 Fig. \ref{v1-au1500Z1A1-grqcv-b}.
While the two types of selection widely seem equivalent for sufficiently 
large $b_0$, it is remarkable that for $b_0 \rightarrow 0$ the {\it ERAT}
selections converge at all energies, except the lowest one ($0.12A$ GeV,
upper left panel) to a smaller value of $v_{11}$.
Rather than making a difference in $b_0$ resolution power responsible for
this, it is tempting to conclude that very high multiplicity events are less
central than events with maximal stopping {\it ERAT}.
The sorting according to {\it MUL} is based on the idea that the
participant/spectator ratio is increased with increasing multiplicity because
the hotter participants emit more particles.
However, for very central collisions, where little spectator matter is
present, this is no longer a sufficient impact parameter selection criterion.
There stronger stopping (larger {\it ERAT}) is correlated with higher achieved
densities.
This in turn leads in the expansion phase to more achieved cooling, hence more
clusterization.
This picture is supported by simulations with transport codes varying the
achieved density by varying the stiffness of the EoS, and by data varying with
system size the achieved stopping \cite{reisdorf10}.

Our observation also shows that sorting data according to maximal multiplicity
\cite{lehaut10} does not in the present energy regime lead to the same
stopping as sorting them according to  {\it ERAT} first, and then
  picking out a fixed cross section sample with the highest stopping
  \cite{reisdorf10}.
The latter procedure is not an 'autocorrelation' as maintained in
\cite{lehaut10}, but a legitimate procedure searching for maximal (not
'typical') stopping and hence, presumably, maximal achieved density.

\begin{figure}
\hspace{\fill}
\epsfig{file=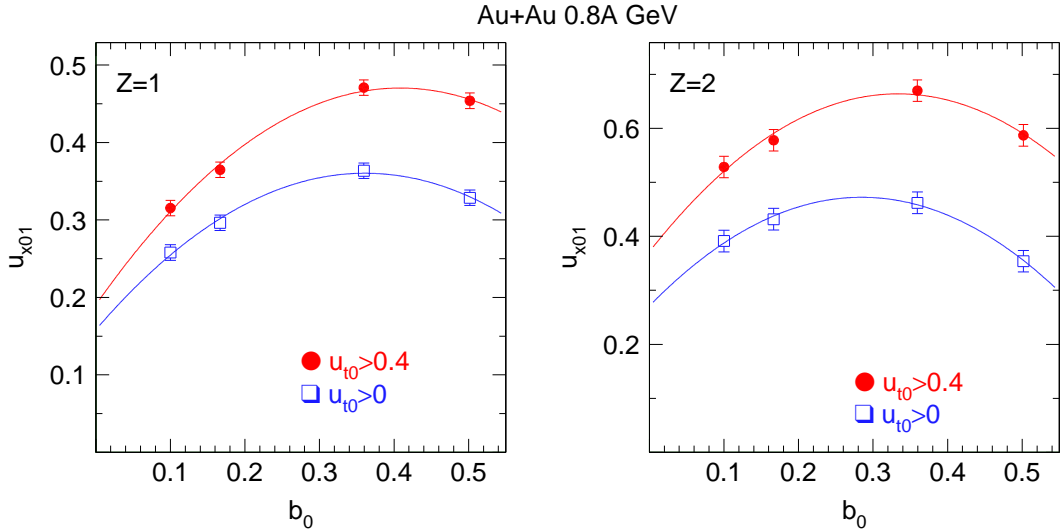,width=150mm}
\hspace{\fill}
\caption{
%
Directed flow of $Z=1$ (left) and $Z=2$ (right) fragments in Au+Au 
collisions at $0.8A$ GeV.
 Shown is the centrality ({\it ERAT}) dependence of the mid-rapidity
slope parameter $u_{x01}$ with (full red circles) and without 
transverse 4-velocities $u_{t0}$ below 0.4  cut off.
}
\label{ux-au800u0Z1A0-grcv}
\end{figure}

The low transverse momentum suppression that we make for charge {\it and}
mass identified particles is resulting from our apparatus limitations
(see section \ref{apparatus}).
As this limitation is not present in our charge (but not mass) separated data,
we can get an idea of the effect caused by transverse momentum cuts by looking
at charge one and two data in our study.
We show an example in Fig. \ref{ux-au800u0Z1A0-grcv} for Au on Au collisions at
$0.8A$ GeV (centrality $0.25<b_0<0.45$) both with and without the $u_{t0}>0.4$
constraint.
Here we choose the transverse momentum weighted flow $u_{x0}$
 (the corresponding mid-rapidity slope is dubbed $u_{x01}$) for reasons that
 will become clear when we try (subsection \ref{v1compa}) to compare with
older data in the literature.
As can be seen, flow is significantly larger when low transverse momenta are
suppressed.
The flow of particles that are only charge separated is difficult to
interpret because it is a superposition from various isotopes with very
different flow and hence is heavily influenced by the relative yields of these
isotopes (see section \ref{iqmd}, Fig. \ref{v1-au400Z1A0-grcv-c2}).
Therefore in the next subsection we take a look at the flow of individual
isotopes.

\subsection{LCP dependences of directed flow}\label{v1LCP}

\begin{figure}
\hspace{\fill}
\epsfig{file=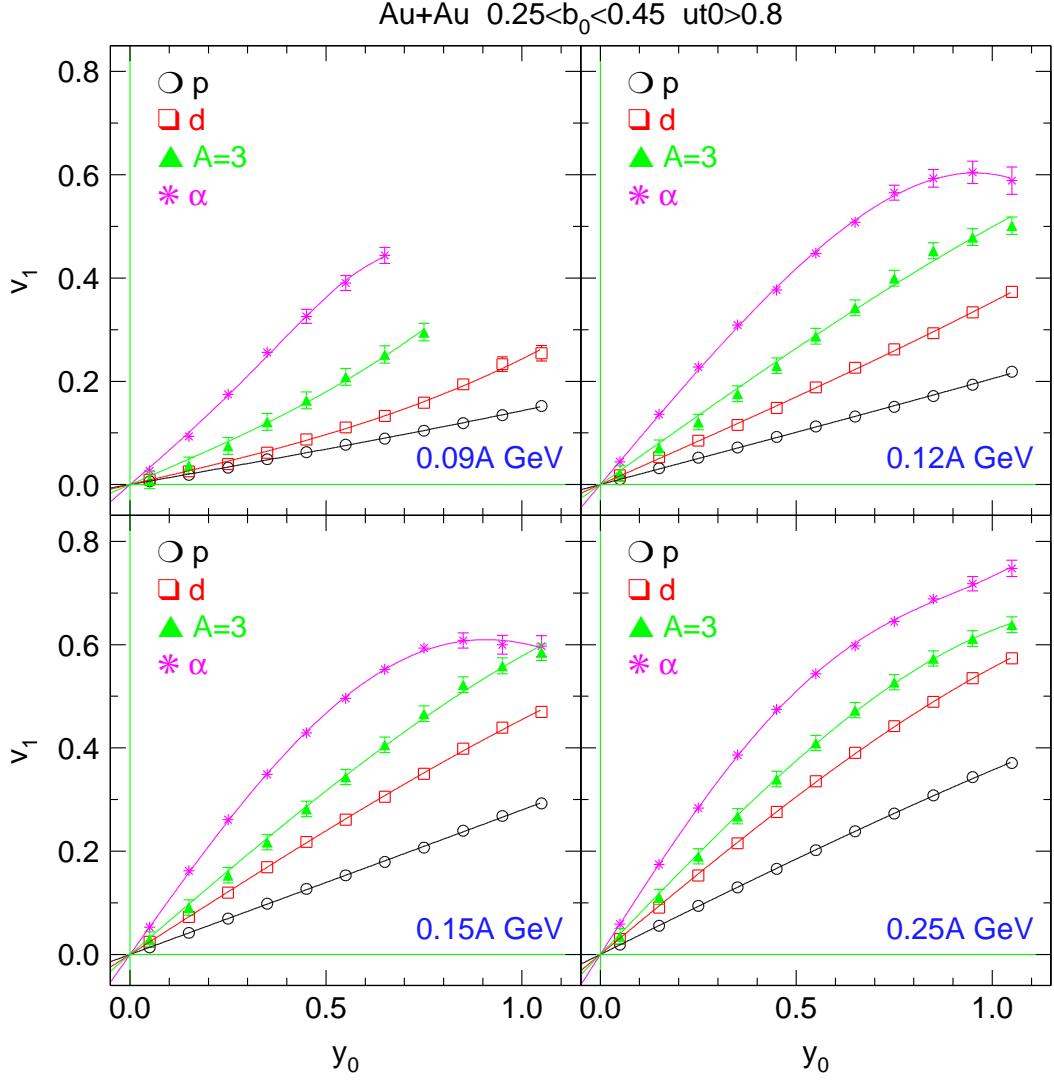,width=150mm}
\hspace{\fill}
\caption{
%
Directed flow $v_1$ of protons (black circles), deuterons (red  squares),
mass-three clusters (filled green triangles) and alpha-particles
(pink asteriks) in Au+Au collisions at $0.09A$, $0.12A$, $0.15A$,
 $0.25A$ GeV for
centrality $0.25<b_0<0.45$.
The mass-three flow is obtained as an equal-weight average of $^3$H and $^3$He.
Low transverse 4-velocities $u_{t0}< 0.8$ are cut out. 
}
\label{v1-auc2-jjscv}
\end{figure}

\begin{figure}
\hspace{\fill}
\epsfig{file=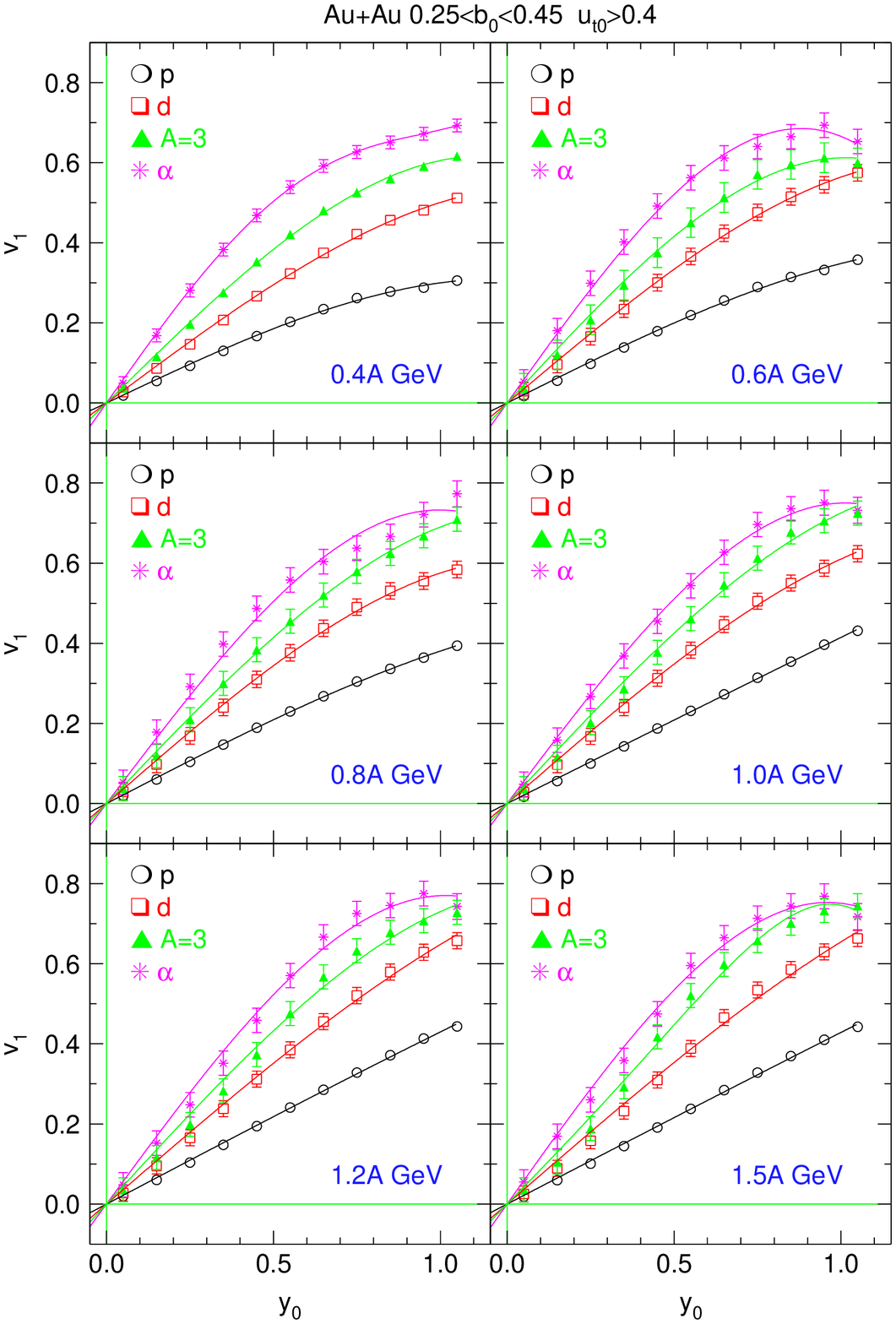,width=150mm}
\hspace{\fill}
\caption{
%
Directed flow $v_1$ of protons (black circles), deuterons (red  squares),
mass-three clusters (filled green triangles) and alpha-particles
(pink asterisks) in Au+Au collisions at $0.4A$, $0.6A$, $0.8A$, $1.0A$,
$1.2A$, $1.5A$ GeV for
centrality $0.25<b_0<0.45$.
The mass-three flow is obtained as an equal-weight average of $^3$H and $^3$He.
Low transverse 4-velocities $u_{t0}< 0.4$ are cut out. 
}
\label{v1-au1500c2-grqc14v-A}
\end{figure}

\begin{figure}
\hspace{\fill}
\epsfig{file=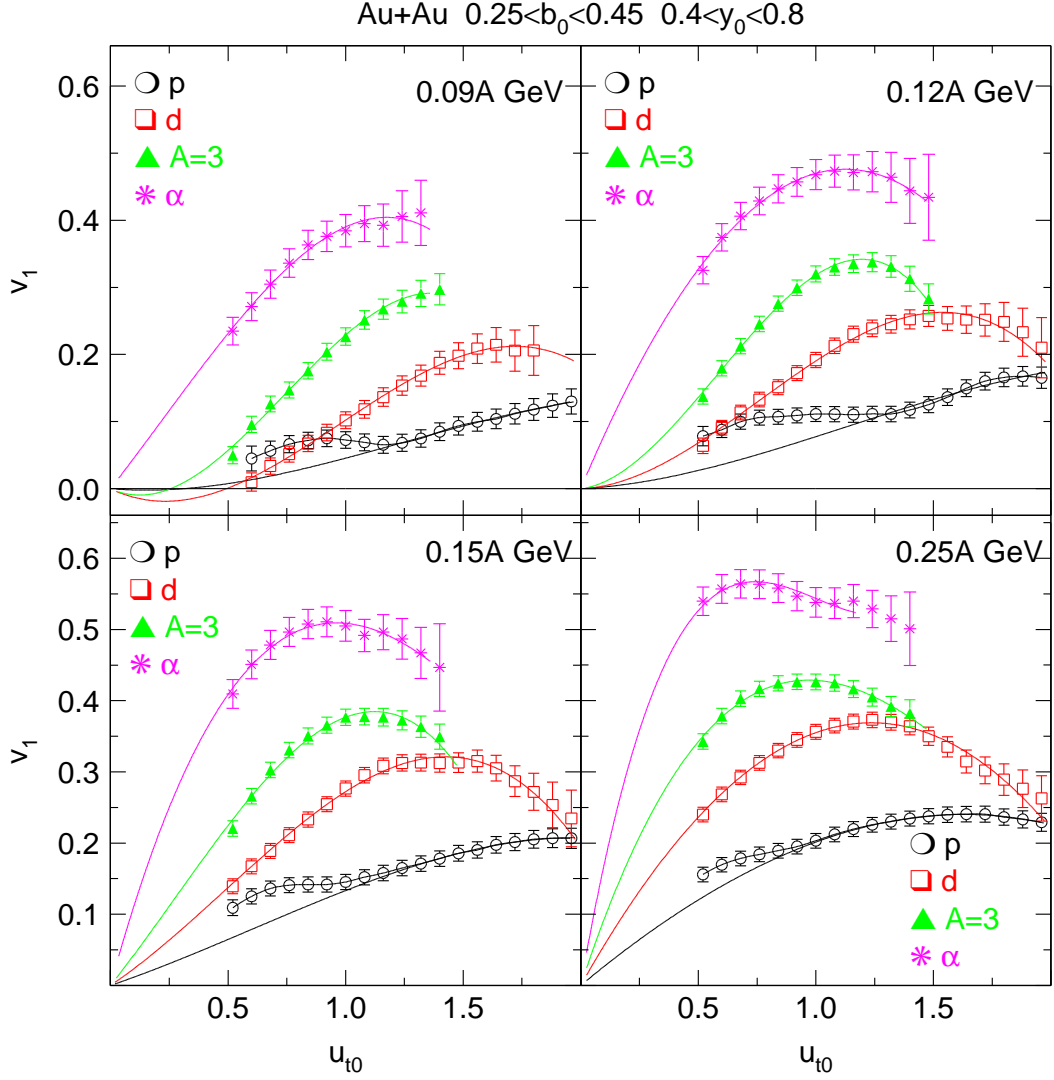,width=150mm}
\hspace{\fill}
\caption{
%
Directed flow $v_1(u_{t0})$ of protons (black circles), deuterons (red
squares), mass three fragments (green full triangles) and alpha particles
(pink asterisks)
in Au+Au collisions at beam energies of $0.09A$, $0.12A$, $0.15A$
and $0.25A$ GeV indicated in the various panels. 
The centrality is $0.25<b_0<0.45$.
The longitudinal rapidity is constrained to the interval $0.4<y_0<0.8$.
All smooth curves are fits to the data assuming 
$v_1(u_{t0})=v_{t11}\cdot u_{t0}+ v_{t12}\cdot u_{t0}^2
 + v_{t13}\cdot u_{t0}^3$.
These fits are used to extrapolate to $u_{t0}=0$.
For the proton data only the high momentum parts have been taken into
account for the fits.
}
\label{v1ut-au90c2-jjscv-A4}
\end{figure}

\begin{figure}
\hspace{\fill}
\epsfig{file=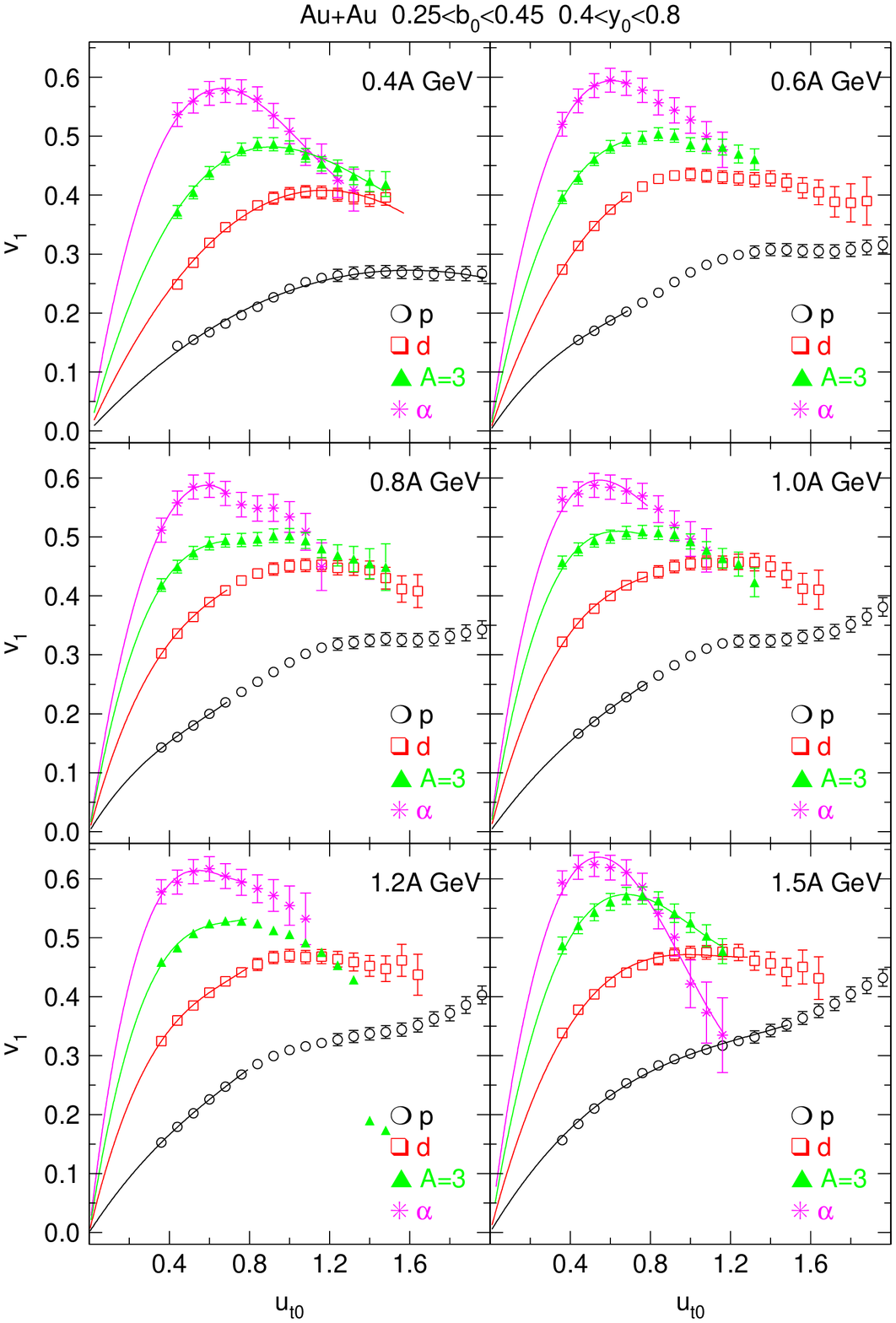,width=150mm}
\hspace{\fill}
\caption{
%
Directed flow $v_1(u_{t0})$ of protons (black circles), deuterons (red
squares), mass three fragments (green full triangles) and alpha particles
(pink asterisks)
in Au+Au collisions at beam energies of $0.4A$, $0.6A$, $0.8A$, $1.0A$, $1.2A$
and $1.5A$ GeV indicated in the various panels. 
The centrality is $0.25<b_0<0.45$.
The longitudinal rapidity is constrained to the interval $0.4<y_0<0.8$.
All smooth curves are fits to the data assuming 
$v_1(u_{t0})=v_{t11}\cdot u_{t0}+ v_{t12}\cdot u_{t0}^2
 + v_{t13}\cdot u_{t0}^3$.
These fits are used to extrapolate to $u_{t0}=0$.
}
\label{v1ut-au400c2-grqcv-6} 
\end{figure}

We show the ejectile mass dependence of directed flow for centrality 
$0.25<b_0<0.45$ as a function of rapidity $y_0$ (integrated starting from 
$u_{t0}>0.8$ for the low energy runs up to $E=0.25A$ GeV and 
$u_{t0}>0.4$ for the higher beam energies) in Figs. \ref{v1-auc2-jjscv} 
and \ref{v1-au1500c2-grqc14v-A} and as function of transverse 
four-velocity $u_{t0}$
in a rapidity interval $0.4<y_0<0.8$ in Figs. \ref{v1ut-au90c2-jjscv-A4}
and \ref{v1ut-au400c2-grqcv-6}.
In these plots the mass three data are an average of the isospin pair
$^3$H/$^3$He.
The more subtle differences between $^3$H and $^3$He will be discussed later.

The data span the whole energy range measured for the Au on Au system
($0.09A$ GeV to $1.5A$ GeV) and hence contains a rather detailed information
that represents a challenge for any theoretical attempts to reproduce all of
them.
One observes a gradual evolution of the shapes of the curves with the mass
of the ejectiles:
The variation with either $y_0$ or $u_{t0}$ is seen to be more pronounced as
the mass is increased.
The general increase of flow with mass is not new of course 
\cite{partlan95,doss87,barrette99}
 and is expected even in pure hydrodynamics treatments \cite{schmidt93}
 where it is explained
qualitatively in terms of random (thermal) motion, which is in velocity space
largest for the lightest particles, on top of a common flow.
This interpretation clearly does not then lead to (nucleon) number scaling
which has been claimed to be proof of deconfinement of the constituents
(constituent quarks at RHIC energies) \cite{phenix07a}).
It is obvious that the curves for the different masses (A=1 to 4) are not
equidistant along the ordinate and the different shapes as both the function 
of rapidity and transverse momentum (watch the location of the maxima) show
that a perfect number scaling, even with transverse kinetic energy instead of
$u_{t0}$, would not emerge. 
We shall see later that the same conclusion also holds for elliptic flow.
Of course due to cuts in the $u_{t0}$ ($y_0$ dependence) and in $y_0$ 
($u_{t0}$ dependence) a global scaling, if present, is not tested here.
Qualitatively, we expect the purely hydrodynamic minded view to be 
an oversimplification of reality since
 the system is not governed by ideal hydrodynamics, see namely
the incomplete stopping and the stopping hierarchy of clusters observed in
\cite{reisdorf10,reisdorf04a}.

The smooth curves plotted in Figs. \ref{v1ut-au90c2-jjscv-A4} and
\ref{v1ut-au400c2-grqcv-6} are just polynomial fits with maximally three
 parameters such as
$v_1(u_{t0})=v_{t11}\cdot u_{t0}+ v_{t12}\cdot u_{t0}^2
 + v_{t13}\cdot u_{t0}^3$.
These fits can be extrapolated down to $u_{t0}=0$ giving an idea of how the
flow data might converge to zero at low transverse momenta.
However, we find that the proton $v_1(u_{t0})$  data at the lowest incident
beam energies (Fig. \ref{v1ut-au90c2-jjscv-A4}) seem to have an additional
puzzling structure.
One tentative explanation is that the bump is caused by protons emitted from
heavier clusters, the proton then carrying a fraction of the flow of their
parents.
This would mean that late decays (evaporation) 
must be included in simulation codes.
This could be of some non-academic importance for efforts to deduce isospin 
dependences of the nuclear EoS from neutron-proton difference flows
\cite{leifels93,russotto10} as it is the slow particles that are generally found to be sensitive
to mean fields.
Evaporation could also have a sizeable influence on the $\alpha$ data for
the lower beam energies.


\subsection{System size dependences of directed flow}\label{v1syst} 

\begin{figure}
\hspace{\fill}
\epsfig{file=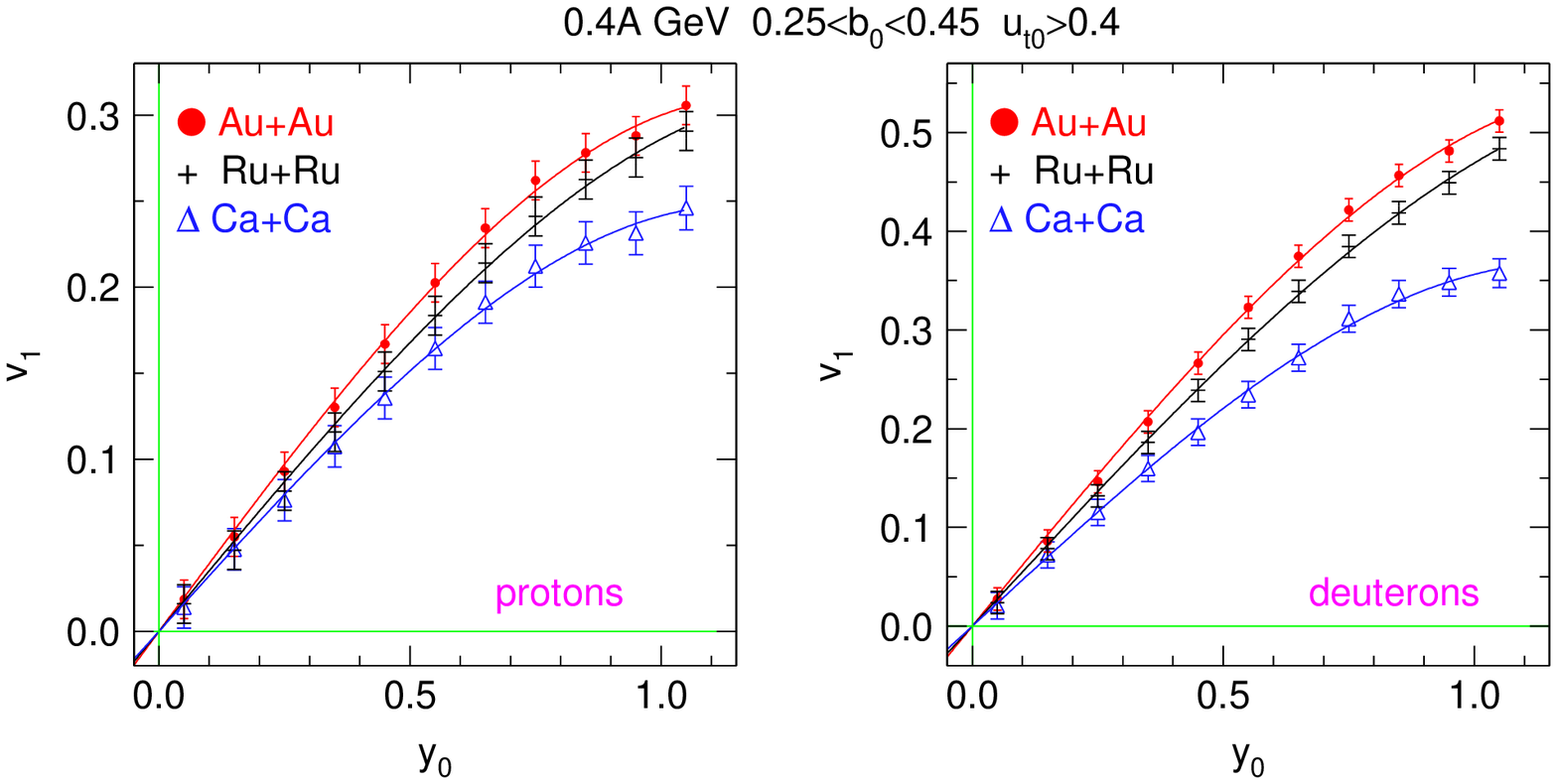,width=150mm}
\hspace{\fill}
\caption{
%
Left:
Directed flow $v_1$ of protons (left) and deuterons (right)
in collisions of $^{40}$Ca + $^{40}$Ca (blue open triangles),
$^{96}$Ru + $^{96}$Ru (black crosses) and $^{197}$Au+$^{197}$Au
(red dots) at incident energy of $0.4A$ GeV and centrality $0.25<b_0<0.45$.
The smooth curves are fits to $v_{11}\, y_0+v_{13}\, y_0^3$.
A cut $u_{t0}>0.4$ is applied.
}
\label{v1-ca400Z1A1-grqv}
\end{figure}

\begin{figure}
\hspace{\fill}
\epsfig{file=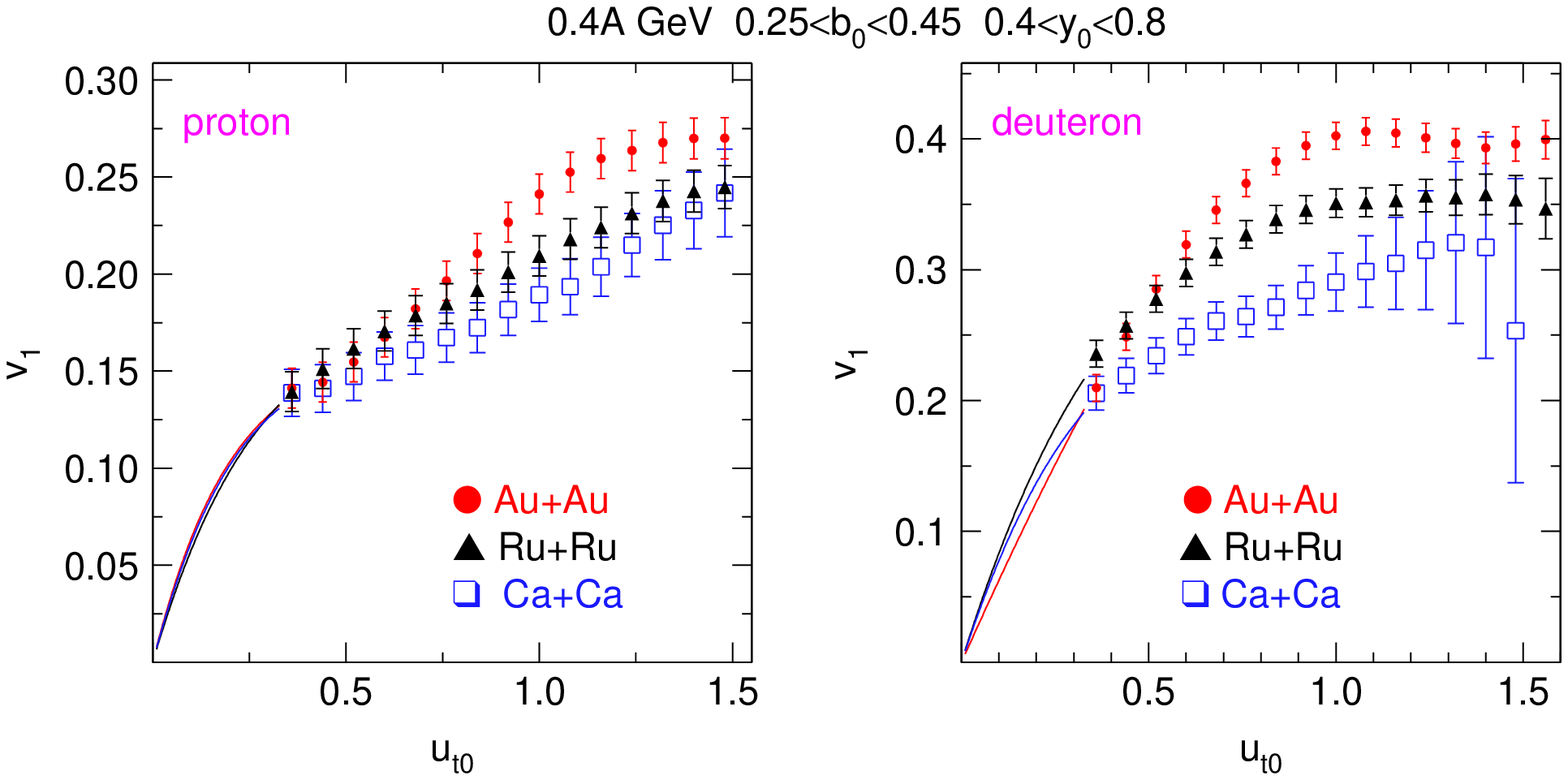,width=150mm}
\hspace{\fill}
\caption{
%
Directed flow $v_1(u_{t0})$ of protons (left) and deuterons (right)
in collisions of $^{40}$Ca + $^{40}$Ca (blue open squares),
$^{96}$Ru + $^{96}$Ru (full black triangles) and $^{197}$Au+$^{197}$Au
(red dots) at incident energy of $0.4A$ GeV and centrality $0.25<b_0<0.45$.
The smooth curves are extrapolations from fits
  $v_{t11}\cdot u_{t0}+v_{t12}\cdot u_{t0}^2+v_{t13}\cdot u_{t0}^3$
 to the data in the range below $u_{t0}=1.2$.
A cut $0.4<y_0<0.8$ is applied. In going from proton to deuteron flow there is
a change of scale by a factor 1.44.
}
\label{v1ut-ca4000Z1A1-grqv}
\end{figure}

\begin{figure}
\hspace{\fill}
\epsfig{file=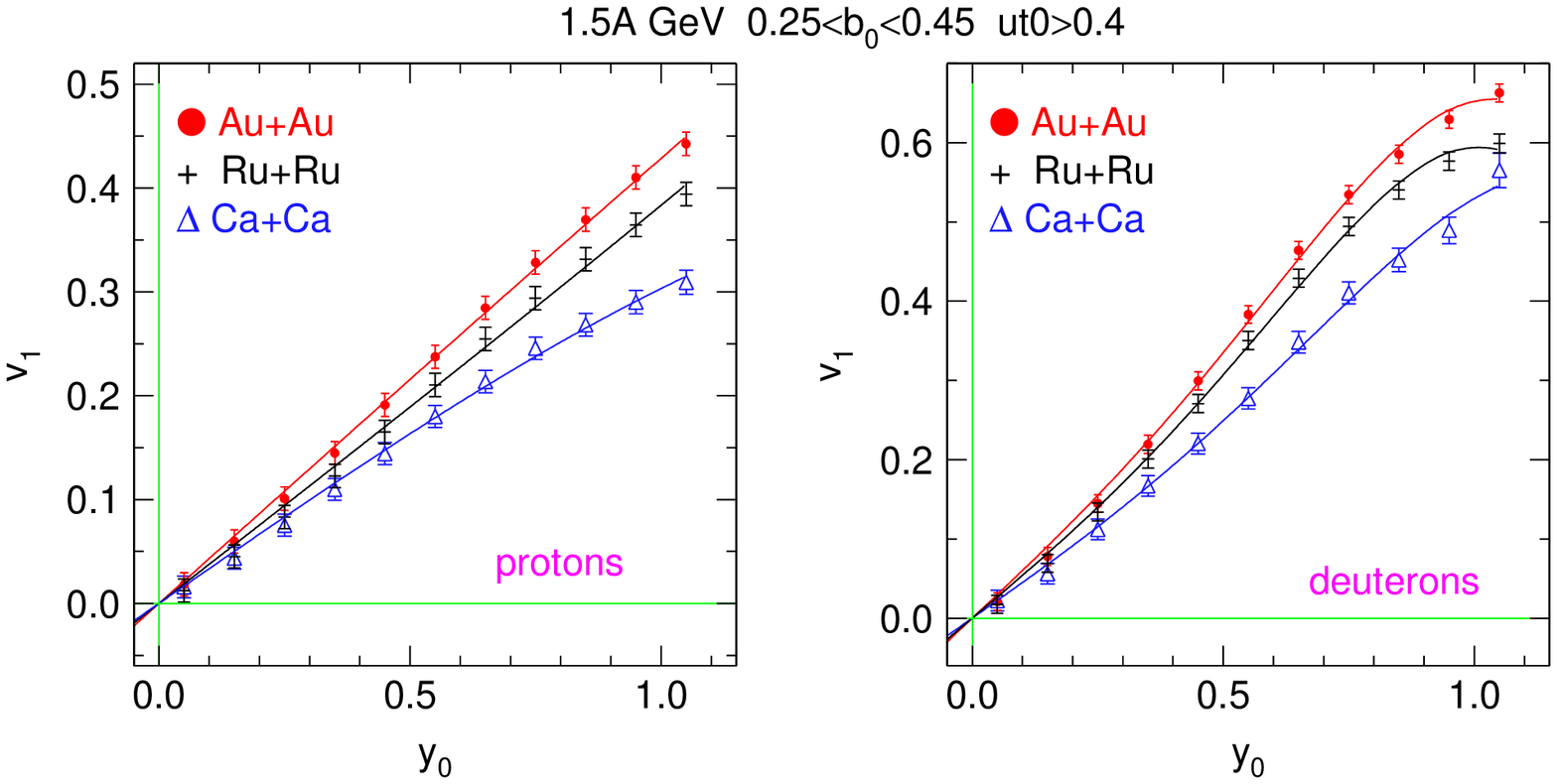,width=150mm}
\hspace{\fill}
\caption{
%
Directed flow $v_1$ of protons (left) and deuterons (right)
in collisions of $^{40}$Ca + $^{40}$Ca (blue open triangles),
$^{96}$Ru + $^{96}$Ru (black crosses) and $^{197}$Au+$^{197}$Au
(red dots) at incident energy of $1.5A$ GeV and centrality $0.25<b_0<0.45$.
The smooth curves are fits to $v_{11}\, y_0+v_{13}\, y_0^3+v_{15}\, y_o^5$.
A cut $u_{t0}>0.4$ is applied.
}
\label{v1-ca1500Z1A1-grqcv}
\end{figure}

\begin{figure}
\hspace{\fill}
\epsfig{file=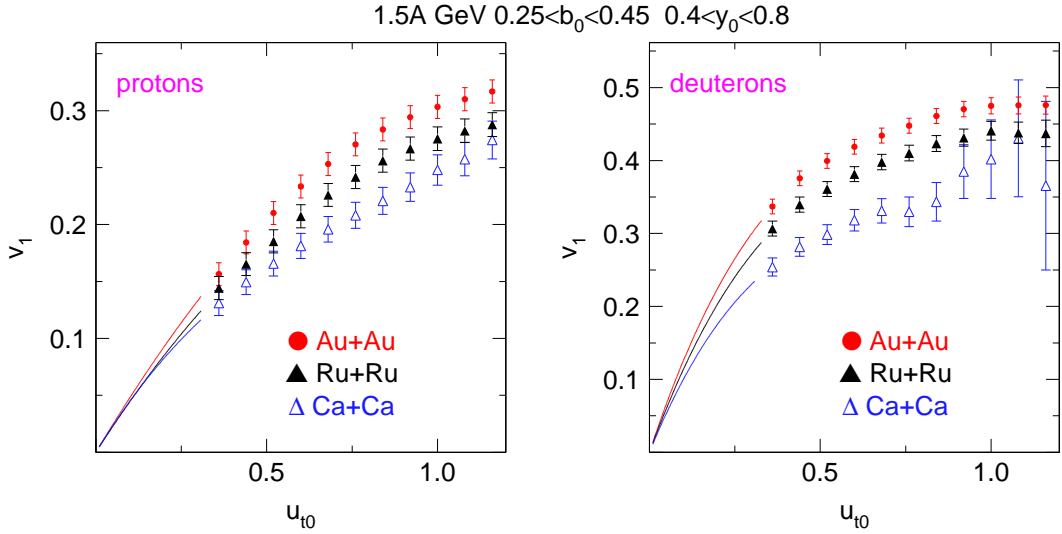,width=150mm}
\hspace{\fill}
\caption{
%
Directed flow $v_1(u_{t0})$ of protons (left) and deuterons (right)
in collisions of $^{40}$Ca + $^{40}$Ca (blue open triangles),
$^{96}$Ru + $^{96}$Ru (full black triangles) and $^{197}$Au+$^{197}$Au
(red dots) at incident energy of $1.5A$ GeV and centrality $0.25<b_0<0.45$.
The smooth curves are extrapolations from fits  $v_{t11}\, u_{t0}+v_{t12}\,
u_{t0}^2+v_{t13}\, u_{t0}^3$ to the data in the range below $u_{t0}=1$.
A cut $0.4<y_0<0.8$ is applied. In going from proton to deuteron flow there is
a change of scale by a factor 1.55.
}
\label{v1ut-ca1500Z1A1-grqcv}
\end{figure}

System size dependences of directed flow were first observed by the Plastic
Ball group \cite{gutbrod89r}.
A sample of our present data is shown in Figs. \ref{v1-ca400Z1A1-grqv} to
\ref{v1ut-ca1500Z1A1-grqcv}. 
The systems studied are indicated in the figures and their captions.
The incident energies are  $0.4A$ and $1.5A$ GeV,
the chosen centrality is $0.25<b_0<0.45$. 
For identified protons and deuterons
both the rapidity dependences and the transverse momentum dependences are
addressed. 
For the latter a rapidity interval $0.4<y_0<0.8$ was selected. 

\vspace{5mm}

A few remarks can be made at this stage:\\
1) Correlations between directed flow and stopping have been suggested
earlier, \cite{reisdorf04a}. 
By now extensive stopping data  for the SIS energy regime exist
\cite{reisdorf10}. 
It will therefore be a special challenge for future transport model
simulations to try to reproduce simultaneously the system size dependences
of stopping and of directed flow.\\
2) The system size variations are of similar magnitude as predicted
variations from modifying the stiffness of the EoS (see section \ref{iqmd}):
(partial) transparency can mockup a soft EoS.\\
3) As before, we note the absence of simple-minded (nucleon) number scaling
in the SIS energy regime for any system size. Looking at the data for
many incident energies, we find ratios of $1.45\pm 0.10$ for deuteron to
proton flow. Since a switch to (transverse) kinetic energies for the
abscissa would not modify the ordinates ($v_1$), this conclusion holds also
for the flow scaling as proposed by the RHIC community (which however was for
elliptic flow, see later, section \ref{v2}). 

\subsection{Excitation functions for directed flow}\label{v1excit}

\begin{figure}
\hspace{\fill}
\epsfig{file=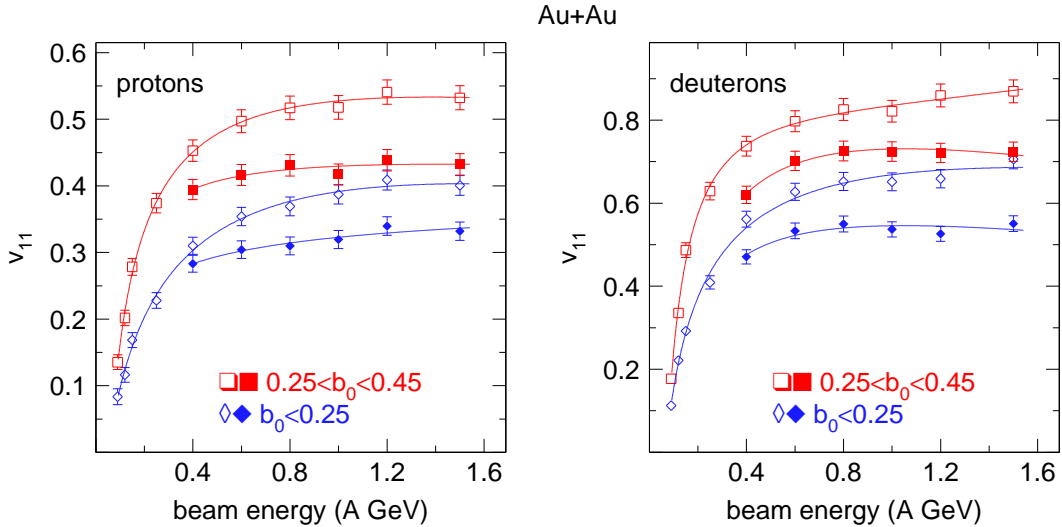,width=150mm}
\hspace{\fill}
\caption{
%
Excitation function of directed flow $v_{11}$ of protons (left) and
deuterons (right) in Au+Au collisions.
 $v_{11}$ was determined from a fit 
$v_1(y_0) = v_{11}\cdot y_0 + v_{13}\cdot y_0^3$ in the range $-1.1<y_0<1.1$.
The various selections on the centrality $b_0$  are indicated.
A cut $u_{t0}>0.4$ (0.8) is applied to the data with full (open) symbols.
Note the ordinate scales differing by approximately a factor 1.5.
}
\label{v11-auZ1A1-2}
\end{figure}

Excitation functions for the midrapidity slope of protons and deuterons
in the Au on Au system are shown for two different centralities
($b_0<0.25$ and $0.25<b_0<0.45$) in figure \ref{v11-auZ1A1-2}.
In order to be able to joinup the higher energy data with the low
energy data we apply a cut $u_{t0}>0.8$.
A range limited to the high energies is also shown with a cut $u_{t0}>0.4$.
The most striking feature is the steep rise up to $0.4A$ GeV, followed
by a rather flat behaviour beyond this energy.
We stress that this shape is linked to the use of scale-free quantities,
$v_1$ and $y_0$.
Repulsive pressure starts developing close to $0.1A$ GeV, increasing
first at a non-linear rate with incident energy.
This regime is followed by a scenario where pressure augments at about
the rate of the energy brought in causing little change in $v_{11}$.
The asymmetry caused by the sidepush is obviously larger if low
transverse momenta are suppressed: this follows from the more
detailed presentations shown before in subsections \ref{v1c} and \ref{v1LCP}.
Comparing protons with deuterons, we note the change of the
ordinate scale by approximately a factor 1.5, far from both one (no change)
 and two (nucleon number scaling).

\subsection{Directed flow and isospin}\label{v1iso}

The observed flow of particles is strongly dependent on their mass as we
have seen before.
In our present study, confined to charged LCP's, there is only one pair,
$^3$H/$^3$He, that can potentially give information on isospin dependences
of flow of ejectiles having the same mass.
Another way to try to catch information on isospin is to vary the isospin of
the systems.

Before presenting some of our data relevant to the isospin degree of
freedom, we feel it is necessary to point out in more detail some of our
apparatus response affecting the results.

\begin{figure}
\epsfig{file=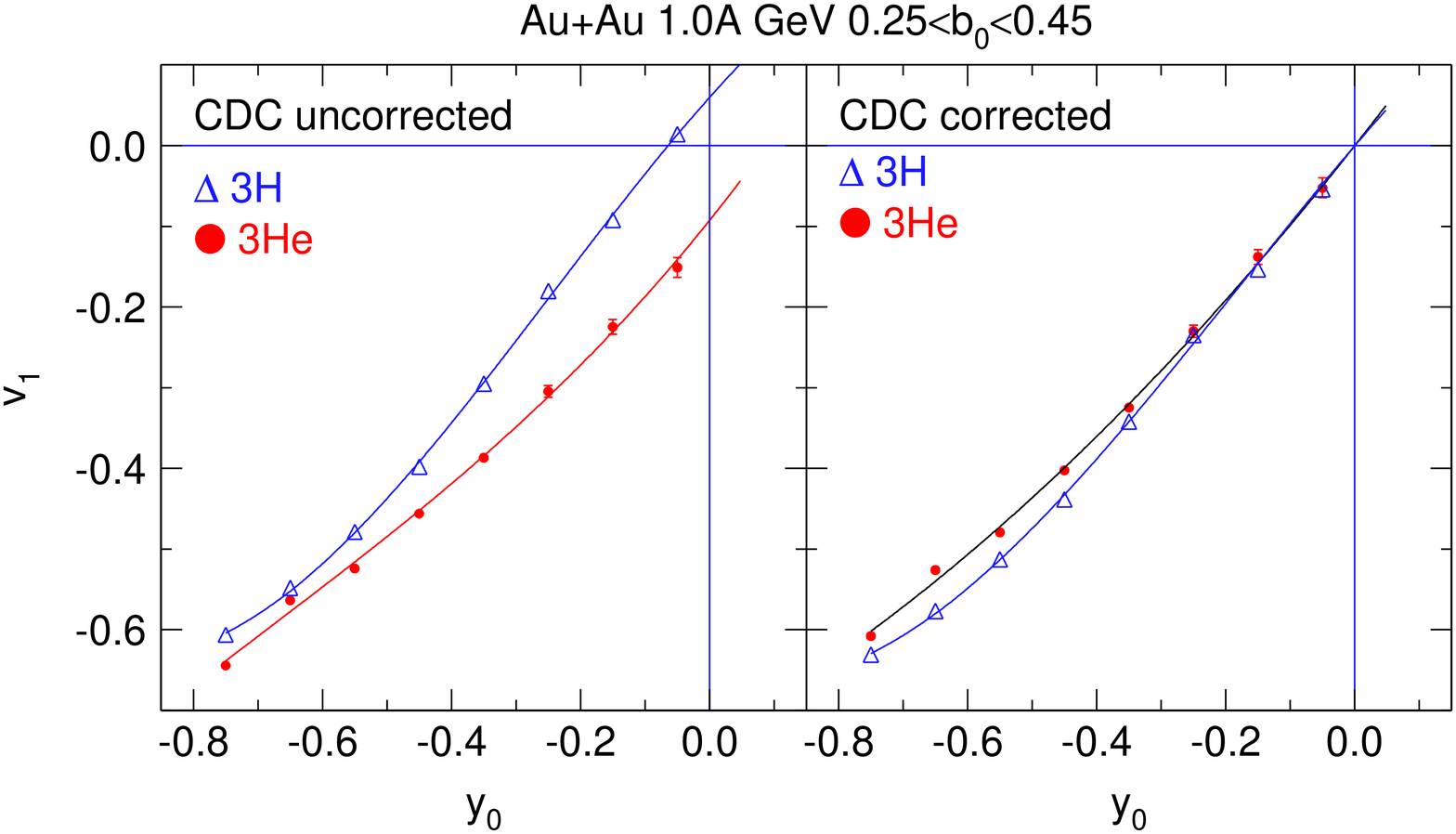,width=150mm}
\vspace{4mm}\\
\epsfig{file=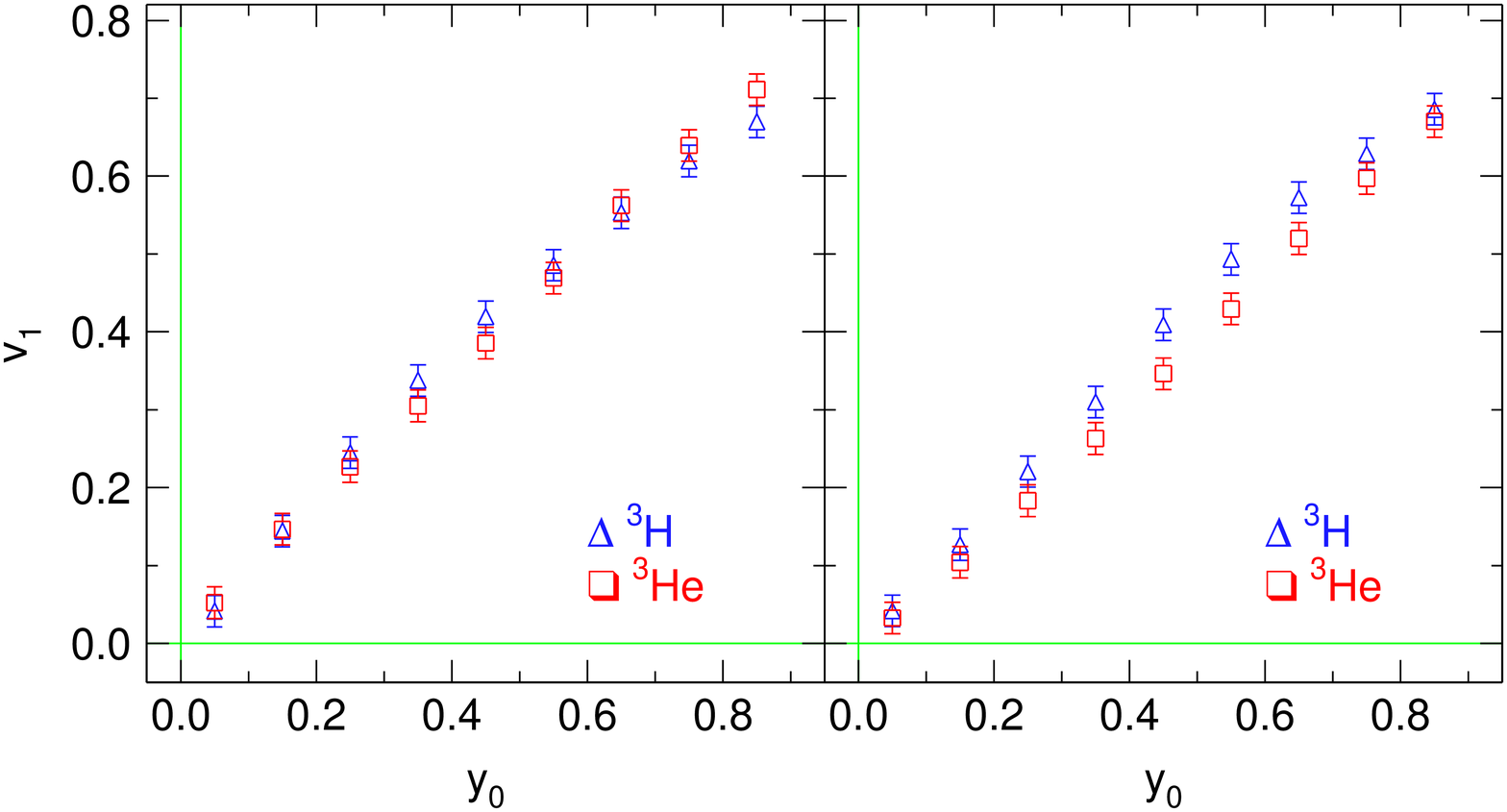,width=150mm}
\caption{
%
Comparison of the rapidity dependence of directed flow of $^3$H and $^3$He
 in Au+Au collisions at $1.0A$ GeV incident energy and
centrality $0.25<b_0<0,45$ using the CDC filter.
Upper left panel: uncorrected data using the CDC filter.
Upper right panel: corrected data using the CDC filter.
Lower left panel: uncorrected data using the full apparatus after 
symmetrization.
Lower right panel: corrected data using the full apparatus after
symmetrization.
}
\label{v1-au1000Z2A3c}
\end{figure}

Apparatus distortions influence  the measured $v_1$ and $v_2$ values
and herefore have to be corrected for.
The corrections turn out to be critical for small isospin differences
such as observed between $^3$H and $^3$He. 
For the symmetric systems that we study here, $v_1(y_0)$ should be asymmetric
with respect to midrapidity ($y_0=0$), in particular $v_1(0)$ should be
zero.
As can be seen in Fig.~\ref{v1-au1000Z2A3c}, showing data obtained with use of
the CDC (which covers predominantly the backward hemissphere),
$v_1(y_0)$ does not  cross the  origin of the axes. 
We have checked that this mid-rapidity offset or '$v_1$-shift'
was {\it not} due to deviations from azimuthal isotropy in the laboratory
reference system which were below the $5\%$ level and could be shown to
produce, on the average over many events, only a negligible second order
 effect on the $\phi$ distributions with respect to the reaction plane.
A systematic investigation of the  $v_1$-shift
 showed that it depended
on particle type, centrality and system size in a way suggesting that
it was correlated with the track density difference in the
'flow' quadrant $Q_1$ and the 'antiflow' quadrant $Q_3$.
While this could be simulated using our GEANT based implementation
of the apparatus response, a sufficiently accurate quantitative
reproduction of the offset at mid-rapidity was not achieved.
We  therefore opted for an empirical method to correct for
the distortion, using the very sensitive requirement of antisymmetry
with respect to midrapidity.
\vspace{4mm}\\
The quadrant method, defined in section \ref{apparatus},
 is well suited  for this.
We correct the $Q_i$ ($i=1,2,4,3$) by replacing them
 by $Q_i+c_iQ_i$.
The coefficients $c_i$ are fixed by three conditions.
First, as we could reproduce total charges in central collisions to within
$5\%$ in 25 system-energies \cite{reisdorf10} we require that after correction
the total sum $Q_0$ is unchanged.
Second, we assume a linear interpolation of the correction for $Q_{24}$,
i.e. $c_{24}=0.5(c_1+c_3)$ and finally we demand
$v_1=0$ at $y_0=0$, i.e. $Q_1(1+c_1))=Q_3(1+c_3)$ at $y_0=0$.
This is done separately for each particle, centrality and beam energy.
A check of the $v_1$-shift as a function of $p_t$  (equivalently $u_{t0}$)
or of polar angle showed that satisfactory results were obtained
assuming the distortions were proportional to $p_t$ in the range covered by
the CDC, and independent of $p_t$ in the range covered by the Plastic Wall.
\vspace{4mm}\\
In the upper right panel of Fig. \ref{v1-au1000Z2A3c} we show the effect of the
 correction for the mass three isospin pair.
Clearly, the surviving isospin-effect is small.
A slightly steeper slope of $^3$H, visible in the figure,
 was also found in the forward (Helitron-Plastic Wall) part of the data. 
Also, as we shall see in subsection \ref{v2iso}, the $v_2$ data, which are
 affected by the corrections just described,
 also support  more repulsion for $^3$H than $^3$He.
This gives us confidence in the consistency of the corrections.
\vspace{4mm}\\
The two lower panels show the results when combining the data of all the
subdetectors of FOPI.
In the left lower panels we ignore the $v_1$-shift problem but symmetrize
the data as outlined in section \ref{apparatus} with 
Fig. \ref{uty-au1000c2Z1A1};
in the lower right panel the data from all the detectors have been first
 corrected to eliminate the $v_1$-shifts and subsequently were symmetrized.
This is the procedure adopted for all the data shown in this work.

\begin{figure}
\hspace{\fill}
\epsfig{file=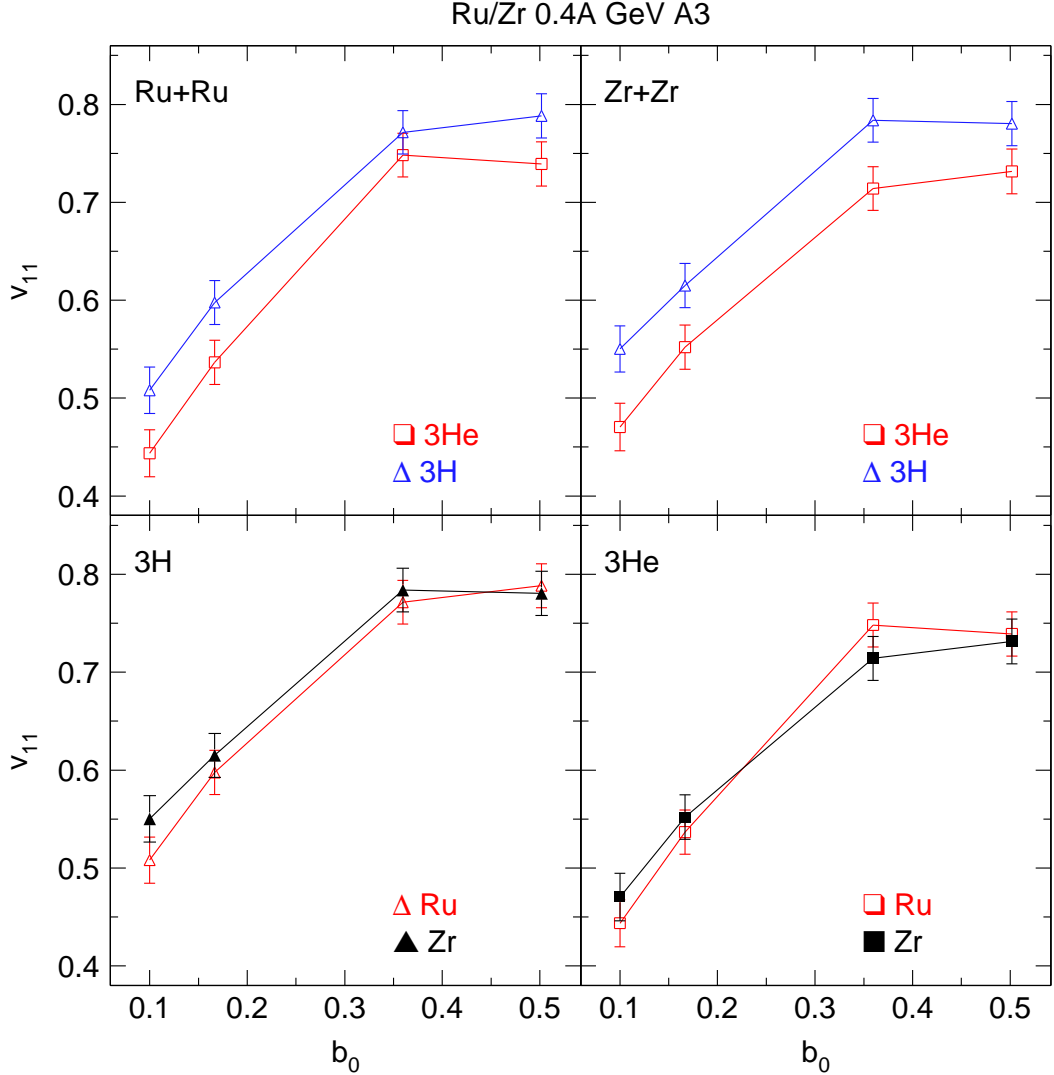,width=150mm}
\hspace{\fill}
\caption{
%
Centrality dependence of
midrapidity slope $v_{11}$ of directed flow of the probes $^3$He (squares)
 and $^3$H (triangles)
 in the systems Ru+Ru and Zr+Zr at $0.4A$ GeV incident energy.
In the upper two panels we compare the two probes in a fixed system as 
indicated in the panels, in the lower two panels we compare the same
probe ($^3$H left, $^3$He right) in two different systems.
}
\label{v11-A3-RuZr}
\end{figure}

The two ways of studying isospin signals
 are used in Fig. \ref{v11-A3-RuZr} comparing flow in the equal mass
systems $^{96}$Ru + $^{96}$Ru ($N/Z=1.182$) and $^{96}$Zr + $^{96}$Zr
($N/Z=1.400$) at an
incident energy of $0.4A$ GeV and a centrality $0.25<b_0<0.45$ with our
usual flat (in $y_0$) cut $u_{t0}>0.4$.
If we look at the same ejectile, we find no convincing (within errors)
difference for the mid-rapidity slopes $v_{11}$ in the two systems
(see the lower panels).
Not shown here are data for all the other LCP and for a higher incident energy
($1.5A$ GeV): these data lead to the same conclusion.
If we compare the two ejectiles $^3$H/$^3$He in the {\it same} system
(upper two panels), we see some evidence for a stronger flow of the
more neutron rich cluster especially in the neutron
richer system.
In view of the systematic uncertainties definite and precise conclusions are
premature, but the data are an encouragement for continued efforts to
improve the statistics and the quality of such observables.

\begin{figure}
\hspace{\fill}
\epsfig{file=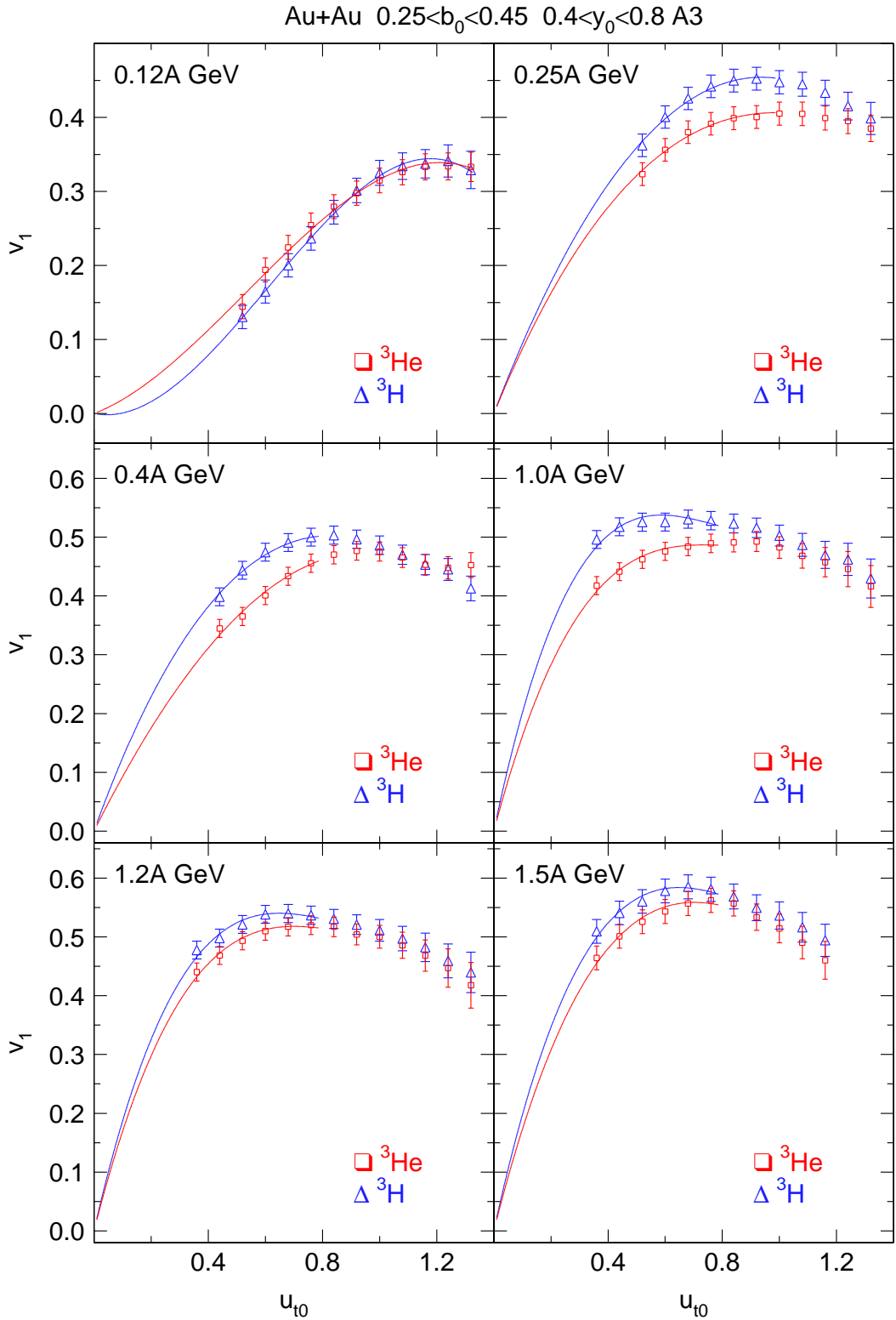,width=150mm}
\hspace{\fill}
\caption{
%
Directed flow $v_1(u_{t0})$ of $^3$He (red open squares) and $^3$H 
(blue open triangles)
in collisions of Au+Au with centrality $0.25<b_0<0.45$ and incident
energies $0.12A$, $0.25A$, $0.4A$, $1.0A$, $1.2A$ and $1.5A$ GeV as indicated 
in the various panels.
The longitudinal rapidity is constrained to the interval $0.4<y_0<0.8$.
All smooth curves are fits to the data assuming 
$v_1(u_{t0})=v_{t11}\cdot u_{t0}+ v_{t12}\cdot u_{t0}^2
 + v_{t13}\cdot u_{t0}^3$.
These fits are used to extrapolate to $u_{t0}=0$.
}
\label{v1ut-au400c2-grqcv-A3a}
\end{figure}

Another way to search for isospin signals in directed flow is shown in
Fig. \ref{v1ut-au400c2-grqcv-A3a} where we plot for Au+Au ($N/Z=1.493$) at six
different incident energies (indicated in the various panels) the
flow $v_1(u_{t0})$ comparing $^3$H (blue triangles) and $^3$He (red squares).
Except for the lowest shown energy, $0.12A$ GeV (upper left panel), we
always find the same pattern: a slightly higher flow of $^3$H in the
range $0.4<u_{t0}<0.8$ with a tendency of the two data curves to merge at the
higher values of $u_{t0}$.
The extrapolations below $u_{t0}=0.4$ have merely orientational value here.

 Closing this subsection, we can say that our data suggest that neutrons
see a more repulsive mean field than protons in neutron rich compressed
systems.
This enhances somewhat the flow of clusters that contain more neutrons than
protons.
Of course it will be interesting to continue studies comparing single
neutrons and protons with neutron sensitive devices
\cite{leifels93,russotto10}.

\subsection{Comparison to other data}\label{v1compa}

The comparison of flow data from different collaborations is a difficult task
as both the centralities and the apparatus cuts or biases need to be
reasonably well matched.
Isotope separated LCP data have been published by the EOS Collaboration in 1995
\cite{partlan95} for the Au+Au system.
The flow quantity $F$ defined in \cite{partlan95} corresponds to our 
midrapidity slope parameter $u_{x01}\cdot p_{cm}$ and hence has the dimension
of a momentum (measured in GeV/c). 
We do not favour this 'half-scaling' (the scaled rapidity is used, but
the momentum projection is not scaled), but use it here to compare directly
to the data in \cite{partlan95}.
In Ref. \cite{partlan95}
$F$ was determined from a linear fit in the range $-0.18<y_0<0.29$ and hence
is technically not strictly equal to our midrapidity slope which was
determined in the larger range $-1.1<y_0<1.1$ using a cubic (in $y_0$)
correction term in addition to the linear term, a procedure we found
to be more robust against systematic errors.
We note further that the recipee of centrality selection via reduced 
charged particle multiplicity bins used in \cite{partlan95} and earlier in
\cite{gustafsson84} does not guarantee a fixed effective $b_0$ 
(or cross section) range as the incident energy is varied.
This will influence the shape of the excitation function somewhat.\\

\begin{figure}
\hspace{\fill}
\epsfig{file=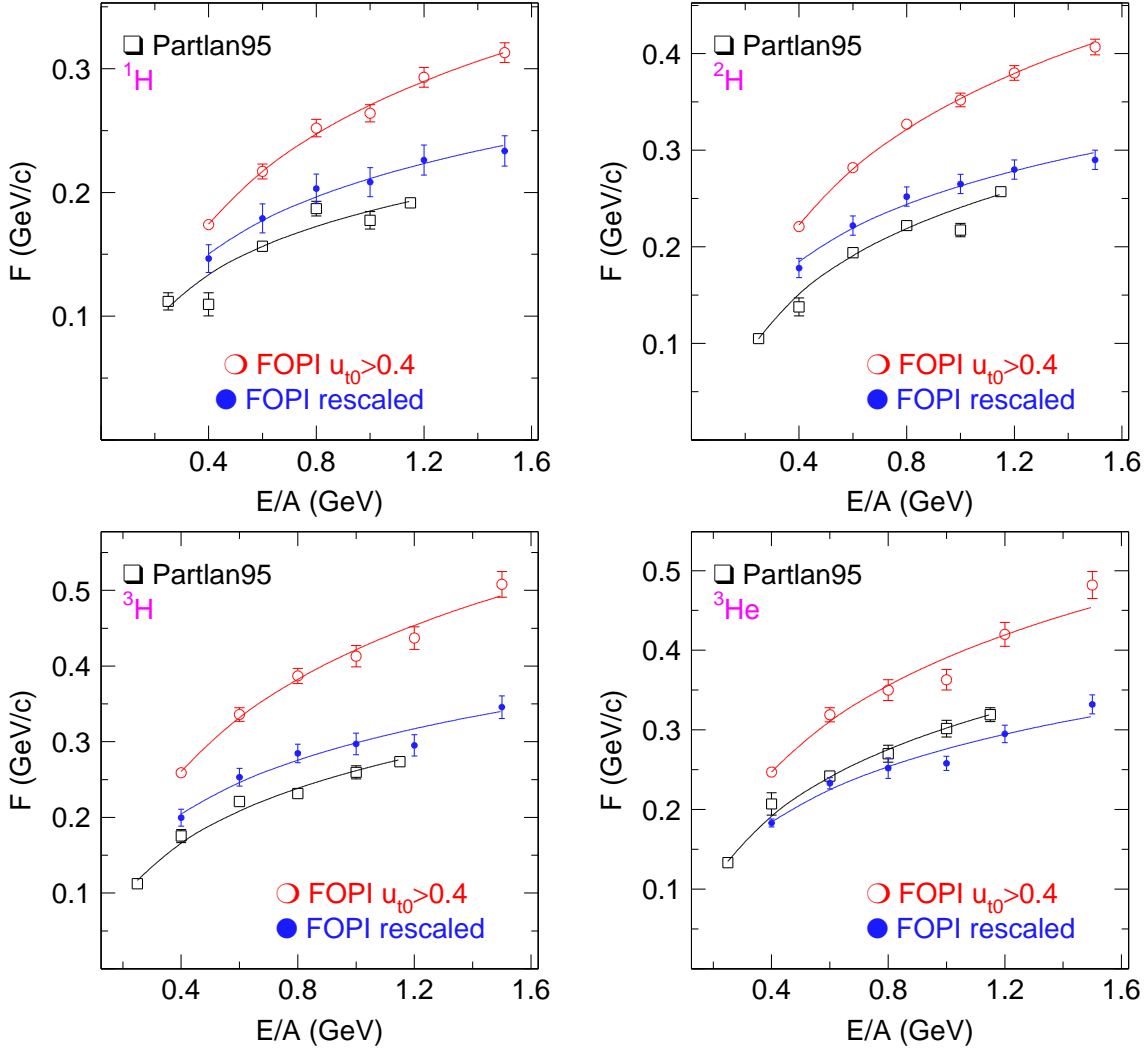,width=160mm}
\hspace{\fill}
\caption{
%
Upper left: comparison of directed flow (midrapidity slopes) of protons
 in collisions
of Au+Au (red, open circles) 
with centrality $0.25<b_0<0.45$ (dubbed 'c2') with the data
 of Ref.~\cite{partlan95} (black open squares).
The latter are multiplicity selected in a bin dubbed 'M3+M4' (not
further characterized in terms of cross sections or effective impact
parameter). 
Upper right: same for deuterons. Lower left: same for tritons.
Lower right: same for $^3$He.
A cut $u_{t0}>0.4$ was applied to the FOPI data (in red).
The influence of this cut is estimated  and the
 corresponding 'rescaled'  curves are plotted as well ( blue dots).
The smooth curves are polynomial fits  to guide the eye.  
}
\label{partlan95-e}
\end{figure}

For separated isotopes our measurements are affected by a transverse momentum
constraint, $u_{t0}>0.4$, which enhances flow as shown earlier using charge (but
not mass) separated data that extend down to lower momenta.
In Fig. \ref{partlan95-e} we compare our flow data for the three hydrogen
 isotopes and for $^3$He with the EOS data \cite{partlan95}. 
Clearly, due to the low-momentum cut our measurements lead to larger flow
values, but come closer to the earlier data if we estimate the influence of
the cut from the information given previously in Fig. \ref{ux-au800u0Z1A0-grcv}.
The remaining difference may be due to imperfect matching of centralities
and a different data fitting range.
Also, for lack of better knowledge, we have assumed that no matching apparatus
filter need to be applied before comparing.
A more disturbing observation is that our cut-corrected estimates for the
isospin pair $^3$H/$^3$He compare differently to the EOS data (see the lower
two panels).
For $^4$He, not shown here, the comparison is qualitatively similar to the
$^3$He case.
In contrast to what we tend to conclude from our data (see for example
Fig.~\ref{v1ut-au400c2-grqcv-A3a}), the EOS data, presumably including low
momenta, suggest a higher integrated $F$ flow of $^3$He compared to $^3$H.
It may be of some consequence that our reduction to zero $u_{t0}$ cuts is
only approximate since it was deduced from charge separated data only
(Fig. \ref{ux-au800u0Z1A0-grcv}).


\section{Elliptic flow}\label{v2}

As announced earlier, we proceed to elliptic flow using a similar presentation
of the various projections as was done for directed flow.
But first we discuss briefly the question of the proper reference frame.

\subsection{Question: to rotate or not to rotate}\label{v2rot}

One can rotate away the $v_1$ component by an appropriate choice of the
so-called 'flow' angle as new axis of symmetry instead of using the beam axis.
This has been done in the early discovery days \cite{gutbrod90f}
and later also in other works \cite{swang96,stoicea04,andronic01}.
The rotation generally leads to larger $v_2$ components.
While this is an instructive exercise, it does not add new data of course
and is associated with a number of drawbacks.
Technically, the Ollitrault \cite{ollitrault98} correction applied to
$v_2$ data is not designed for a rotated system.
The most important drawback however, is the loss of rotational symmetry that
the experimental equipment only has in the beam axis system.
Due to imperfect $4\pi$ coverage this leads to incomplete $\phi$ distributions
in the rotated system and to a loss of othogonality of the various Fourier
components.
A more subtle detail is that the rotation ('flow'-) angle is not really a 
fixed global constant but is, besides being centrality dependent, strictly
speaking also a particle, and even $(y_0,u_{t0})$ dependent quantity
reflecting the multidimensional complexity of $v_1$.
If one compares data with simulations where one wishes to convince oneself
that {\it both} $v_1$ and $v_2$ are reproduced, it is an unnecessary
complication to have to operate with rotations also in the simulation.
For all these reasons we shall therefore stick to the beam axis system paying
the price of a perhaps more complex  interpretation in this system of
reference.
Once {\it all} the unrotated data are reproduced by the simulation,
 one can of course operate with rotations on the simulated events which have
 ideal $4\pi$ coverage, to try to extract possibly 'simple' interpretations.

\subsection{Centrality dependence of elliptic flow}\label{v2c}

\begin{figure}
\hspace{\fill}
\epsfig{file=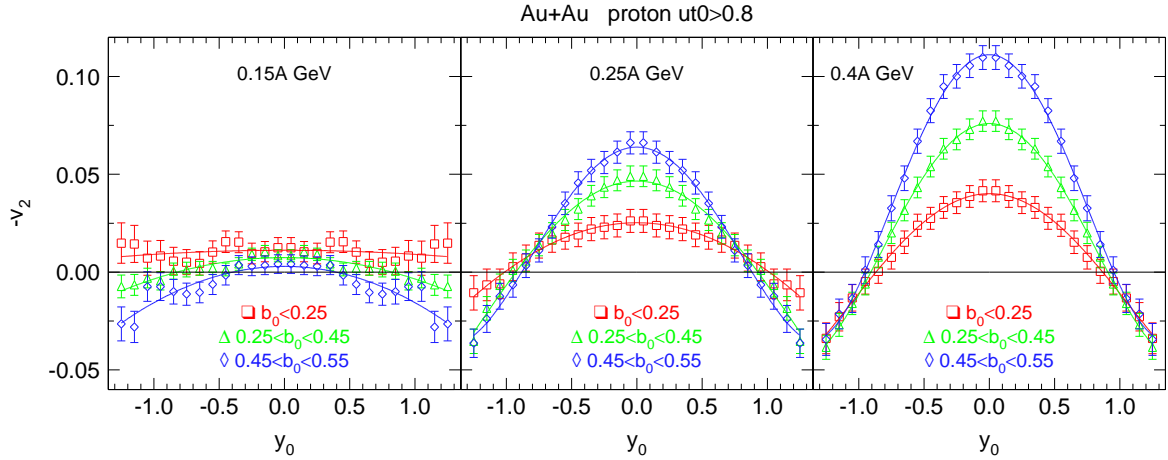,width=160mm}
\hspace{\fill}
\caption{
%
Elliptic flow $-v_2$ of protons in Au+Au collisions  for different
indicated energies and centrality ranges.
The smooth curves are least squares fits to 
$v_2=v_{20} + v_{22}\cdot y_0^2 + v_{24}\cdot y_0^4$.
Transverse 4-velocities $u_{t0}$ below 0.8 are cut off.
}
\label{v2-au250Z1A1-jjscv-3}
\end{figure}

\begin{figure}
\hspace{\fill}
\epsfig{file=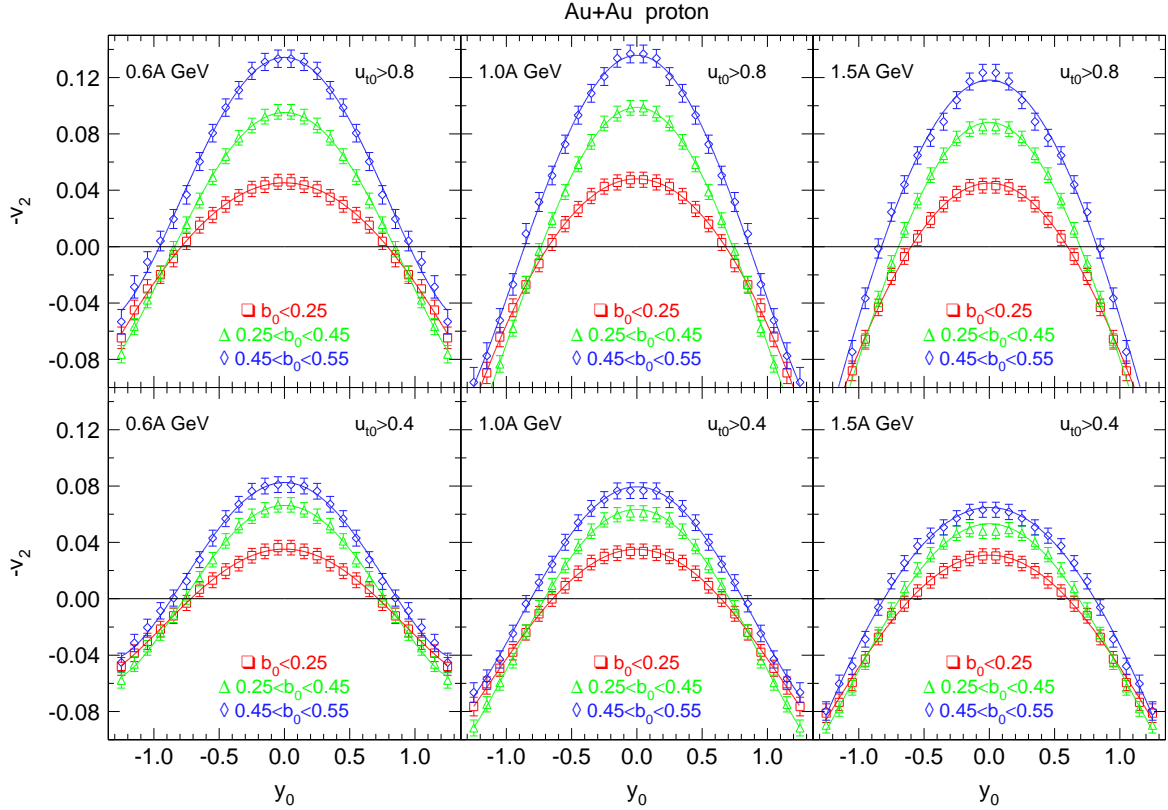,width=160mm}
\hspace{\fill}
\caption{
%
Elliptic flow $-v_2$ of protons in Au+Au collisions  for different
indicated energies and centrality ranges.
The smooth curves are least squares fits to 
$v_2=v_{20} + v_{22}\cdot y_0^2 + v_{24}\cdot y_0^4$.
In the upper three panels
transverse 4-velocities $u_{t0}$ below 0.8 are cut off.
The cut off is located at 0.4 for the lower three panels.
}
\label{v2-au600u8Z1A1-grqcv-6}
\end{figure}

We show in Figs. \ref{v2-au250Z1A1-jjscv-3} and \ref{v2-au600u8Z1A1-grqcv-6} the
rapidity dependence of proton elliptic flow for six different indicated
incident energies spanning one order of magnitude: $0.15A$ to $1.5A$ GeV.
In each panel the data for three centralities,
$b_0<0.25$,
$0.25<b_0<0.45$ and $0.45<b_0<0.55$ are given.
The shown energy regime is characterized by preferred out-of-plane emission
at midrapidity $(y_0=0)$ \footnote{recall however the sign change  at still
  lower energies close to the 'transition energy' \cite{andronic01}}: 
we plot $(-v_2)$ here and in the sequel to associate this effect with a
positive sign.
We remind of the simple interpretation  $Q_{24}/Q_0-1/2$ in terms of the
quadrants.
In the pioneering experimental papers \cite{gutbrod90f,gutbrod89} the term
'squeeze-out' was coined describing the scenario in more visual language.
Even earlier 'squeeze-out' was used in a theoretical paper \cite{stoecker82}
with predictive power.
The strong centrality dependence seen in the data suggests, alternatively,
that we are observing an increasing shadowing due to the presence of
spectator matter {\it in} the reaction plane in less central collisions.
While this view point stresses the interaction between participants and
spectators, the other viewpoint, squeeze-out, sees more the
participant-participant interaction: already compressed ('shocked') matter
being pushed out of the way by still incoming matter.
The common feature of 'shadowing' and 'squeeze-out' is the fact that the 
expansion
time scale and the passing time scale are comparable, a feature that is unique
to the present energy regime, see also subsection \ref{v2excit}.
This creates sensitivity to properties tightly connected with the nuclear EoS,
namely the sound propagation velocity (see section \ref{iqmd}).

There are two other striking observations:

a) The spectacular variation in the low energy regime,
Fig. \ref{v2-au250Z1A1-jjscv-3}, followed by the relatively steady behaviour
 at the higher energy, Fig. \ref{v2-au600u8Z1A1-grqcv-6}.
Note that in order to be able to follow the evolution in both regimes under
similar constraints, we have used a common cut $u_{t0}>0.8$ (upper panels in 
 Fig. \ref{v2-au600u8Z1A1-grqcv-6}, the lower panels show data for a cut
$u_{t0}> 0.4$).

b) The quasi-parabolic shapes leading to a sign change at large rapidities.
The smooth curves are three parameter fits with even polynomials as indicated
in the captions.
Clearly, this observed shape is associated with using the beam axis as
symmetry axis: 
we shall see, however, in section \ref{iqmd}, that the full rapidity range for
$v_2(y_0)$,  not just the mid-rapidity part, is sensitive to the EoS
allowing to make a more convincing and robust comparison to simulations than
using just the value at $y_0=0$.
In works such as \cite{swang96}, where a rotation of the symmetry axis was
performed minimizing the cross correlations to directed flow, one finds 
'squeeze-out' over a broader rapidity range, but at the cost of a rapidity
dependent rotation ('flow') angle.
As we shall see, we will be able to reproduce both $v_1(y_0)$ and $v_2(y_0)$
with the same EoS.

\begin{figure}
\hspace*{\fill}
\epsfig{file=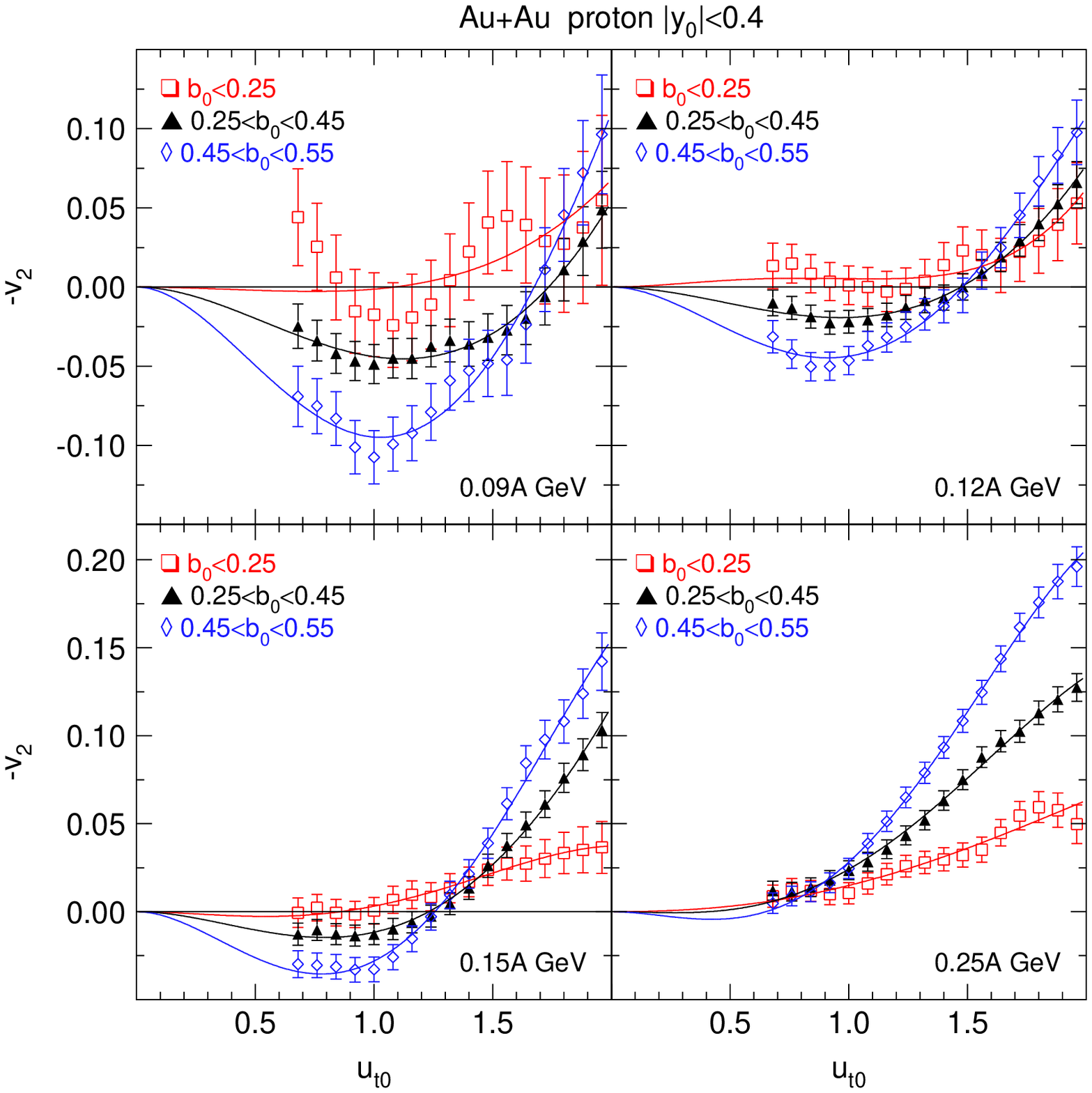,width=150mm}
\hspace*{\fill}
\caption{
%
Dependence of elliptic flow of protons on scaled moment $u_{t0}$
for three indicated centrality bins in Au on Au collisions.
The incident energies are $0.09A$ GeV (upper left), $0.12A$ GeV (upper right),
$0.15A$ GeV (lower left) and $0.25A$ GeV (lower right). 
The rapidity interval is constrained to $|y_0|<0.4$.
The smooth curves are 3-parameter fits to 
$v_2(u_{t0})= v_{t22}\cdot u_{t0}^2 + v_{t23}\cdot u_{t0}^3 +
 v_{t24}\cdot u_{t0}^4$. 
}
\label{v2ut-au250Z1A1-jjscv-4}
\end{figure}

\begin{figure}
\hspace*{\fill}
\epsfig{file=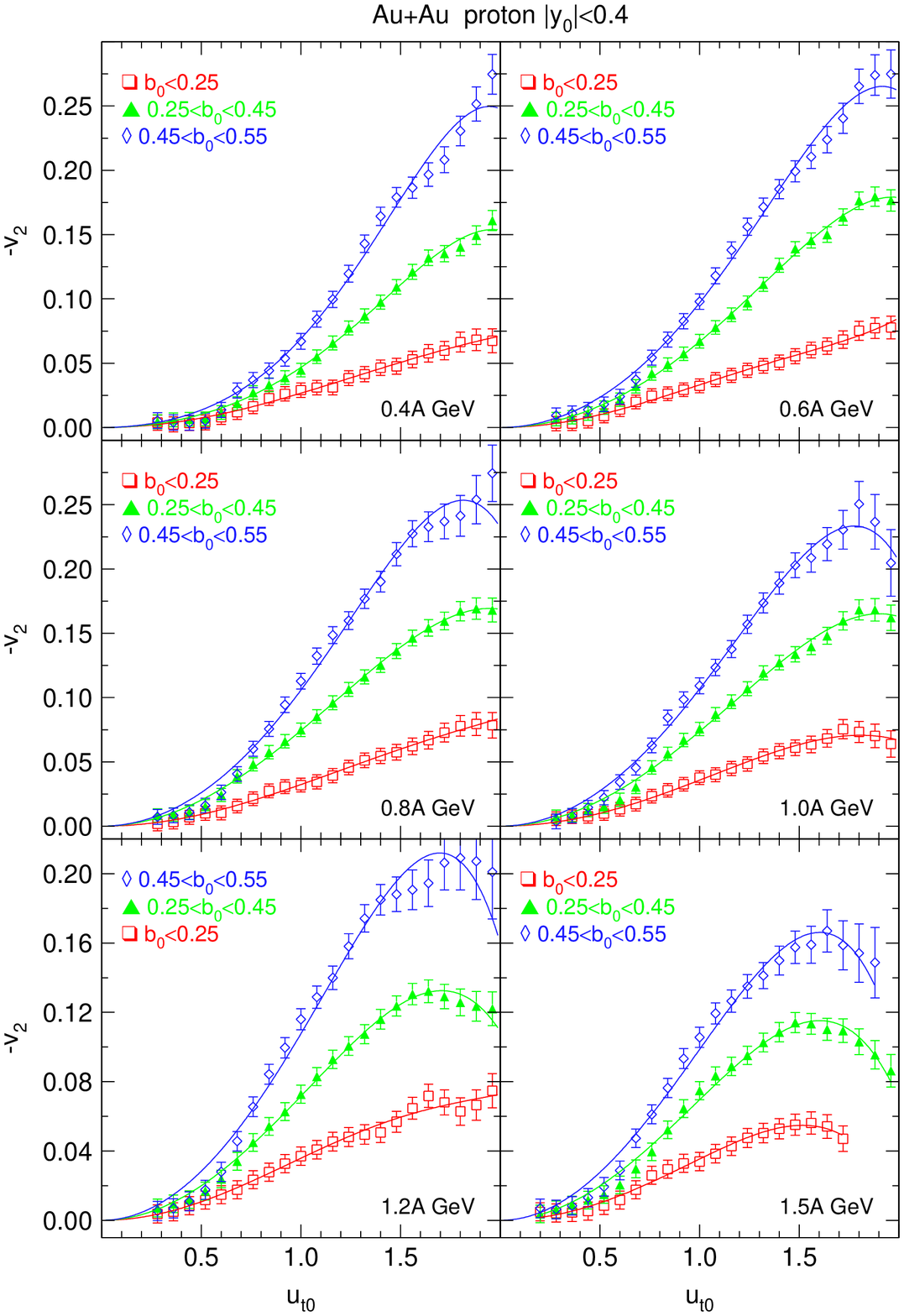,width=150mm}
\hspace*{\fill}
\caption{
%
Dependence of elliptic flow of protons on scaled moment $u_{t0}$
for three indicated centrality bins.
Left: Au+Au at $0.4A,\: 0.8A,\: 1.2A$ GeV, right: at $0.6A,\: 1.0,\: 1.5$ GeV.
The smooth curves are 3-parameter fits to 
$v_2(u_{t0})= v_{t22}\cdot u_{t0}^2 + v_{t24}\cdot u_{t0}^4 + 
v_{t26}\cdot u_{t0}^6$. 
}
\label{v2ut-au400Z1A1-grqcv-6}
\end{figure}

Additional information on the elliptic flow of protons is given in terms of
its momentum dependence in the various panels of 
Figs. \ref{v2ut-au250Z1A1-jjscv-4}
and \ref{v2ut-au400Z1A1-grqcv-6} for the same three centralities as before.
A rapidity window $|y_0|<0.4$ is taken, the smooth curves are three-parameter
fits to
$v_2(u_{t0})= v_{22}\cdot u_{t0}^2 + v_{23}\cdot u_{t0}^3 + v_{24}\cdot u_{t0}^4$
(lower energies) and to
$v_2(u_{t0})= v_{22}\cdot u_{t0}^2 + v_{24}\cdot u_{t0}^4 + v_{26}\cdot u_{t0}^6$
(higher energies). 
Again, the large differences between the low and the high energy regimes are
evident.
At the lowest three incident energies there is a momentum 
and centrality dependent sign change.
At $0.09A$ GeV, and more weakly at $0.12A$ GeV, the most central
collisions seem to display rather erratic fluctuations which, however, could
benefit from smaller error bars.
Such 'critical' fluctuations could be resulting from a scenario where
repulsive and attractive forces of almost equal strength on the average
strongly compete.
Further data from our earlier study of the complex behaviour in the vicinity
of the transition energy are available \cite{andronic01}.\\
At the energies beyond $0.25A$ GeV a more regular pattern has emerged.
The variation with centrality is strong, stressing the need to
match centralities very carefully when comparing different experiments
or comparing to a simulation.
One notices a gradual tendency of the shape changes to be more 'compact'
as the energy is raised.
One obvious reason for this is that, due to higher multiplicity,
the phase space extension, in the scaled units used here, has to
shrink since the total energy must be conserved.

\subsection{LCP-dependence of elliptic flow}\label{v2LCP}
\begin{figure}
\hspace{\fill}
\epsfig{file=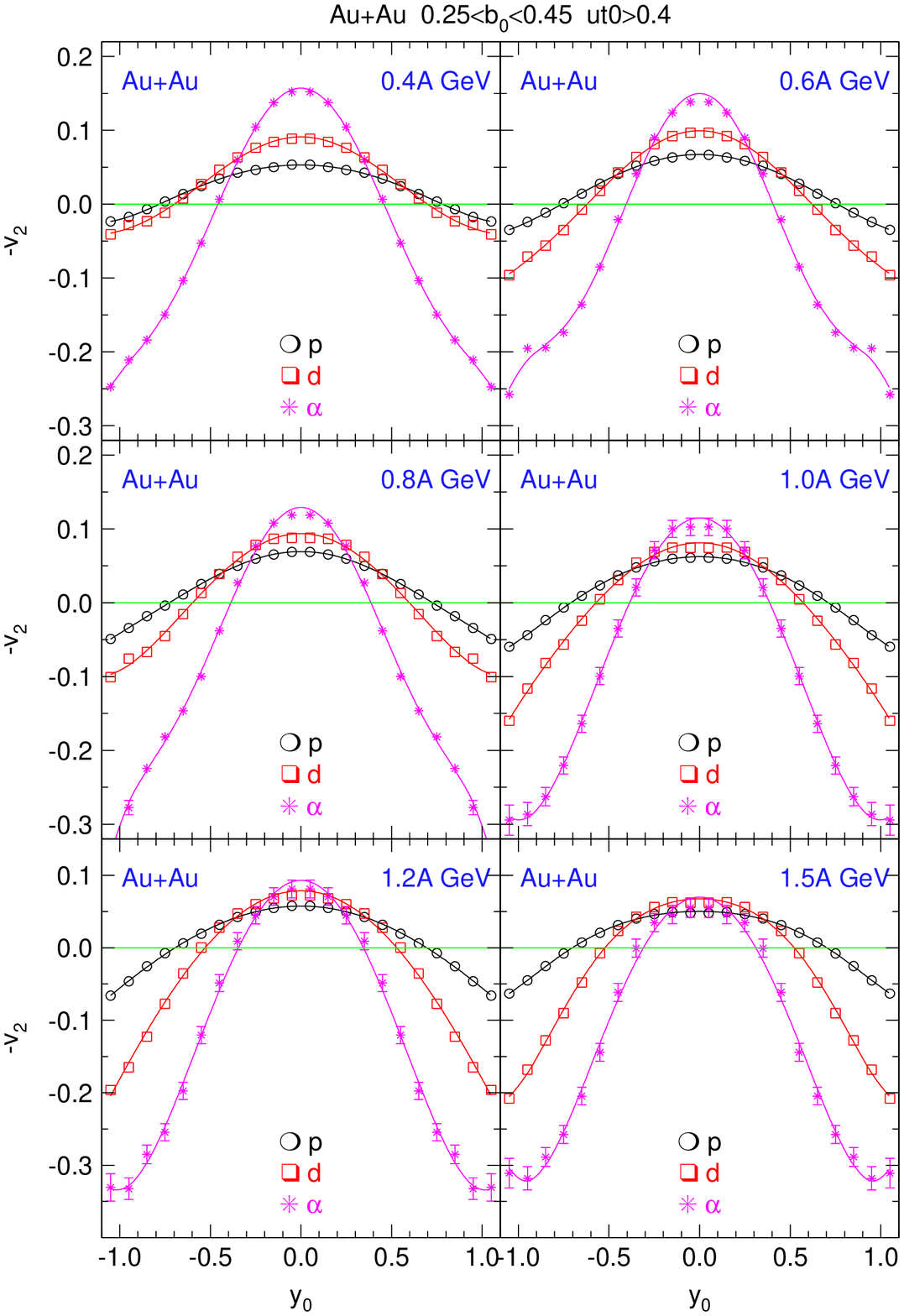,width=150mm}
\hspace{\fill}
\caption{
%
Elliptic flow $-v_2(y_0)$ of protons (black circles), deuterons (red squares)
and alpha particles (pink asterisks)
 in Au+Au collisions for different
indicated incident energies and centrality $0.25<b_0<0.45$.
Transverse 4-velocities $u_{t0}$ below 0.4 are cut off.
The smooth curves are even polynomial fits.
}
\label{v2-au1500c2-grqc14v-6}
\end{figure}

Moving to the cluster mass dependence of elliptic flow, we show in
 Fig. \ref{v2-au1500c2-grqc14v-6} a sample of the rapidity dependences  for
protons (black circles), deuterons (red squares) and alpha particles (pink
asterisks) together with even polynomial fits (smooth curves).
The chosen centrality is $0.25<b_0<0.45$. 
We present the higher energy data (0.4 to $1.5A$ GeV) with our usual
constraint $u_{t0}>0.4$.
Again  quasi-parabolic shapes are seen, but they are more compact in
the rapidity direction as the cluster mass is increased.
Whereas there is a clear mass hierarchy near midrapidity at the lower energy
end ($0.4A$ GeV, upper left panel), this hierachy is almost gone at $1.5A$
GeV (lower right panel), but the increasing compactness is preserved.
This shows that limiting the flow information to mid-rapidity could be
misleading when looking for possible interpretations.

\begin{figure}
\hspace{\fill}
\epsfig{file=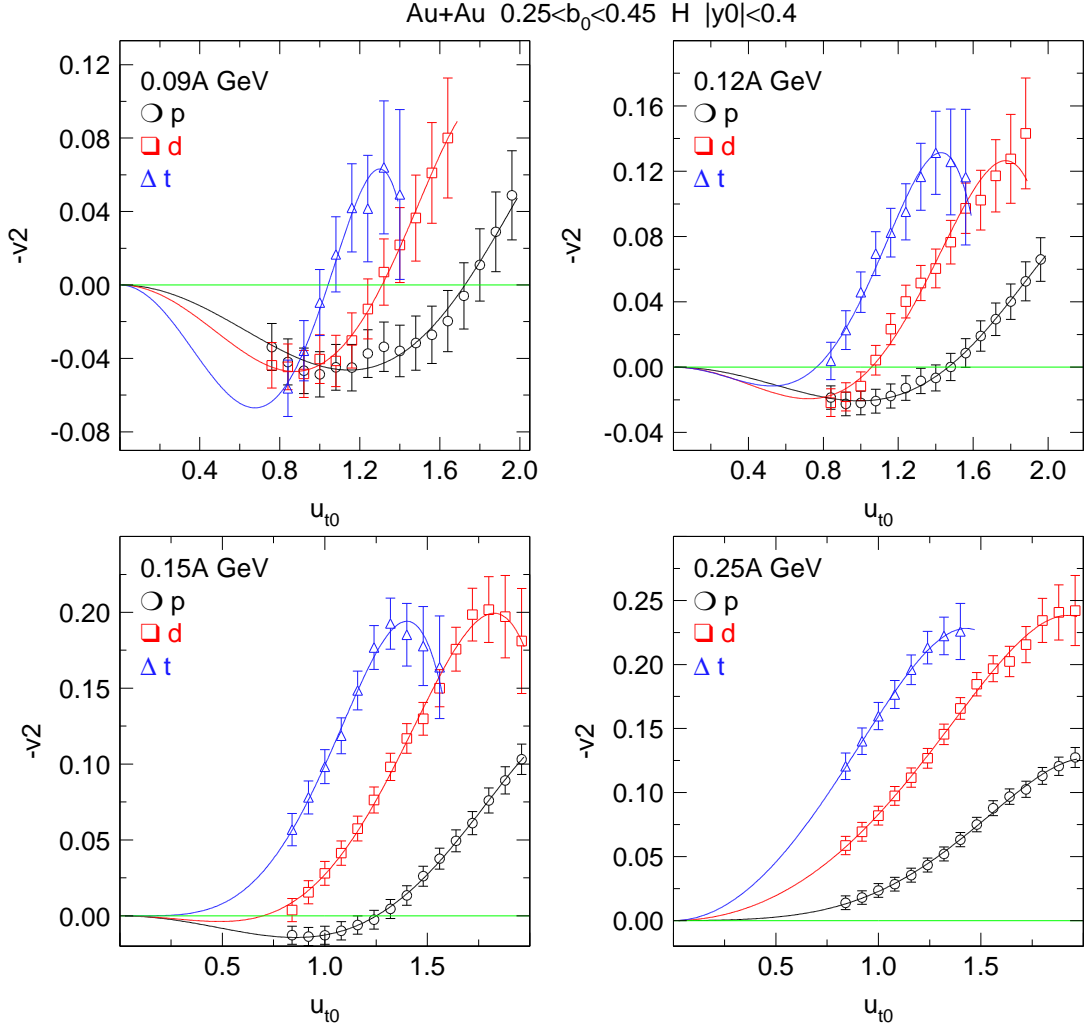,width=150mm}
\hspace{\fill}
\caption{
%
Left: elliptic flow $-v_2(u_{t0})$ of hydrogen isotopes for Au+Au at 
various indicated incident energies
with centrality $0.25<b_0<0.45$ and in the rapidity window $|y_0|<0.4$.
The smooth curves are three-parameter polynomial fits including even (2,4,6) 
 exponent terms.
}
\label{v2ut-au90c2-jjscv}
\end{figure}

\begin{figure}
\hspace{\fill}
\epsfig{file=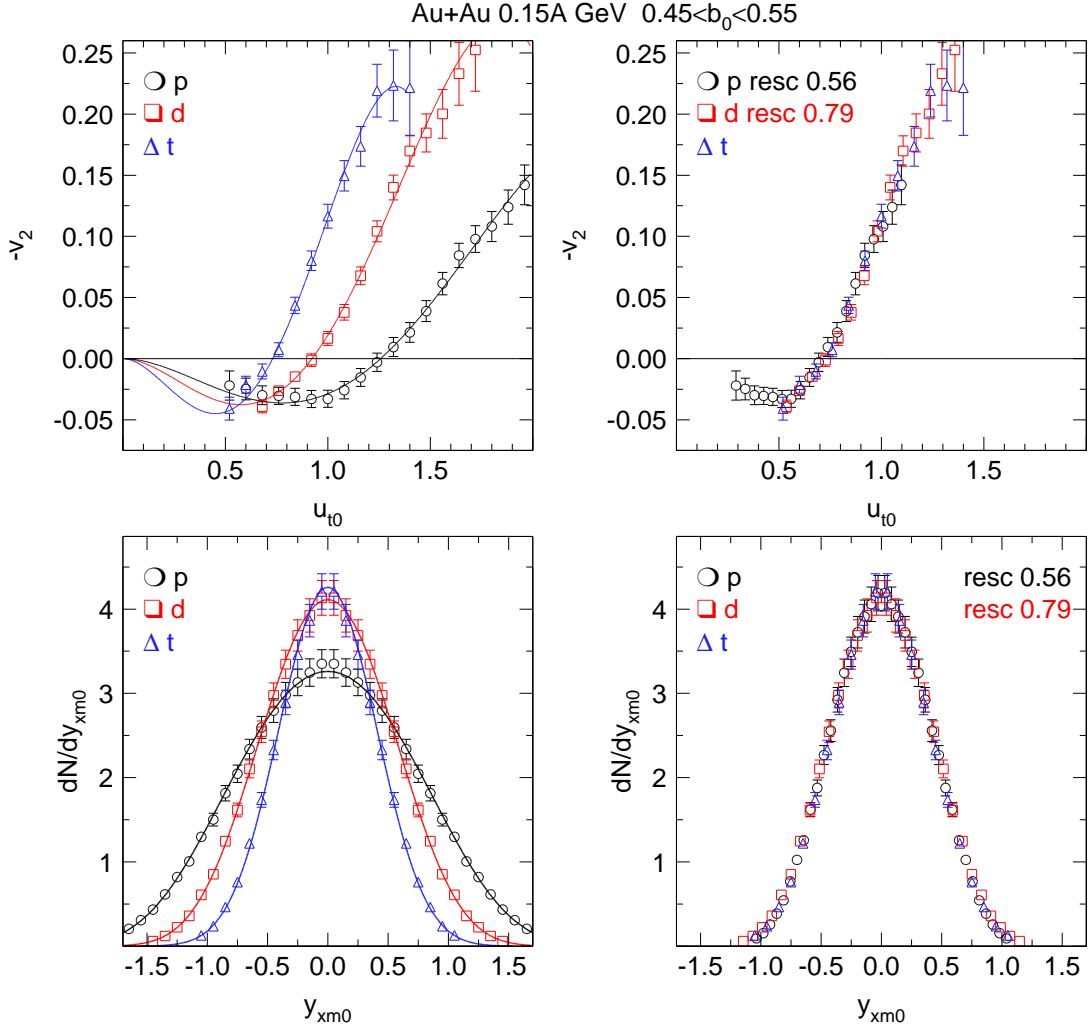,width=150mm}
\hspace{\fill}
\caption{
%
Upper left:
elliptic flow $-v_2(u_{t0})$ of hydrogen isotopes for Au+Au at $0.15A$ GeV
with centrality $0.45<b_0<0.55$ and in the rapidity window $|y_0|<0.4$.
The smooth curves are three-parameter polynomial fits including a
quadratic, cubic and quartic term. Flow is seen to change sign for
$u_{t0}=$ 1.26 (protons p), 0.92 (deuterons d) and 0.73 (tritons, t
triangles).
Lower left:
 Transverse rapidity distributions \cite{reisdorf10}
of the same particles under the same energy and centrality conditions
 constrained within a longitudinal rapidity window $|y_{z0}|<0.5$.
In the right panels the abscissa for the proton and deuteron data are
rescaled by a factor (indicated in the panel) allowing to superpose them on
the triton data. At this beam energy this rescaling of the rapidity axes
 is the same for the
rapidity distributions (which are in addition renormalized to the same
area, lower right) and the flow (upper right).
}
\label{v2ut-au150Z1A1-jjscv-4}
\end{figure}

Elliptic flow data for the lower energies ($0.09A$ to $0.4A$ GeV)
have been shown in an earlier publication of our Collaboration 
\cite{andronic01}, but it is useful to complement the data given there
with more details on the transverse momentum dependences, limited here
by the longitudinal rapidity cut $|y_0|<0.4$ and taken in the beam axis
reference system.
This is done in Figs. \ref{v2ut-au90c2-jjscv} and \ref{v2ut-au150Z1A1-jjscv-4}.
As before for protons varying the centrality, we also find, at a fixed
centrality, varying the particle mass, a fast evolution of the flow with
incident energy.
As observed in \cite{andronic01}, there is a sign change at some transverse
four-velocity that depends on the particle type.
The even polynomial fits (constrained to zero for $u_{t0}=0$) shown in the
figures offer educated guesses of the flow behaviour down to zero momentum.
At $0.25A$ GeV it seems that a fully repulsive regime is reached.

In Fig. \ref{v2ut-au150Z1A1-jjscv-4} we illustrate again the sign change
 of $v_2$
at some (scaled) transverse 4-velocity $u_{t0}$ characteristic for the mass
of the emitted  cluster (upper left panel).
It is possible to superimpose the flow data for the three particles
if we introduce a rescaling of the abscissa by a constant factor, 0.56 for
the proton data, 0.79 for the deuteron data, with the triton data left
unchanged. (The ordinates ($-v_2$) are left unchanged).
In the lower two panels of the same figure we show the effect of the 
same rescaling of the
abscissa for the reconstructed contrained {\it transverse} rapidity, $y_{xm0}$, distributions
\cite{reisdorf10} of the hydrogen isotopes.
'Constrained', denoted by  the index '$m$' for $m$id-rapidity, means that a cut
$|y_0|\equiv|y_{z0}|<0.5$ has been applied.
 (Here also the
ordinates have been renormalized to give the same integrated yields).
This suggests that a random, perhaps thermal, motion is superimposed on
a common flow profile at this incident energy.

\begin{figure}
\hspace{\fill}
\epsfig{file=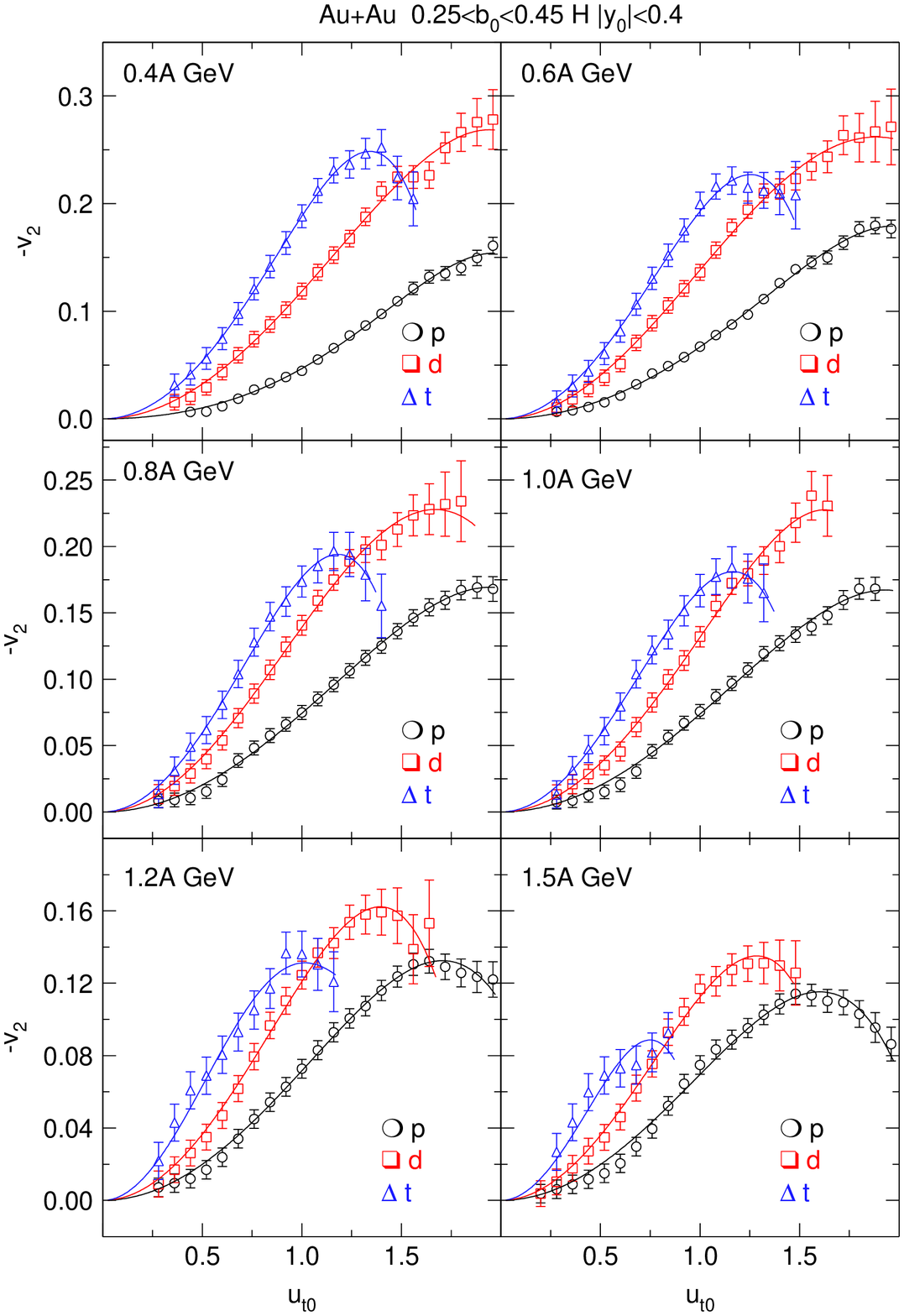,width=150mm}
\hspace{\fill}
\caption{
%
Left: elliptic flow $-v_2(u_{t0})$ of hydrogen isotopes for Au+Au at 
various indicated incident energies
with centrality $0.25<b_0<0.45$ and in the rapidity window $|y_0|<0.4$.
The smooth curves are three-parameter polynomial fits including even (2,4,6) 
 exponent terms.
}
\label{v2ut-au400c2-grqcv}
\end{figure}
\begin{figure}
\hspace{\fill}
\epsfig{file=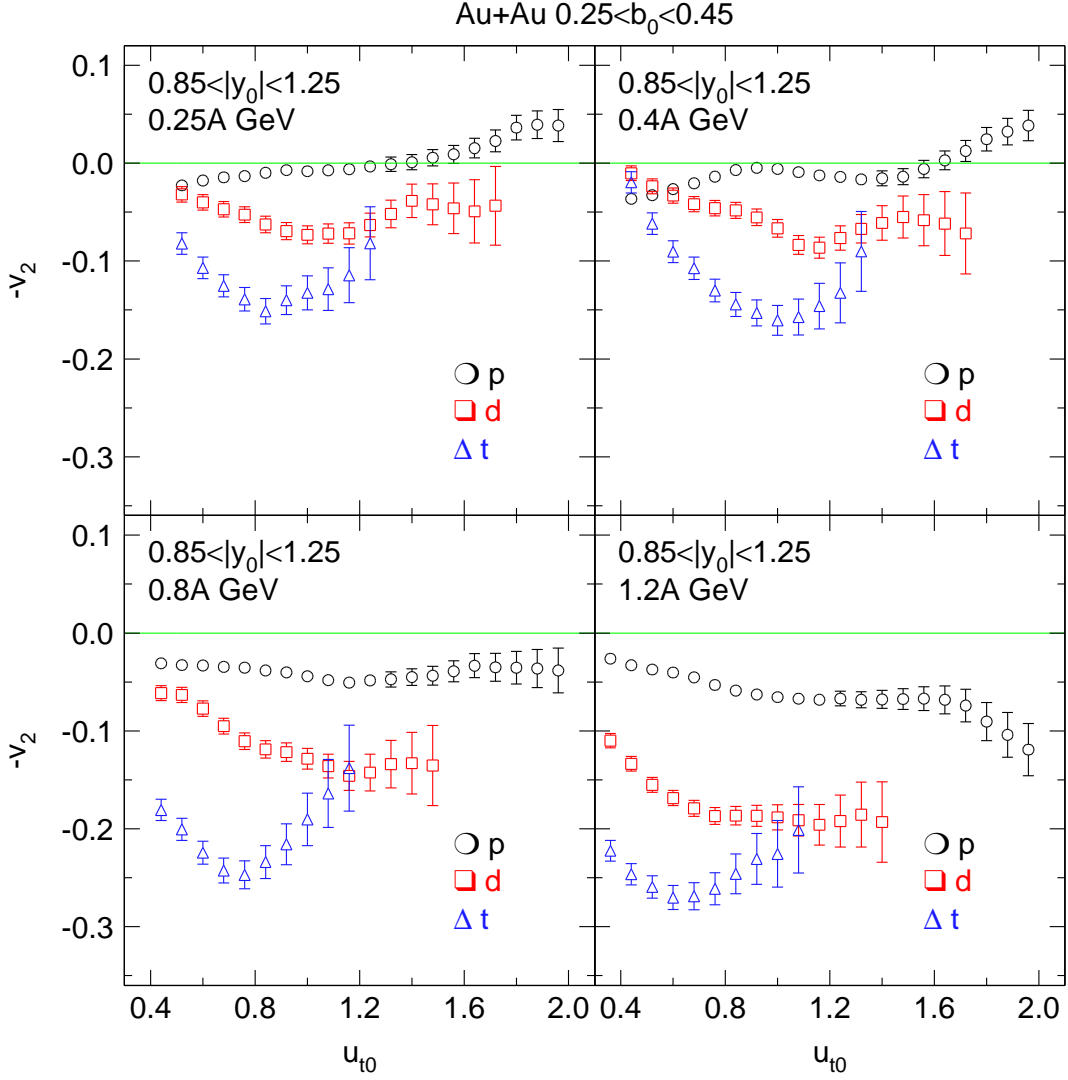,width=150mm}
\hspace{\fill}
\caption{
%
Left: elliptic flow $-v_2(u_{t0})$ of hydrogen isotopes for Au+Au at 
various indicated incident energies
with centrality $0.25<b_0<0.45$ and in the rapidity window 
$0.85<|y_0|<1.25$.
}
\label{v2ut-au400c2-y01}
\end{figure}

The transverse four-velocity dependences of elliptic flow, $v_2(u_{t0})$,
for the hydrogen isotopes at the higher incident energies ($0.4A$ to $1.5A$
GeV) are shown in Figs. \ref{v2ut-au400c2-grqcv} and \ref{v2ut-au400c2-y01},
again, for centrality $0.25<b_0<0.45$.
In Fig. \ref{v2ut-au400c2-grqcv} a midrapidity constraint $|y_0|<0.4$ was
 applied, while in Fig. \ref{v2ut-au400c2-y01} we show another instructive cut
 close to projectile rapidity, $0.85<y_0<1.25$.
The sign of the mass hierarchy is inverted here.

Attempts to extend 'simple' scalings of such data, as we showed in 
Fig. \ref{v2ut-au150Z1A1-jjscv-4} for beam energy $0.15A$ GeV, to higher beam
energies, were moderately successful.
In general, one can superimpose the $v_2(u_{t0})$ functions of the
three hydrogen isotopes if one uses momentum scaling factors
 in the ranges $1.5\pm 0.1$
(deuterons), resp. $2.0\pm0.2$ (tritons).
 At close inspection one finds that these scale factors vary somewhat
within the indicated limits and, in contrast to the $0.15A$ GeV data, are
larger than the factors needed to 'scale' the transverse rapidity spectra
in a way similar to the case shown in the lower panels of 
Fig. \ref{v2ut-au150Z1A1-jjscv-4}.

Exploring the use of 'number scaling' (a nucleon number scaling here)
in analogy to the (constituent) quark number scaling at RHIC \cite{phenix07a},
i.e. plotting $v_2/A$ versus the transverse kinetic energy per nucleon
failed significantly and hence no such plots are presented here.
As nucleons can be safely considered to be 'deconfined' at maximum pressure
from the clusters they end up with at freeze out, one can as safely
conclude that deconfinement does not necessarily lead to number scaling of
observed elliptic flow.
In the present energy regime  from hydrodynamic scenarios
 (common flow velocity profile
superimposed by random motion due to a common 'local' temperature) one does
not expect number scaling, although a mass hierarchy is predicted
\cite{schmidt93}.
As we shall see in the next section, microscopic transport theory is
fairly successful in describing flow of various light clusters.

\subsection{System size dependence of elliptic flow}\label{v2syst}

\begin{figure}
\hspace{\fill}
\epsfig{file=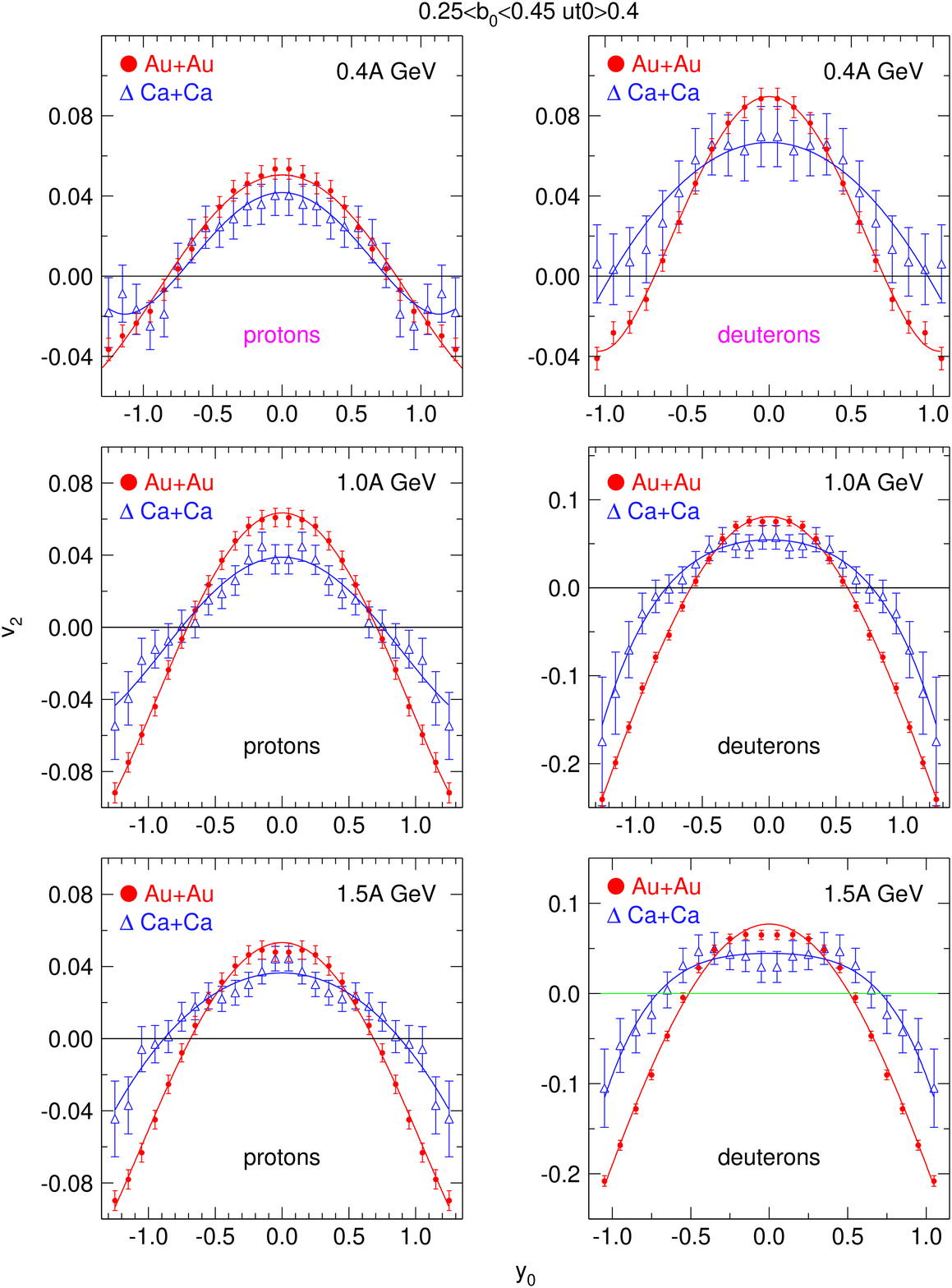,width=160mm}
\hspace{\fill}
\caption{
%
Elliptic flow $-v_2(y_0)$ of protons (left) and deuterons (right) 
 in $^{40}$Ca+$^{40}$Ca (blue open triangles),
 and Au+Au (red dots) 
collisions for $0.4A$ (top), $1.0A$ (middle) and $1.5A$ GeV (bottom)
 incident energy
 and centrality $0.25<b_0<0.45$.
A cut $u_{t0}>0.4$ is applied.
The smooth curves are fits of $v_{20} + v_{22}\,y_o^2 + v_{24}\,y_o^4$ to the
data. 
}
\label{v2-ca1500Z1A1-grqcv-6}
\end{figure}

We study the system size dependence by comparing the rapidity dependence
of the elliptic flow in $^{40}$Ca + $^{40}$Ca and Au+Au collisions.
This is shown in Fig. \ref{v2-ca1500Z1A1-grqcv-6} for protons (left panels)
and deuterons (right panels) at three different energies 
(top to bottom: $0.4A$, $1.0A$ and $1.5A$ GeV) and centrality
$0.25<b_0<0.45$.
We observe a rather strong effect of the size of the system (our Ru+Ru data,
not plotted here, are intermediate between the two systems).
Not only the value at midrapidity is affected, but the whole shape of the
curves: smooth fits of  $v_{20} + v_{22}\,y_o^2 + v_{24}\,y_o^4$ are
included in the figures.
The heavier system is characterized by more 'compact' shapes.
As we will see in section \ref{iqmd}, this corresponds to the action of
more repulsive mean fields.
Using the stopping data from Ref. \cite{reisdorf10} as additional information,
we can say that the larger densities reached in the heavier system,
as implied by the larger stopping, generate elliptic flow characteristic of
a 'stiffer' system at compression time.
As we shall see, the effect caused by the increased transparency of the
lighter system is stronger than the difference between the so-called
'stiff' and 'soft' EoS options.
 
\subsection{Excitation functions for elliptic flow}\label{v2excit}

We have published earlier \cite{andronic05} an excitation function for
the elliptic flow of $Z=1$ particles.
It has by now been joined up smoothly \cite{andronic06} to complementary
lower energy data by the ALADIN-INDRA Collaboration.

\begin{figure}
\hspace*{\fill}
\epsfig{file=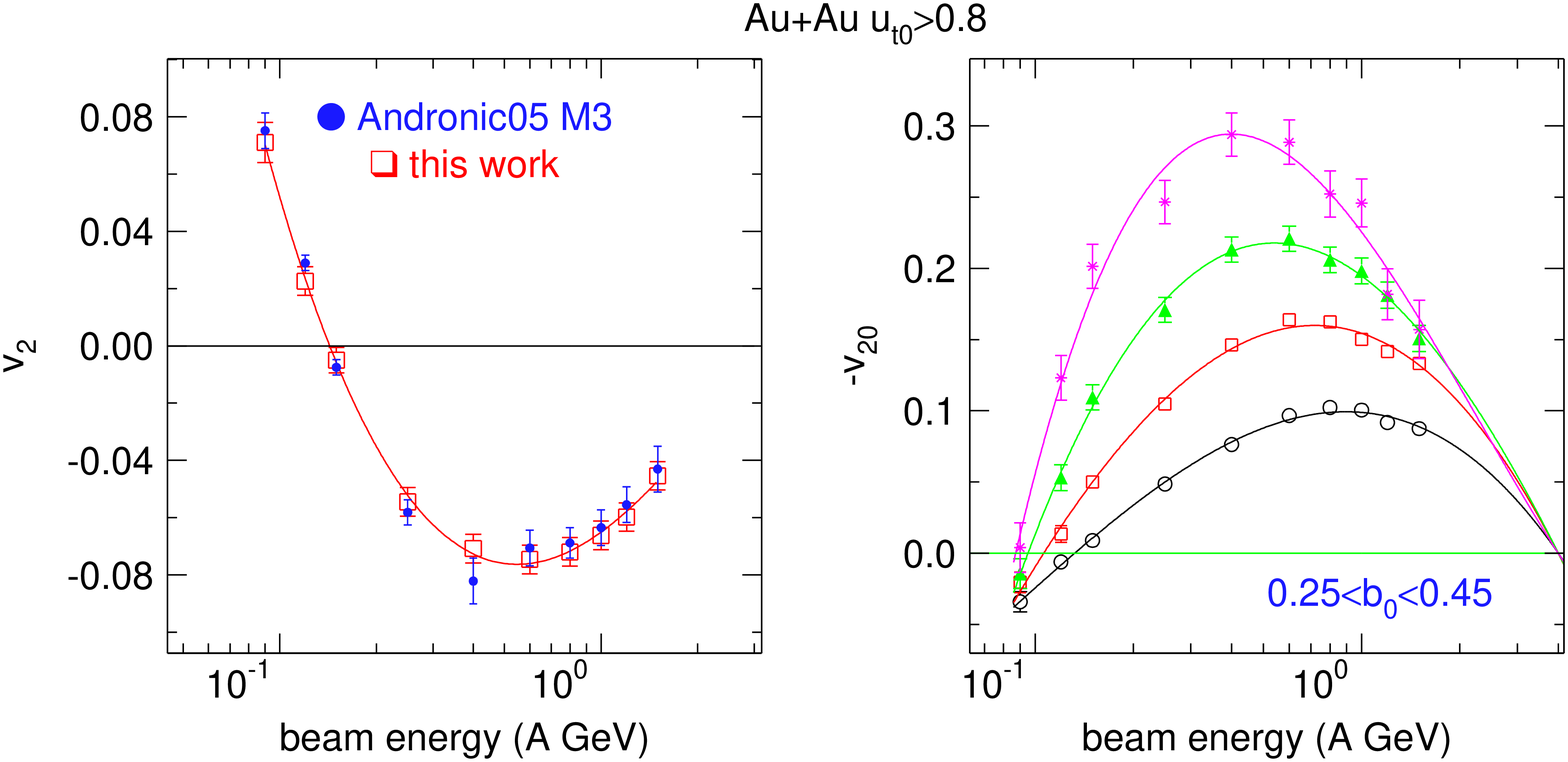,width=150mm}
\hspace*{\fill}
\caption{
%
Excitation functions of
elliptic flow  in Au+Au collisions for centrality $0.25<b_0<0.45$.
Left panel:
Elliptic flow $v_2$ of Z=1 nucleonic particles in Au+Au collisions as a function
of incident energy.
The present  data  (red open squares) are compared to the data
published earlier \cite{andronic05} (blue dots). 
Right panel: Elliptic flow $-v_{20}$
Protons: open black circles, deuterons: open red squares, mass 3: full green
triangles, $\alpha$'s: pink asterisks.
The smooth curves are polynomial fits
along a logarithmic abscissa constrained to traverse zero at $4A$ GeV
 \cite{pinkenburg99}.
A low $p_t$ cut was applied ($u_{t0}>0.8$).
This cut allows to join up the data from two different (low and high beam
energy) runs.
}
\label{v20q-andro}
\end{figure}
We show the earlier FOPI data again here in the left panel of 
Fig. \ref{v20q-andro}
(blue dots) together with our present evaluation which was done with a
completely independent software and is seen to be in excellent
agreement with \cite{andronic06}.
The nominal impact parameter ranges are very similar: 
$b_0$=0.41-0.56 (for M3), $b_0$=0.45-0.55 (this work), both selected by
multiplicity binning.
Also, the same low momentum cut, $u_{t0}>0.8$ was applied in both evaluations.
The smooth curve is a fit of a third degree polynomial in terms of
$\log (E/u)$ to the present data that reproduces the $v_2$ data with an average
accuracy of 0.002.
This comparison also confirms that the  quadrant method, section
\ref{apparatus} and \cite{reisdorf07} is equivalent to taking $<cos(2\phi)>$,
at least at SIS energies.
However, the quadrant method could be used to design a way to correct for
apparatus response asymmetries.
As it turns out, due to the relative flatness of the $v_2(y_0)$ dependences,
Figs. \ref{v2-au600u8Z1A1-grqcv-6}, \ref{v2-au1500c2-grqc14v-6} and 
\ref{v2-ca1500Z1A1-grqcv-6}, the corrections to $v_2$ at mid-rapidity, our
present parameter $v_{20}$, are virtually negligible.

With our present particle identified data, and switching back to our sign
change for positive 'squeeze-outs', we can now plot (in the right panel of
Fig. \ref{v20q-andro}) the separate excitation functions for masses one to four.
To be able to join up the low and high energy data we again need to apply
the common cut $u_{t0}>0.8$.
The smooth curves are polynomial fits along a logarithmic abscissa constrained
to traverse zero at $4A$ GeV to be compatible with the proton data of
\cite{pinkenburg99}.
With this procedure, assuming for lack of better knowledge that the zero
crossing at the high energy end is independent of particle type, we can
summarize the complete range of energies where the phenomenon of
'squeeze-out' is present.
Note that the various curves are not 'scalable' in a simple way along the
$-v_{20}$ ordinate: the locations of the maxima vary with the mass of the
 ejectile.
This finding could be influenced by the strong cut in $u_{t0}$ which affects
the heavier particles more strongly.
More data between $2A$ GeV and $4A$ GeV would be useful: 
as shown before in this section and in \cite{andronic01} the onset of
squeeze-out at the low energy end was found to be complex.


\subsection{Elliptic flow and isospin}\label{v2iso}
 
\begin{figure}
\hspace{\fill}
\epsfig{file=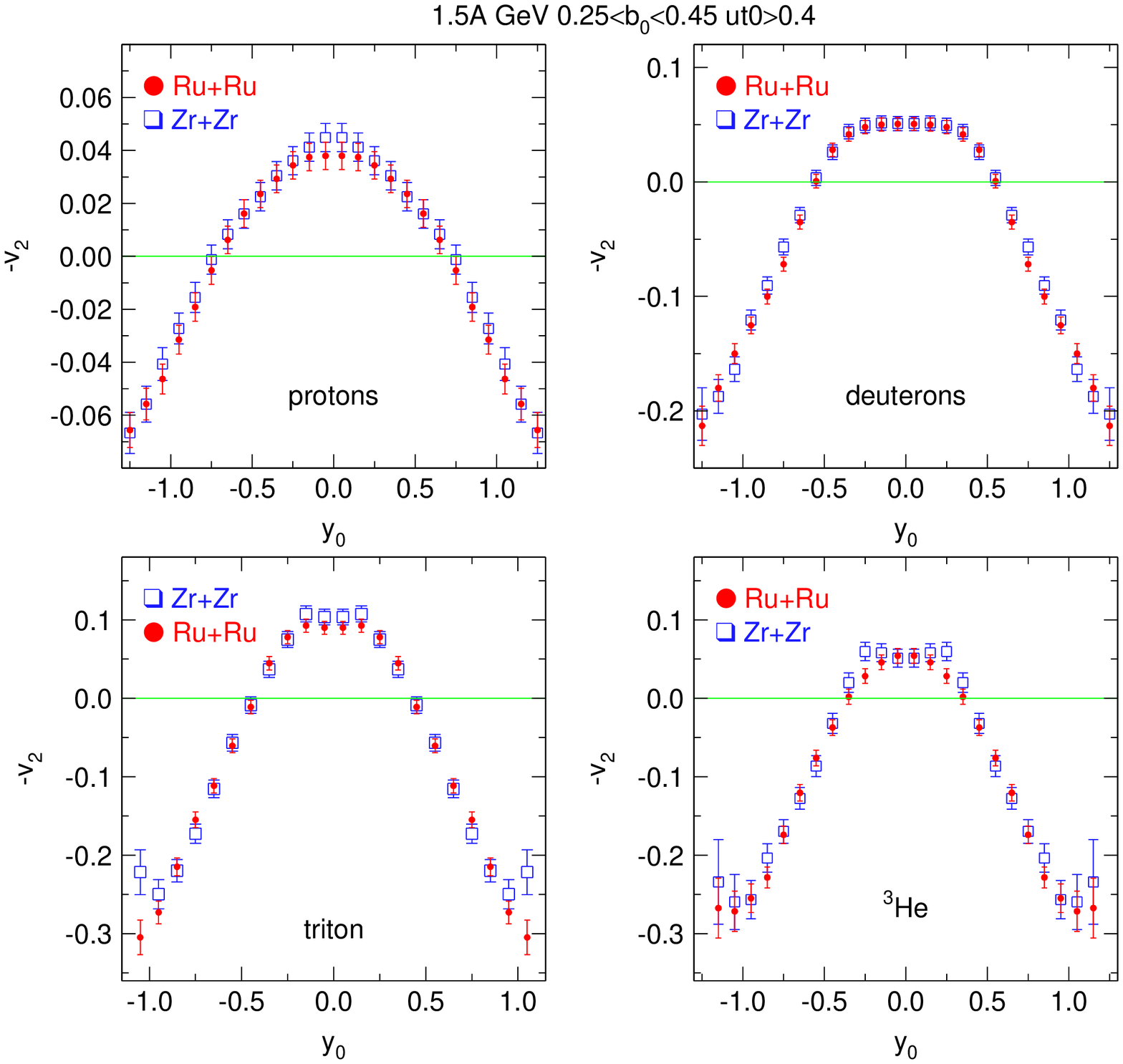,width=160mm}
\hspace{\fill}
\caption{
%
Elliptic flow $-v_2(y_0)$ of protons (upper left),  deuterons (upper right),
tritons (lower left) and $^3$He (lower right) 
 in $^{96}$Zr+$^{96}$Zr (blue open squares) and $^{96}$Ru + $^{96}$Ru (red dots) 
 collisions for $1.5A$ GeV incident energy
 and centrality $0.25<b_0<0.45$.
A cut $u_{t0}>0.4$ is applied.
}
\label{v2-zrzr1500Z1A1-qrqcv}
\end{figure}

\begin{figure}
\hspace{\fill}
\epsfig{file=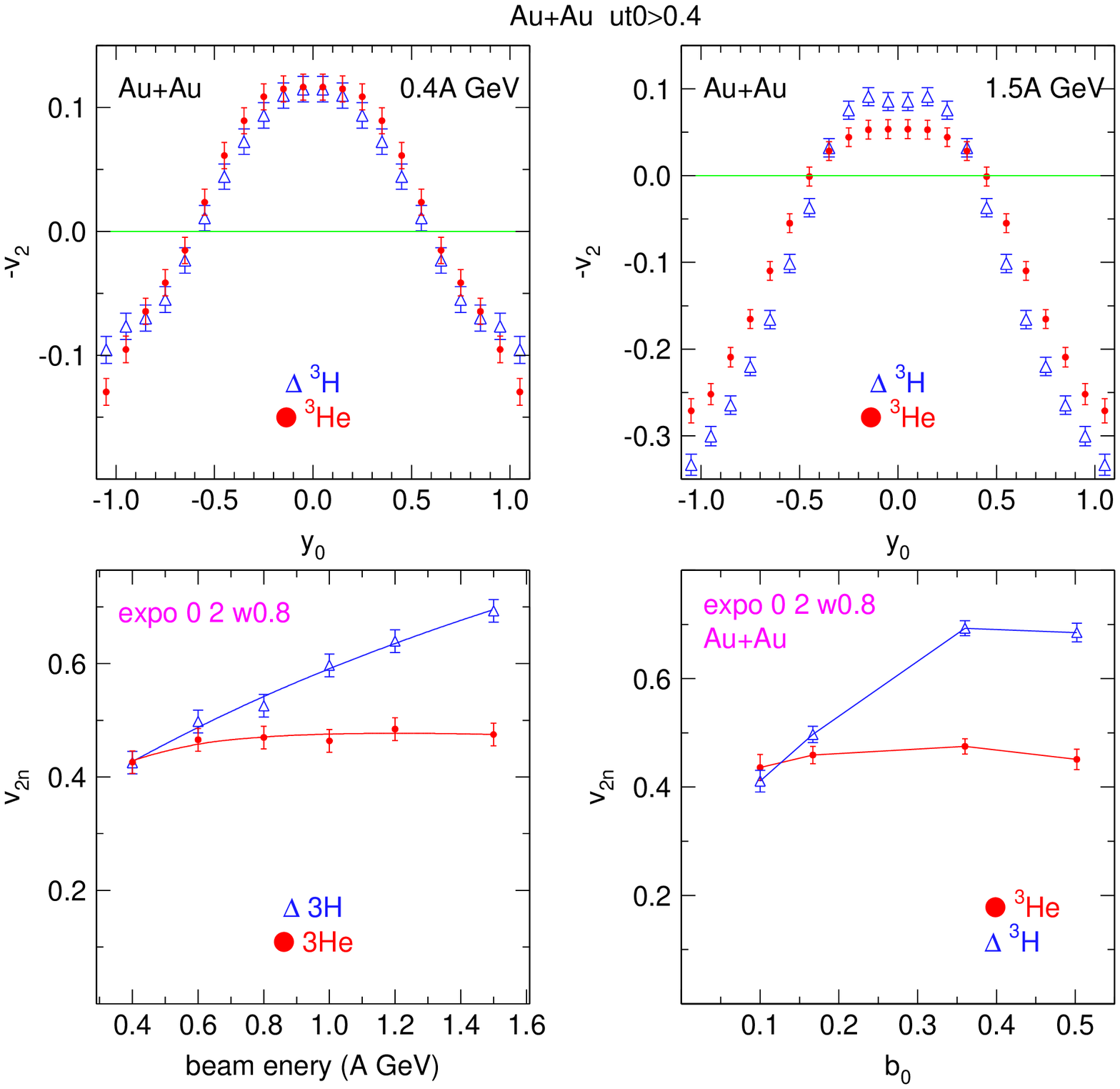,width=160mm}
\hspace{\fill}
\caption{
%
Upper panels:
elliptic flow $-v_2(y_0)$ of $^3H$ (blue) and $^3He$ 
(red) in Au+Au collisions for $0.4A$ and $1.5A$ GeV
 incident energies and centrality $0.25<b_0<0.45$.
Lower panels: elliptic flow characterizer $v_{2n}$ (see text)
 for $^3$H (blue triangles)
and $^3$He (red dots). 
Lower left: beam energy dependence for centrality $0.25<b_0<0.45$,
lower right: centrality dependence for $1.5A$ GeV incident energy. 
}
\label{v2-au1500c2-grqc14v-4}
\end{figure}

\begin{figure}
\hspace{\fill}
\epsfig{file=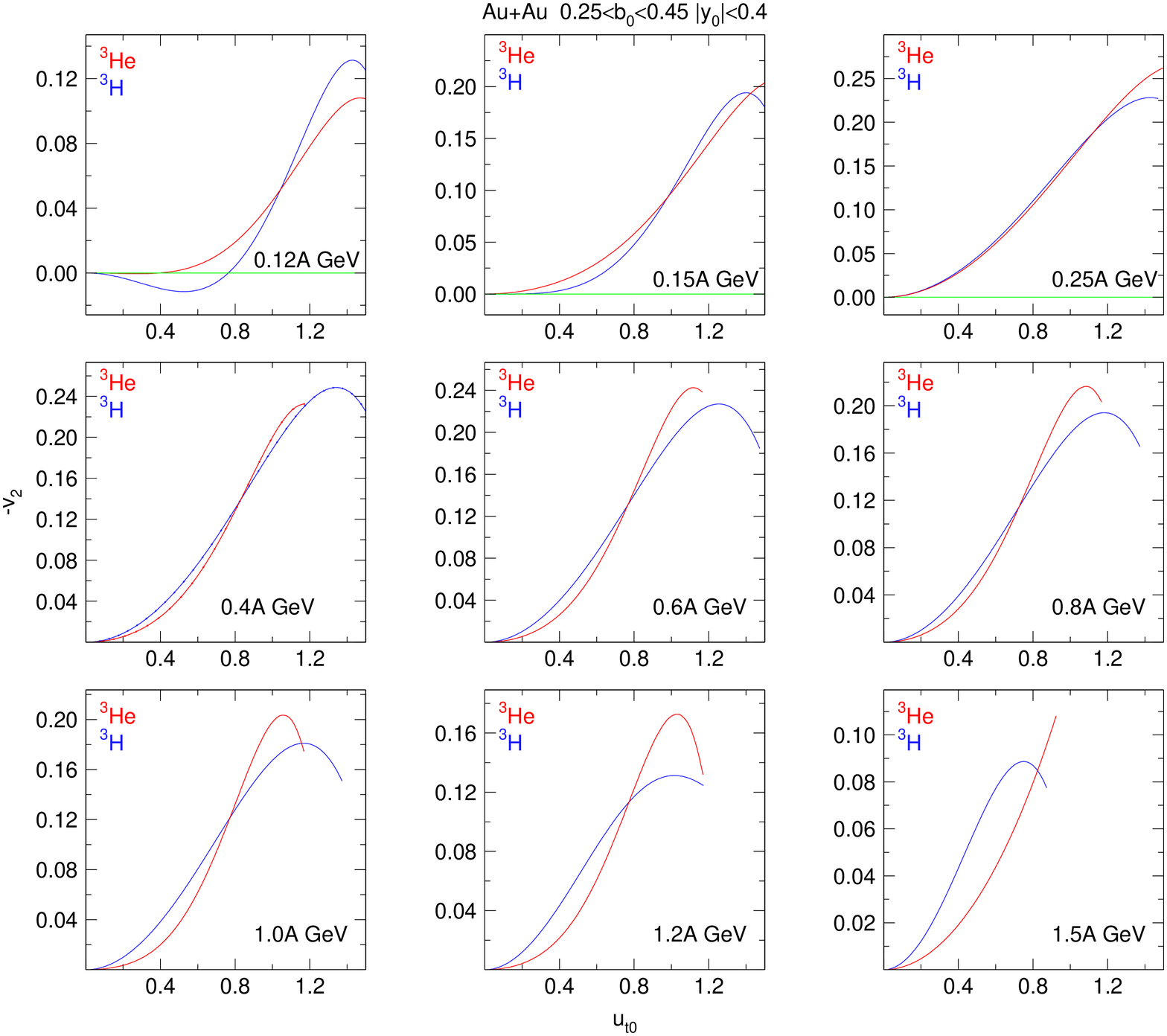,width=160mm}
\hspace{\fill}
\caption{
%
Elliptic flow $-v_2(u_{t0})$ of $^3H$ (blue) and $^3He$ 
(red) in Au+Au collisions for different
indicated incident energies and centrality $0.25<b_0<0.45$.
The smooth curves were obtained from three parameter fits of even polynomials
(excluding the constant term) to the data. The low momentum parts 
(below $u_{t0}=0.6$ for the upper three panels, below 0.4 for the other
panels) are extrapolations of the data.
}
\label{v2ut-au1500c2-grqc14v-A3}
\end{figure}

To study the isospin dependence of elliptic flow we  proceed in analogy to
our data analysis for directed flow and study either differences between
the flow of $^3$H and $^3$He in a fixed $N>Z$ system like Au+Au, 
Figs. \ref{v2-au1500c2-grqc14v-4} and \ref{v2ut-au1500c2-grqc14v-A3},
or look separately at various fixed ejectiles varying the isospin asymmetry
of the system, Fig. \ref{v2-zrzr1500Z1A1-qrqcv}.  

\begin{figure}
\hspace{\fill}
\epsfig{file=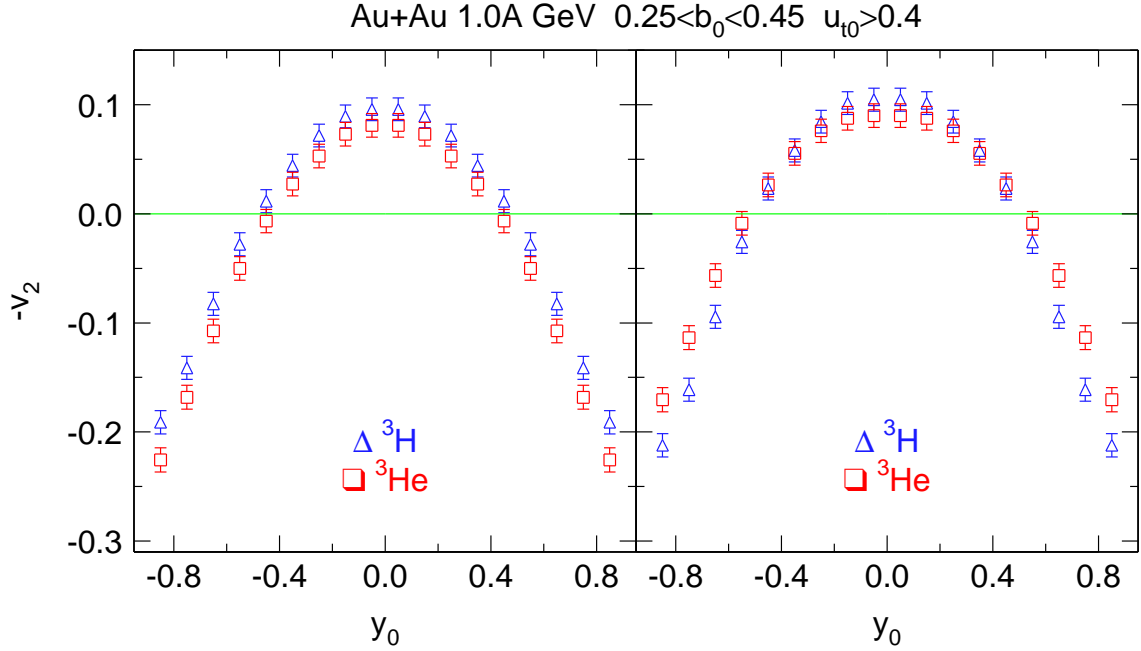,width=160mm}
\hspace{\fill}
\caption{
%
Comparison of the elliptic flow $-v_2(y_0)$ of $^3$H (blue triangles) and
 $^3$He (red squares) 
  in Au+Au collisions for $1.0A$ GeV incident energy
 and centrality $0.25<b_0<0.45$.
A cut $u_{t0}>0.4$ is applied.
Left panel: the data after applying the forward-backward symmetry constraint.
Right panel: same data after additional $v_1$-shift corrections.
}
\label{v2-au1000c2A3-2a}
\end{figure}

However, before we proceed and, since the isospin effects are rather small,
we show again in Fig. \ref{v2-au1000c2A3-2a} the subtle influence of the
 corrections
to the apparatus distortions, here for the $v_2(y_0)$ functions.
The left panel compares the flows of the isospin pair $^3$H/$^3$He after
forward/backward symmetrization (see earlier, Fig. \ref{uty-au1000c2Z1A1}), but
ignoring the shift problems that were illustrated in
 Fig. \ref{v1-au1000Z2A3c}.
In the right panel the final data after full correction are shown.
As we will see later in section \ref{iqmd}, the small resulting differences
over the entire rapidity interval are typical for a small increase of
repulsive forces when switching from $^3$He to $^3$H, a conclusion in line
with our earlier observations using directed flow, section \ref{v1}.

Similar to our observations for directed flow, we see no significant
differences in the rapidity dependences $v_2(y_0)$ for the various
particles (protons, deuterons, tritons and $^3$He in
Fig. \ref{v2-zrzr1500Z1A1-qrqcv}) when switching from $^{96}$Zr + $^{96}$Zr to
$^{96}$Ru + $^{96}$Ru. 
The shown data are for $1.5A$ GeV incident energy with centrality
$0.25<b_0<0.45$ and the general constraint $u_{t0}>0.4$.

We see some interesting effects when we compare the isospin pair of mass three
in Au + Au collisions, Fig. \ref{v2-au1500c2-grqc14v-4}.
While the rapidity dependences for $0.4A$ GeV beams are still rather similar
(upper left panel), a notable difference emerges at $1.5A$ GeV (upper right
panel).
The shape differences between $^3$H and $^3$He data (for $0.25<b_0<0.45$),
larger $|v_2|$ values at midrapidity for $^3$H coupled to a more compact
shape as function of rapidity, corresponds to effectively more
repulsive field gradients.

In order to quantize the differences into one parameter, dubbed $v_{2n}$ and
shown in the lower two panels (left, an excitation function for fixed
centrality, right, centrality dependence for fixed beam enegy),
we have fitted the $v_2(y_0)$ data in a limited range, $|y_0|<0.8$,
setting $-v_2(y_0)=v_{20} + v_{22}\cdot y_0^2$ and defining
$v_{2n}=v_{20}+|v_{22}|$.
The parameter $v_{2n}$ increases both with the value {\it and} the
curvature near mid-rapidity.
With this parameter the difference between the two isotopes grows steadily
with incident energy, being virtually zero at $0.4A$ GeV (lower left panel),
but has the puzzling behaviour to disappear also at high energy ($1.5A$ GeV)
if the centrality is lowered (lower right panel).
An interesting speculation at this point, explaining perhaps the large
isospin difference at higher energies and larger impact parameters, could be
that the effect is dominated by momentum rather than density dependence
\cite{giordano10}.
 In \cite{giordano10} the simulations for Au+Au were limited
to $0.4A$ and $0.6A$ GeV and to our knowledge no covariant formalism was
used, i.e. momentum dependences implied by Lorentz invariance
 were not included.

Using smooth fits to $v_2(u_{t0})$ data similar to those presented in
Fig. \ref{v2ut-au400c2-grqcv}, we are able to show for the isospin pair in 
Fig. \ref{v2ut-au1500c2-grqc14v-A3} the gradual evolution with energy starting
 from $0.12A$ GeV all the way to $1.5A$ GeV.
Although the systematic error (0.1) of the original data for $^3$H and $^3$He 
is rather large (and caused mainly by the 
uncertainty of the correction of symmetry violations by the apparatus),
the evolution of the curves fitted to the data is surprisingly regular
through-out this large span of energy and clearly shows the need for a
systematic data taking.
It will be an interesting task for the future to try to reproduce
such data with transport codes dedicated to isospin dependences
and capable of reproducing cluster yields.
In view of the subtleties of such evolutions in isospin effects
a confirmation of the present data would also be useful. 
 
\subsection{Comparison to other data}\label{v2compa}

\begin{figure}
\begin{minipage}{80mm}
\epsfig{file=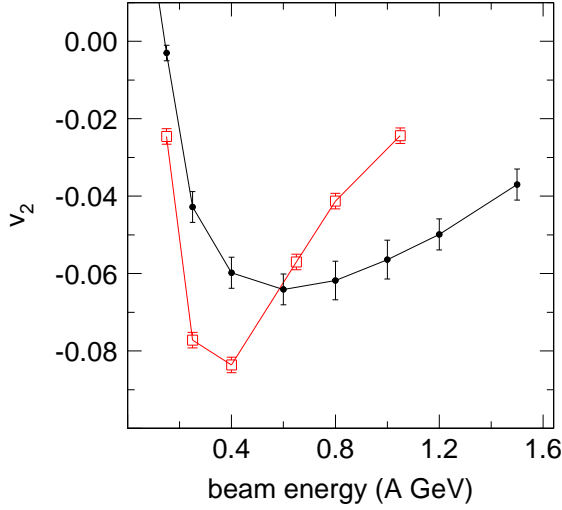,width=78mm}
\end{minipage}
%
\begin{minipage}{80mm}
\caption{
Elliptic flow data for Au+Au collisions as function of beam energy.
Open red squares: data from Ref. \cite{gutbrod90f} for $Z=1$ and 2 particles
in the flow reference frame.
Closed black dots: present data in the beam axis reference frame for $Z=1$ 
and centrality $0.45<b_0<0.55$.
Both types of data are uncorrected for reaction plane resolution.
}
\label{gutbrod90}
\end{minipage}
\end{figure}

In Fig. \ref{gutbrod90} we compare our data for $Z=1$ and centrality
$0.45<b_0<0.55$ to those of Ref. \cite{gutbrod90f} which were the
first 'squeeze-out' data to appear in the refereed literature.
Originally these data were parameterized in terms of the 
'number squeeze-out' parameter $R_n$ which can easily be converted to
the nowadays more common $v_2$ using the simple relation
$2v_2=(1-R_n)/(1+R_n)$.
The authors of Ref. \cite{gutbrod90f} used the flow reference frame
mentioned in the beginning of this section, but to our knowledge no
reaction plane correction was applied.
Therefore, to compare, we have omitted the correction as well (
for the corrected data, see our Fig. \ref{v20q-andro}).
Rotation into the flow axis should maximize the elliptic flow 
(see also \cite{andronic01})
and indeed, for incident energies below $0.5A$ GeV our non-rotated data
show significantly less elliptic flow values.
However, at higher energies our data (despite not including the stronger flow
of $Z=2$ fragments) suggest increasingly more flow than the rotated data.
This is not understood.

\begin{figure}
\hspace{\fill}
\epsfig{file=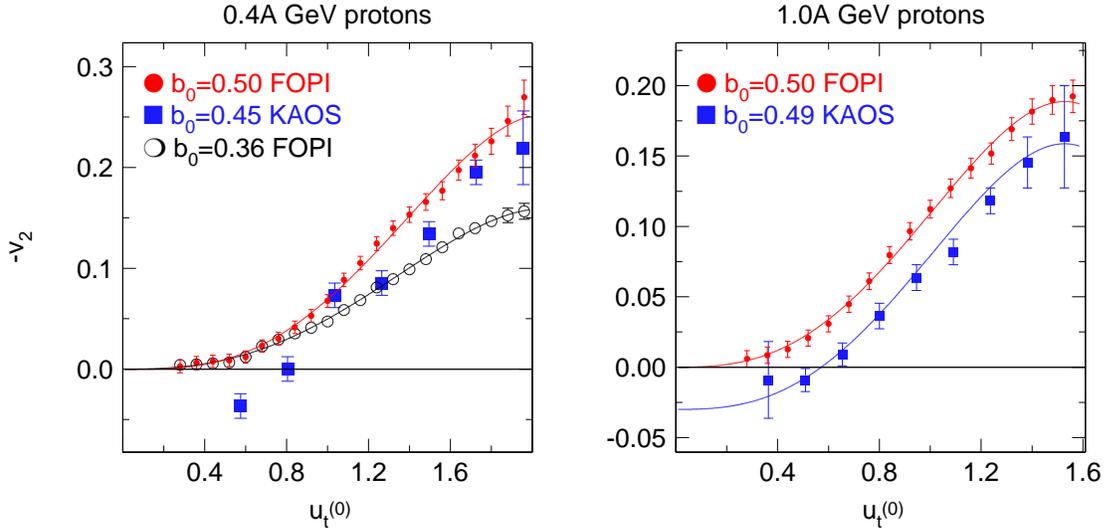,width=150mm}
\hspace{\fill}
\caption{
%
Elliptic flow $-v_{2}(u_{t0})$ of protons  in Au+Au collisions
at $0.4A$ GeV (left panel) and $1.0A$ GeV.
The present data ({\it MUL} selected) are compared to the data
adapted from Ref. \cite{brill96} (KaoS coll.). The two panels show results for
different indicated average centralities.
The KaoS data correspond to blue filled squares.
 The smooth curves to the FOPI data are fits 
 with a polynomial $v_{t22}\cdot u^2_{t0} + v_{t23}\cdot u^3_{t0} +
v_{t24}\cdot u^4_{t0}$. 
In the right panel the FOPI fit is seen to also describe the KaoS
data if it is downshifted by 0.03.
}
\label{brill96-v2ut-2}
\end{figure}

In Fig. \ref{brill96-v2ut-2} we compare to proton flow data from the KaoS
Collaboration \cite{brill96}.
These data, besides the kaon data \cite{sturm01,foerster07},
 played an important role
in some past theoretical efforts to constrain the EoS 
\cite{danielewicz00,danielewicz02}.
We expect the KaoS data to have a better momentum resolution, but the
equally important centrality selection and reaction plane resolution should be
superior in our large acceptance apparatus.

At an incident energy of 0.4A GeV we have unfortunately no direct match of
centralities and show therefore the KaoS data for $b_0\approx 0.45$ together
with our data for two (average) centralities $b_0=0.36$ and $b_0=0.50$.
For large enough momenta there is compatibility of the two sets of data,
but we do not find evidence for a sign change at low $u_{t0}$.
The $1.0A$ data (right panel) match almost perfectly in centrality,
but (see the caption) the KaoS data are systematically downshifted from
the FOPI data and again suggest a sign change at low momenta not seen by us.\\
\vspace{4mm}

\section{Transport model simulations}\label{iqmd}

\begin{figure}
\hspace{\fill}
\epsfig{file=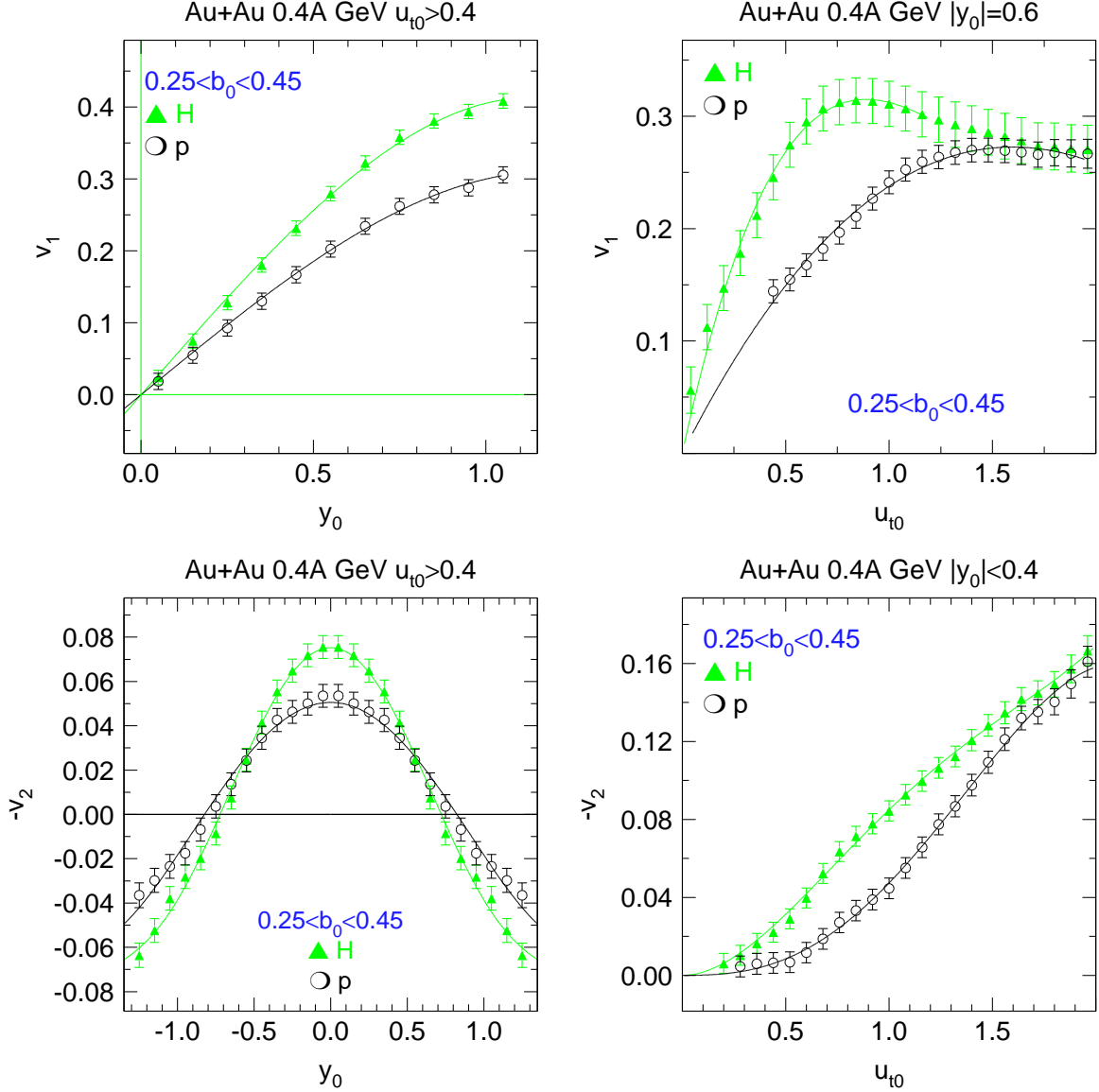,width=160mm}
\hspace{\fill}
\caption{
Comparison of hydrogen and proton flow
 in $0.4A$ GeV Au+Au collisions with centrality
$0.25<b_0<0.45$.
Upper left panel: rapidity dependence of $v_1$ with a constraint $u_{t0}>0.4$.
Upper right panel: $u_{t0}$ dependence of $v_1$ with a constraint 
$0.4<|y_0|<0.8$.
Lower left panel: rapidity dependence of $-v_2$ with a constraint $u_{t0}>0.4$.
Lower right panel: $u_{t0}$ dependence of $-v_2$ with a constraint
 $|y_0|<0.4$.
}
\label{v1-au400Z1A0-grcv-c2}
\end{figure}

Before showing simulations of our data with a transport code, we wish to
stress in Fig. \ref{v1-au400Z1A0-grcv-c2} the importance of having isotope
separated data to do so.
The figure shows for Au on Au collisions the various direct and elliptic flow
projections that we have studied in sections \ref{v1} and \ref{v2}
for protons (black open circles) and for the sum of the three hydrogen
isotopes.
The differences are quite significant and complex.
Clearly, if the relative yield of the various isotopes is not predicted
correctly in the complete three-dimensional momentum space the interpretation
of data that are only charge separated is not reliable.
It is also important to realize this when comparing proton or neutron flow and
hydrogen flow to learn something about isospin dependences.

\begin{figure}
\hspace{\fill}
\epsfig{file=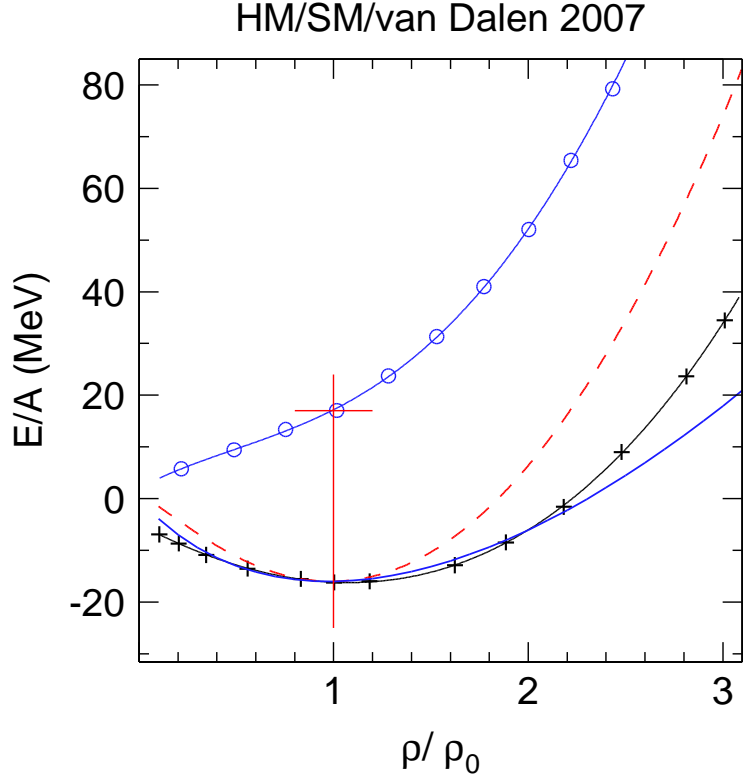,width=120mm}
\hspace{\fill}
\caption{
%
The two equations of state HM (red dashed) and SM (blue continuous)
are compared to symmetric (black crosses) and neutron matter (blue circles) EoS 
obtained using DBHF theory (after \cite{vandalen07}).
}
\label{EOS-vanDalen07-a}
\end{figure}

As transport code we use a quantum molecular dynamics code, IQMD.
The IQMD version we use in the present work largely corresponds to the
description given in Ref.~\cite{hartnack98}. 
More specifically, the following 'standard' parameters were used throughout:
$L=8.66$ fm$^2$ (wave packet width parameter),
 $t=200$ fm (total propagation time), $K=200, 380$ MeV (compressibility of
the momentum dependent soft, resp. stiff EoS), $E_{sym}=25 \rho/\rho_0$ MeV
(symmetry energy, with $\rho_0$ the saturation value of the nuclear density
$\rho$).
The versions with $K=200$, resp. 380 MeV are called IQMD-SM, resp. IQMD-HM.
The clusterization was determined from a separate routine using the minimum
spanning tree method in configuration space with a clusterization radius
$R_c=3$ fm.
\vspace{4mm}\\
We are aware that in the wake of intense nuclear symmetry energy research
more recently 'improved' transport codes have been developed that
address isospin dependences by increasing the relevant parametric flexibility
but not necessarily the general consensus what to conclude:
ImQMD \cite{yzhang08}, IBUU04 \cite{bali04}, UrQMD \cite{qfli11}.
Our motivation to stick for now to the present IQMD version is that we
have used it in two of our earlier papers on pion systematics 
\cite{reisdorf07} and the systematics of central collisions \cite{reisdorf10}
and want here to test IQMD knowing how it performed on the observables
published earlier.
We believe that convincing conclusions on basic nuclear properties require 
a successful simulation of the full set of experimental observables with the
same code using the same physical and technical parameters.
\vspace{4mm}\\
The two options of purely phenomenological {\it cold} nuclear EoS that we
 use are plotted 
in Fig. \ref{EOS-vanDalen07-a} and confronted with a 'microscopic' 
(Dirac-Brueckner-Hartree-Fock, DBHF) calculation \cite{vandalen07}
for symmetric matter.
Also included from the same theoretical work is the {\it cold} EoS for pure
 neutron matter.
These theoretical predictions are fully in-line with what is known so far on
both EoS in the vicinity of the saturation density $\rho_0$.
It is seen that in the density range relevant for SIS energies (up to
$\rho/\rho_0=2.5$) our 'soft' version, SM, is rather close to the theoretical
calculation.
In this density range SM is also close to the EoS dubbed AP4 in 
Refs. \cite{lattimer01,demorest10} which is due to a calculation
\cite{akmal98} based on the Argonne $v_{18}$ two-body potential combined with
the Urbana IX three-body potential and relativistic boost corrections 
(it is dubbed A18+$\delta v$+UIX in Ref.~\cite{akmal98}).
In a recent astrophysical measurement \cite{demorest10} it was shown that
AP4 was compatible with the discovery of a two-solar-mass neutron star.
In order to support such a massive object the nuclear EoS has to be 
sufficiently stiff.
(Actually AP4 has to be shifted down by about three MeV to be in good accord
with the binding energy per nucleon at saturation density $\rho_0$).
\vspace{4mm}\\
The EoS for intermediate values (0 to 1) of the asymmetry parameter
$\delta = (\rho_n -\rho_p)/\rho$ is expected in most of the literature to have
a nearly quadratic dependence on $\delta$,
$E(\rho,\delta)=E_0(\rho) + E_{asy}(\rho)\cdot \delta^2$,
a relation confirmed for example in \cite{vandalen07}.
The $\delta^2$ values available in earthly laboratories are rather modest:
$\delta^2=0.0447$ for the readily available $^{208}$Pb+$^{208}$Pb and
$\delta^2=0.0478$ for a popular radioactive beam system in a foreseeable
future, $^{132}$Sn+$^{124}$Sn.
The binding energy curves for such systems are much closer to 
symmetric nuclear matter than to neutron matter.
\vspace{4mm}\\
We have sorted our simulations by incident energy, as the physics,
and the associated problems understanding it, vary with this externally
diallable parameter.

\subsection{Comparisons at $0.15 A$ GeV.}
\begin{figure}
\epsfig{file=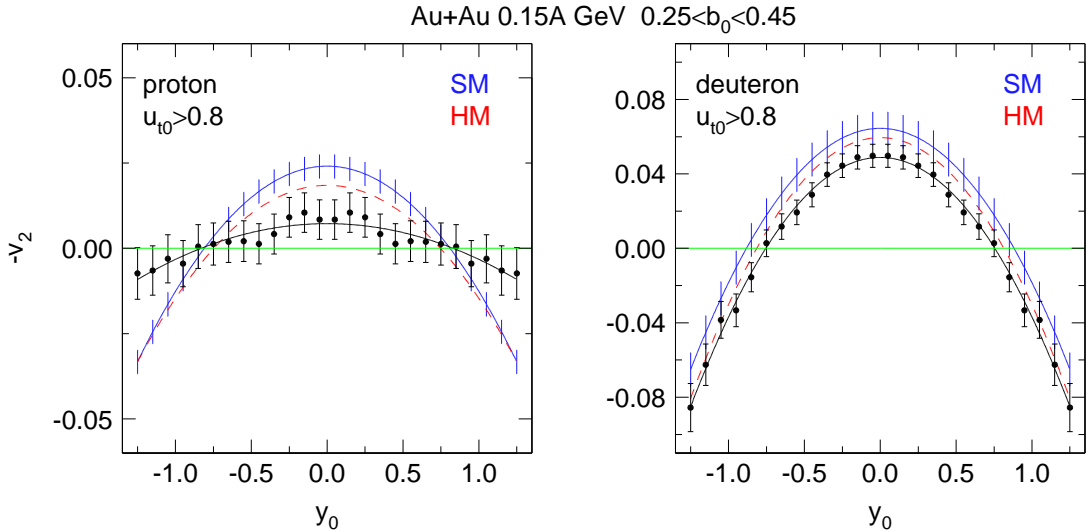,width=150mm}
\caption{
%
Rapidity dependence of
elliptic flow $-v_2$ of protons (left) and deuterons (right)
in Au+Au collisions at $E=0.15A$ GeV for $0.25<b_0<0.45$.
Transverse 4-velocities $u_{t0}$ below 0.8 are cut off.
The curves are fits of $v_{20}+v_{22} y_0^2$ to the  data.
The black dots are the experimental data.
The error bars of the simulated data are statistical assuming the fitted
2-parameter shape is correct and for clarity are indicated only for the
IQMD-SM calculation (blue full curve). The red dashed curve represents the
IQMD-HM calculation.
}
\label{v2-au150c2p-pd}
\end{figure}

We start  at an incident energy still close to the
balance energy, $E=0.15A$ GeV.
In Fig. \ref{v2-au150c2p-pd} we compare the rapidity dependences of elliptic
 flow in Au + Au collisions of centrality $0.25<b_0<0.45$.
Proton data are shown in the left panel, deuteron flow is plotted in the
right panel. 
As for all our low energy data, a constraint $u_{t0}>0.8$ is applied
which is expected to influence the results significantly, but,
as mentioned earlier, preserves the symmetry around mid-rapidity. 
{\it The constraint is of course also applied to the simulation.
In this and all following simulations we always use {\it ERAT}
to select centrality both in the experimental and theoretical events.}
In the figure both the HM and the SM options are shown together with
the data (black dots with systematic errors).
To improve on statistical significance we have fitted
two-parameter parabola's $v_{20} + v_{22} \cdot y_0^2$ to the various
distributions.
At this incident energy 
such shapes spanning beyond the full scaled rapidity gap (-1 to +1)
are seen both in the experiment and the simulation.
  
Some remarks can be made:\\
1) At this low energy  and with the relatively high cut on $u_{t0}$ 
the difference between the various EoS predictions is
rather modest.\\
2) For protons the correct emergence of a positive squeeze-out, i.e.
a  dominant repulsion, is missed and probably influenced by the effective range
of nucleon forces used, that in IQMD depends on partly technical parameters
such as the 'wave packet width' parameter $L$ \cite{hartnack98}.\\
3) Deuteron, and more generally cluster flow gives more prominent signals
and, surprisingly, is better reproduced than proton flow.

\begin{figure}
\hspace{\fill}
\epsfig{file=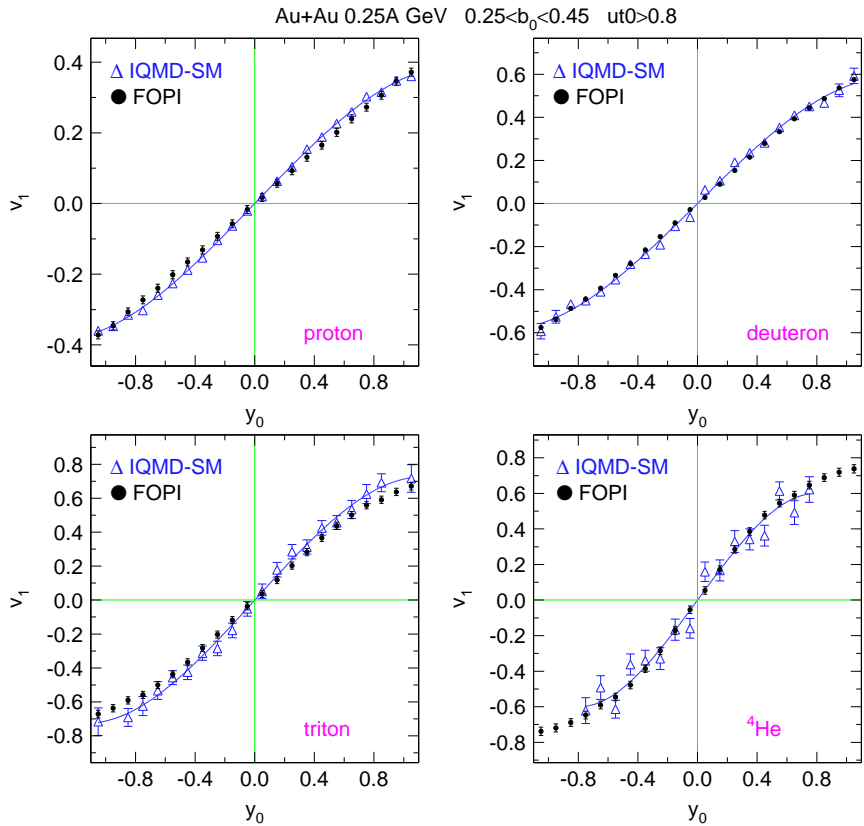,width=120mm}
\hspace{\fill}
\caption{
%
Directed flow $v_1$ of protons, deuterons (upper panels) and of $^3$H and
$^4He$ clusters emitted 
in Au+Au collisions at $E/u=0.25A$ GeV 
and centrality $0.25<b_0<0.45$.
Transverse 4-velocities $u_{t0}$ below 0.8 are cut off.
The small filled black circular dots represent the experimental data.
The smooth curves are least squares fits of 
$v_1= v_{11}\cdot y_0 + v_{13}\cdot y_0^3$
to the simulation results (open blue triangles) 
with  a soft EoS.
}
\label{v1-au250c2-pdta}
\end{figure}

\subsection{Comparisons at $0.25A$ GeV.}

Moving to $0.25A$ GeV where definitely repulsive behaviour is established
in the Au + Au system, we show in Fig. \ref{v1-au250c2-pdta} directed flow data
for protons, deuterons, tritons and $^4$He together with a simulation 
using the SM option.
For the chosen centrality, $0.25<b_0<0.45$ and cutting off momenta
corresponding to $u_{t0}<0.8$, we find an amazing agreement between
experiment and simulation (note the changing ordinate scales in the
various panels).
In view of the fact that absolute yields of such clusters are not
well reproduced by the present IQMD version \cite{reisdorf04b},
 this is not trivial.
Of course, as mentioned earlier, missing the relative yields, the
flows of $Z=1$ or 2 fragments, not separated by mass, would be off.

The success of IQMD at incident energies below $0.4A$ GeV where the
sensitivity to the EoS is still weak can be seen as an important step towards
making sure that some of the technical parameters of the code have
appropriate values.
The possible influence of these parameters was discussed in \cite{hartnack98}
and more recently, for low $E/u$, in \cite{qfli11}.
\subsection{Comparisons at $0.4A$ GeV}

At an energy of $0.4A$ GeV we start considering more seriously the
differences between soft and stiff EoS options and use for that the extended
information of both directed and elliptic flow,
watching the rapidity as well as the transverse momentum dependences.

\begin{figure}
\hspace{\fill}
\epsfig{file=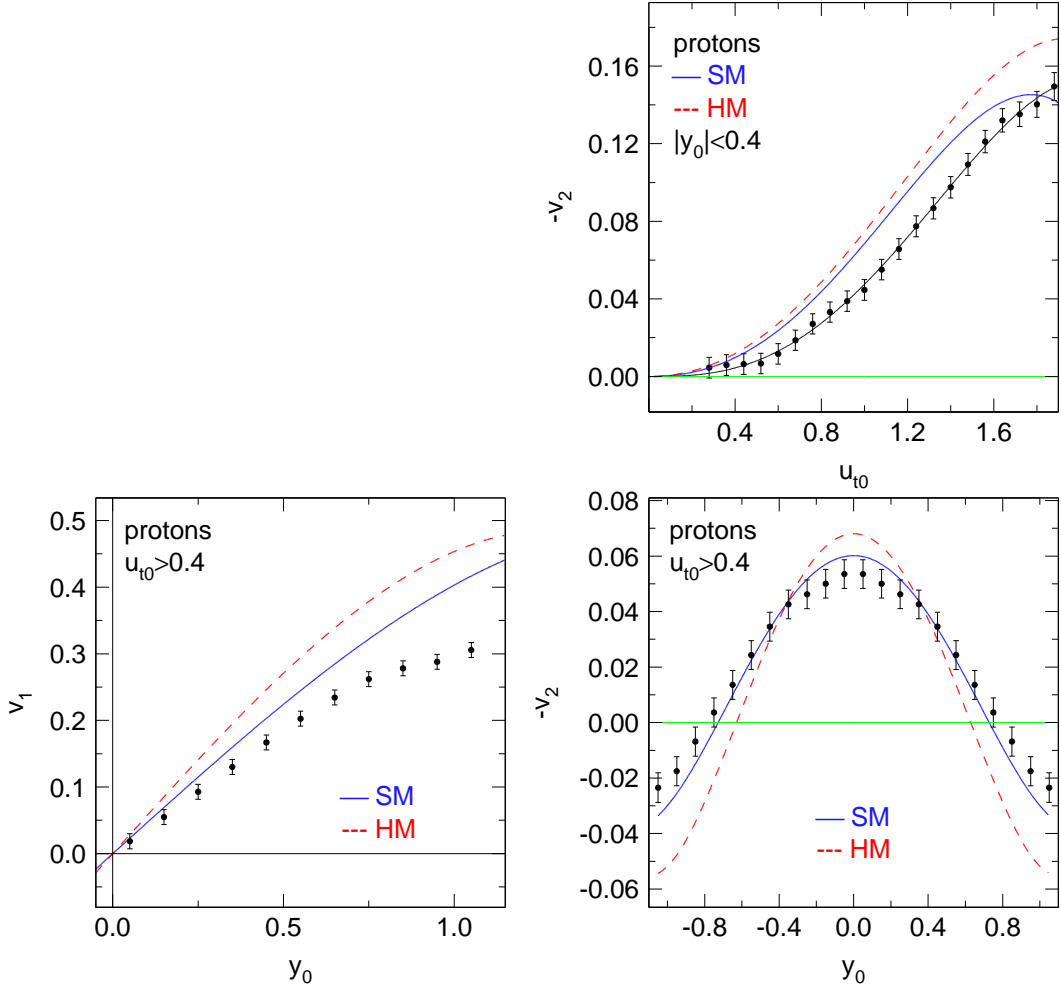,width=150mm}
\hspace{\fill}
\caption{
Flow of protons in Au+Au collisions at $E=0.4A$ GeV
and centrality $0.25<b_0<0.45$.
Smooth fits to simulation data (red dashed: stiff EoS HM, 
blue full: soft EoS SM)
are compared to FOPI data (black filled dots). 
Transverse 4-velocities $u_{t0}$ below 0.4 are cut off.
The $u_{t0}$ dependence of $(-v_2)$ (upper right panel)
is shown within a cut $|y_0|<0.4$.
}
\label{v1-au400c2p}
\end{figure}

\begin{figure}
\hspace{\fill}
\epsfig{file=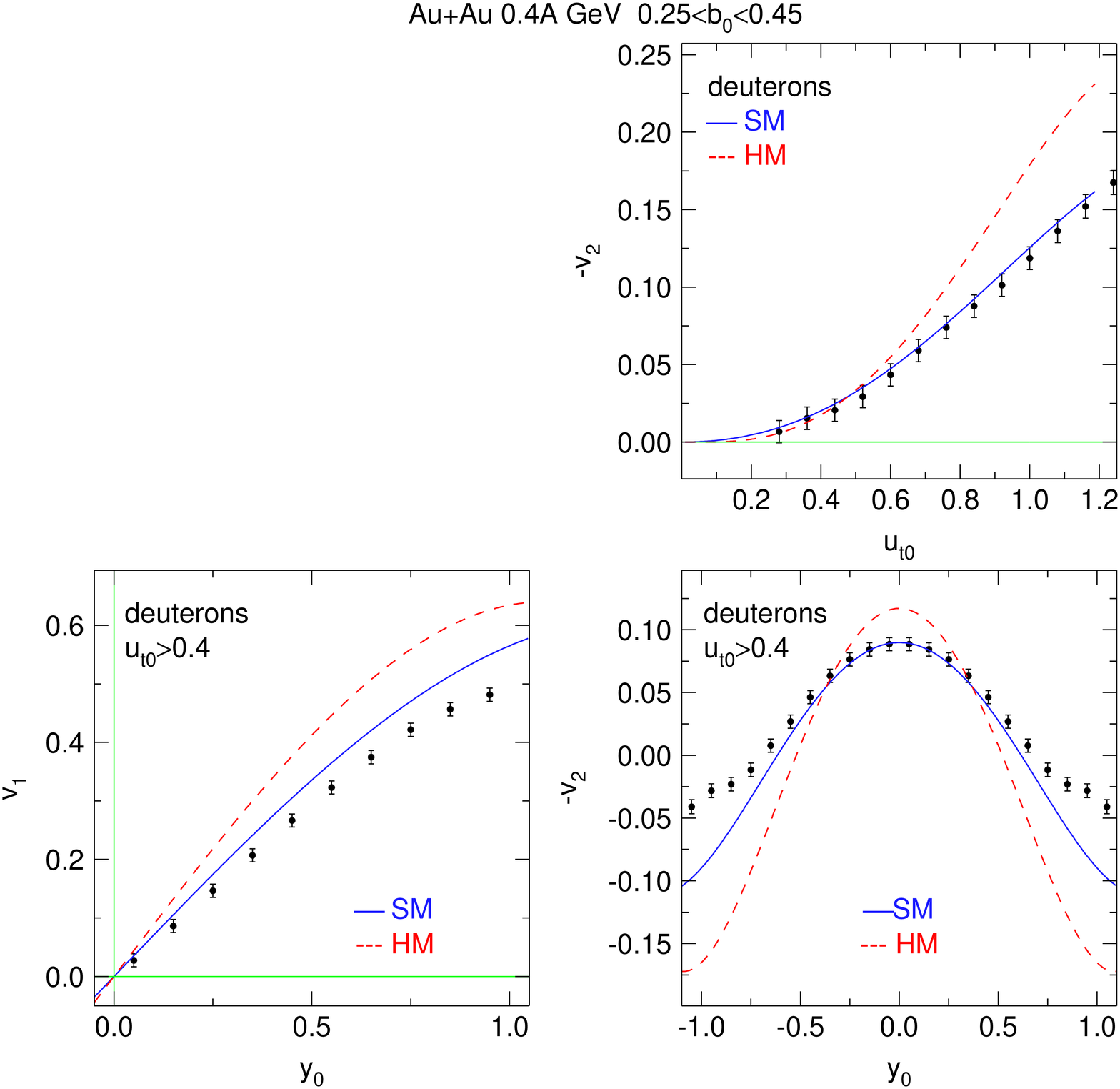,width=150mm}
\hspace{\fill}
\caption{
Flow of deuterons in Au+Au collisions at $E=0.4A$ GeV
and centrality $0.25<b_0<0.45$.
Smooth fits to simulation data (red: stiff EoS HM, blue: soft EoS SM)
are compared to FOPI data (black filled dots). 
Transverse 4-velocities $u_{t0}$ below 0.4 are cut off.
The $u_{t0}$ dependence of $(-v_2)$ (upper right panel)
is shown within a cut $|y_0|<0.4$.
}
\label{v1-au400c2Z1A2-3}
\end{figure}

\begin{figure}
\hspace{\fill}
\epsfig{file=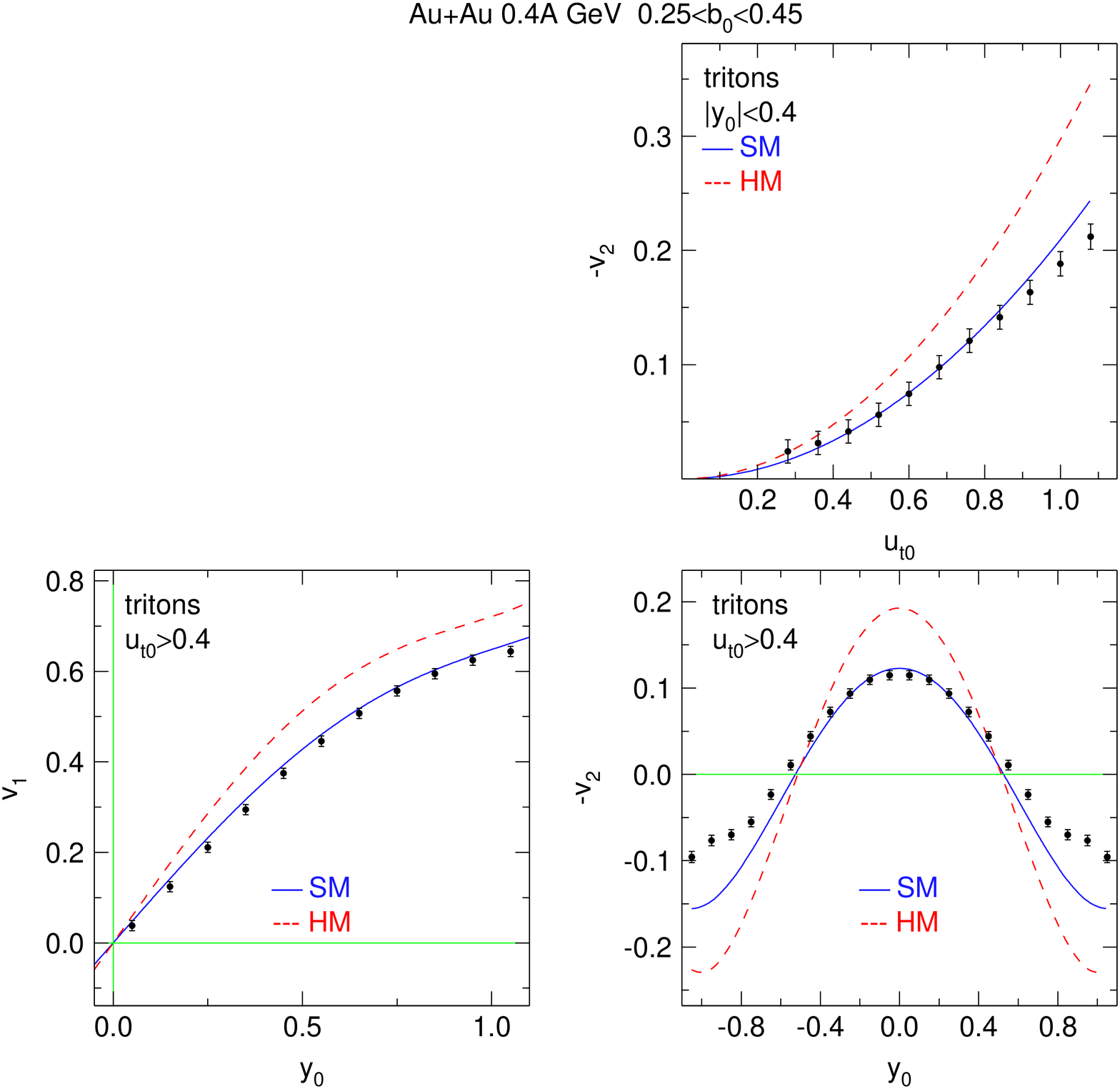,width=150mm}
\hspace{\fill}
\caption{
Flow of tritons in Au+Au collisions at $E=0.4A$ GeV
and centrality $0.25<b_0<0.45$.
Smooth fits to simulation data (red dashed: stiff EoS HM,
 blue full: soft EoS SM)
are compared to FOPI data (black filled dots). 
Transverse 4-velocities $u_{t0}$ below 0.4 are cut off.
The $u_{t0}$ dependence of $(-v_2)$ (upper right panel)
is shown within a cut $|y_0|<0.4$.
}
\label{v1-au400c2t}
\end{figure}

In the three panels of Fig. \ref{v1-au400c2p} this is done for protons,
in Fig. \ref{v1-au400c2Z1A2-3} it is shown for deuterons and in 
Fig. \ref{v1-au400c2t} for tritons.
Again the centrality is $0.25<b_0<0.45$, but at this higher energy we can use
the less restrictive constraint $u_{t0}>0.4$ for the momentum integrated
rapidity dependences a fact which tends to increase the sensitivity to
the 'cold' EoS: low momentum particles are more affected by mean fields.
The momentum dependences (upper right panels) are for the rapidity interval
$|y_0|<0.4$.
In all the panels the smooth curves are polynomial least squares fits using
the same fitting parameters (at most three) and abscissa ranges for data
and simulation. 
As shown specifically for the $v_2(u_{t0})$ data such fits reproduce the
experimental data well within their error bars.

Looking at the figures we can note the following:\\
1) There is a clear preference for the soft version, SM, of the EoS.\\
2) The conclusion is the same for elliptic flow as for directed flow.\\
3) Amazingly, again, the rendering of the proton data is less well achieved
than for the deuteron and it is best for triton clusters.\\
4) Looking especially at $v_2(y_0)$ we see an increasing sensitivity to the
EoS as the mass of the ejectile is increased (right lower panels).

\begin{figure}
\hspace{\fill}
\epsfig{file=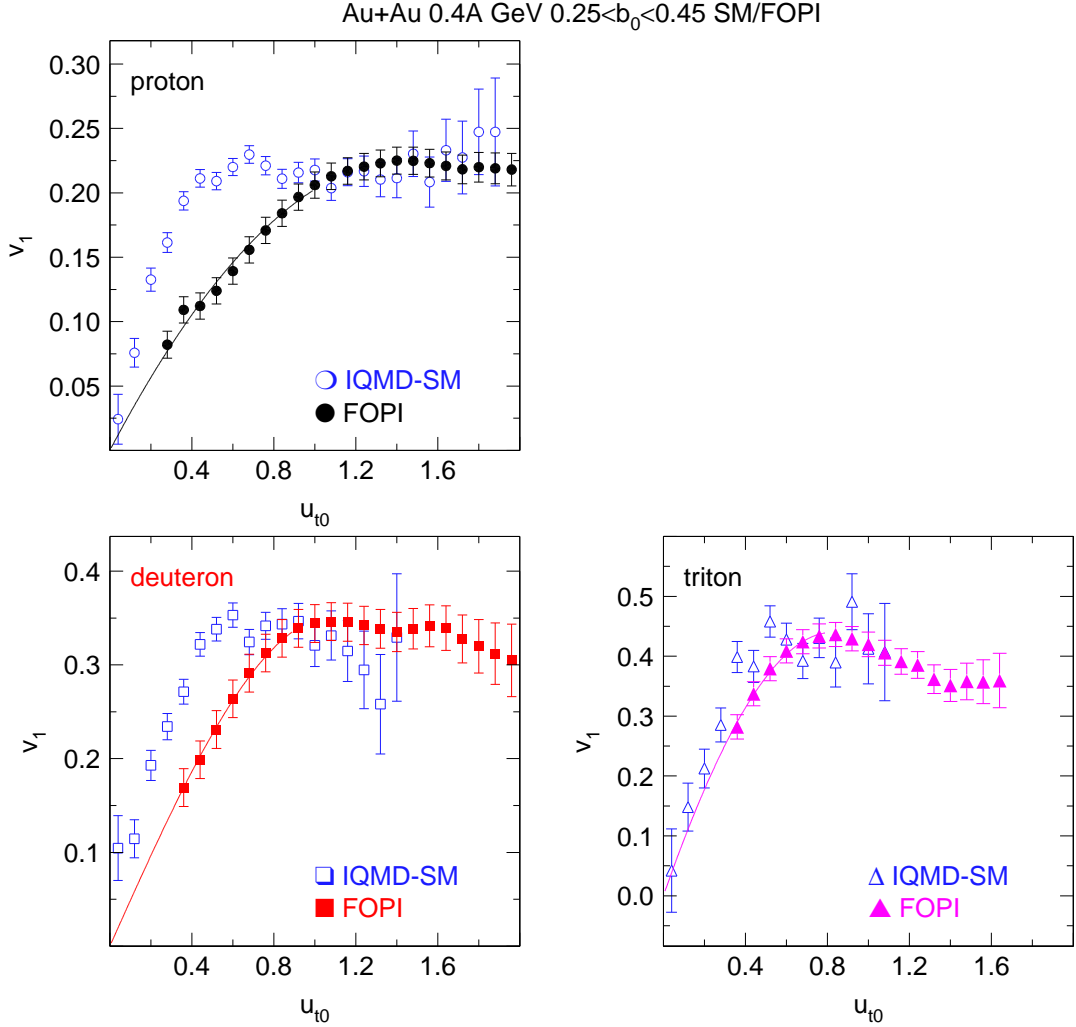,width=150mm}
\hspace{\fill}
\caption{
Flow $v_1(u_{t0})$ of hydrogen isotopes in Au+Au collisions at $E=0.4A$ GeV
and centrality $0.25<b_0<0.45$. The rapidity is constrained to the
interval $0.3<|y_0|<0.7$.
The experimental data (full symbols) are compared to simulations using
IQMD-SM.
Smooth fits to FOPI data are used to extrapolate to $u_{t0}=0$. 
}
\label{v1ut-au400c2pd-3}
\end{figure}

While the comparison of experimental data with the simulation,
in particular the SM version of it, so far looks quite encouraging,
the attentive reader will notice that we have not, in the last
three figures, compared to the momentum dependence of {\it directed} flow.
This is done for the three hydrogen isotopes in Fig. \ref{v1ut-au400c2pd-3}
for the SM option showing an intriguing discrepancy with experiment
that is growing as the emitted particle becomes lighter.
For a centrality $0.25<b_0<0.45$ we show the projection of $v_1$ on the
$u_{t0}$ axis in a rapidity interval $0.3<|y_0|<0.7$.
While for the data the slope of $v_1(u_{t0})$ at low momenta varies
significantly with the mass of the particle, being steepest for the
triton, this effect is completely missing in the simulation.
In particular the gentle slope of $v_1$ for protons is not reproduced.

Searching for reasons why the {\it proton} flow data are more poorly
reproduced by the simulations, it is instructive to come back to 
Fig.~\ref{v1-au400Z1A0-grcv-c2} which compared measured hydrogen ($Z=1$)
 flow with the
measured proton ($A=1$) flow: one sees that the differences between
($Z=1$) and ($A=1$) look qualitatively very much like the differences in
Figs.~\ref{v1-au400c2p} and \ref{v1ut-au400c2pd-3} between the $SM$ calculation
 and the proton data.
This suggests that the lack of sufficient clusterization (rather than an
inappropriate EoS) in the simulation is at the origin of the deficiency of
the theory.
The deuteron and the triton data are less influenced by this effect and
hence better reproduced.

In view of Fig.~\ref{v1ut-au400c2pd-3} another interpretation cannot
 be excluded as well:
It could be that an important quantum mechanical effect is missing at
freeze-out in the (quasi-classical) simulation: the Fermi-motion of
the clusters in the medium, which is expected to be largest for
single nucleons.
As Fermi motion from a flow (i.e. hydrodynamic) point of view is chaotic, it
will tend to 'soften' the visible flow just as thermal motion does.
As a consequence $v_1(y_0)$ for protons (Fig. \ref{v1-au400c2p}) 
is more affected than $v_1(y_0)$ for tritons (Fig. \ref{v1-au400c2t}):
Brownian motion is less imporant for heavier clusters.
These conjectures
 require further investigation as they might shed additional light
on the clusterization mechanism.
As we shall see, this deficiency of the simulation is not unique to
$0.4A$ GeV.
Also, if one looks back at Figs. \ref{v1-au250c2-pdta} to
 \ref{v1-au400c2t} one notices that the
$v_1(y_0)$ data of protons and deuterons are better described by the
simulation for $0.25A$ GeV than for $0.4A$ GeV. 
With the information from Fig. \ref{v1ut-au400c2pd-3} we can say that this
 is probably due to
the higher cut on $u_{t0}$ operated on the $0.25A$ GeV data (0.8 instead of
0.4).

\subsection{Comparisons at $1.0A$ GeV}

\begin{figure}
\hspace{\fill}
\epsfig{file=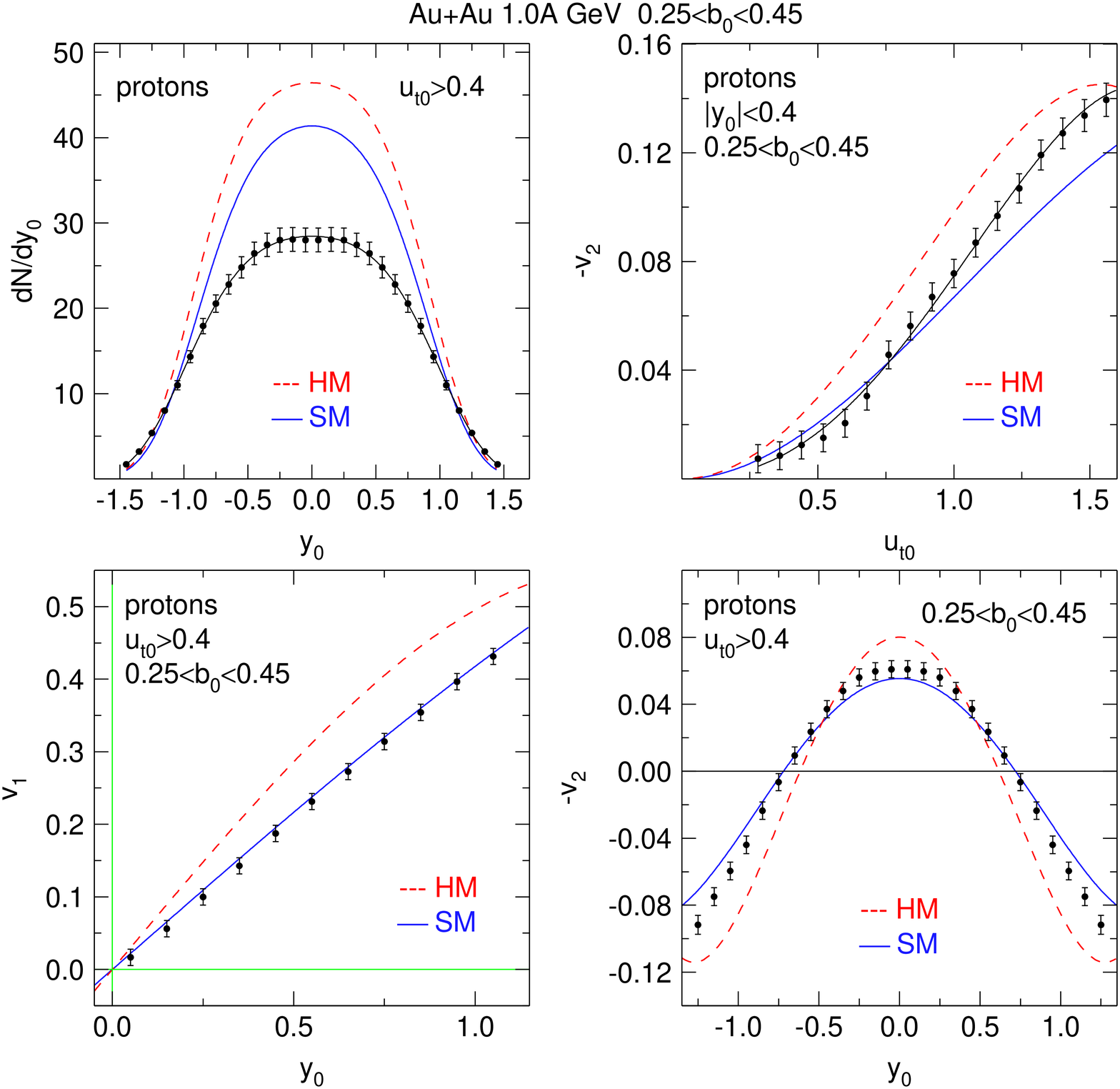,width=150mm}
\hspace{\fill}
\caption{
Flow of protons in Au+Au collisions at $E=1.0A$ GeV
and centrality $0.25<b_0<0.45$.
The simulation data from IQMD-SM are plotted as smooth full curves (blue),
from IQMD-HM as dashed curves (red)
FOPI data as black dots.
The upper right panel shows the $u_{t0}$ dependence in the indicated
$y_0$ bin, the lower two the rapidity dependences integrated over $u_{t0}$,
but constrained to $u_{t0}>0.4$.
Directed flow, $v_1$, is on the left, elliptic flow, $-v_2$, on the right.
The rapidity distributions in the upper left panel are also constrained to
$u_{t0}>0.4$.
}
\label{v1-au1000c2p-4}
\end{figure}

\begin{figure}
\hspace{\fill}
\epsfig{file=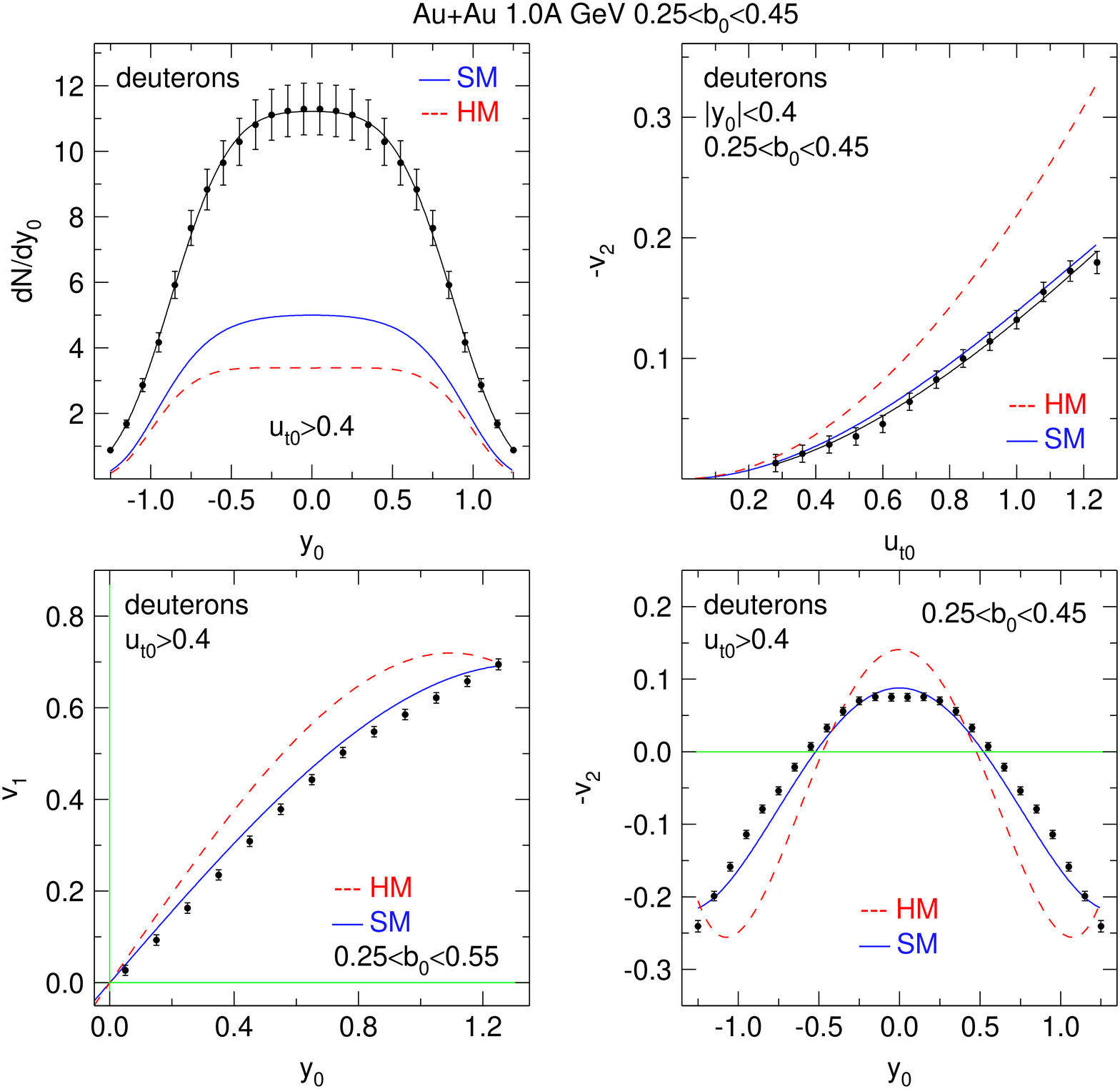,width=150mm}
\hspace{\fill}
\caption{
Flow of deuterons in Au+Au collisions at $E=1.0A$ GeV
and centrality $0.25<b_0<0.45$.
The simulation data from IQMD-SM are plotted as smooth full curves (blue),
from IQMD-HM as dashed curves (red)
FOPI data as black dots.
The upper right panel shows the $u_{t0}$ dependence in the indicated
$y_0$ bin, the lower two the rapidity dependences integrated over $u_{t0}$,
but constrained to $u_{t0}>0.4$.
Directed flow, $v_1$, is on the left, elliptic flow, $-v_2$, on the right.
The rapidity distributions in the upper left panel are also constrained to
$u_{t0}>0.4$.
}
\label{v1-au1000c2d-4}
\end{figure}

\begin{figure}
\hspace{\fill}
\epsfig{file=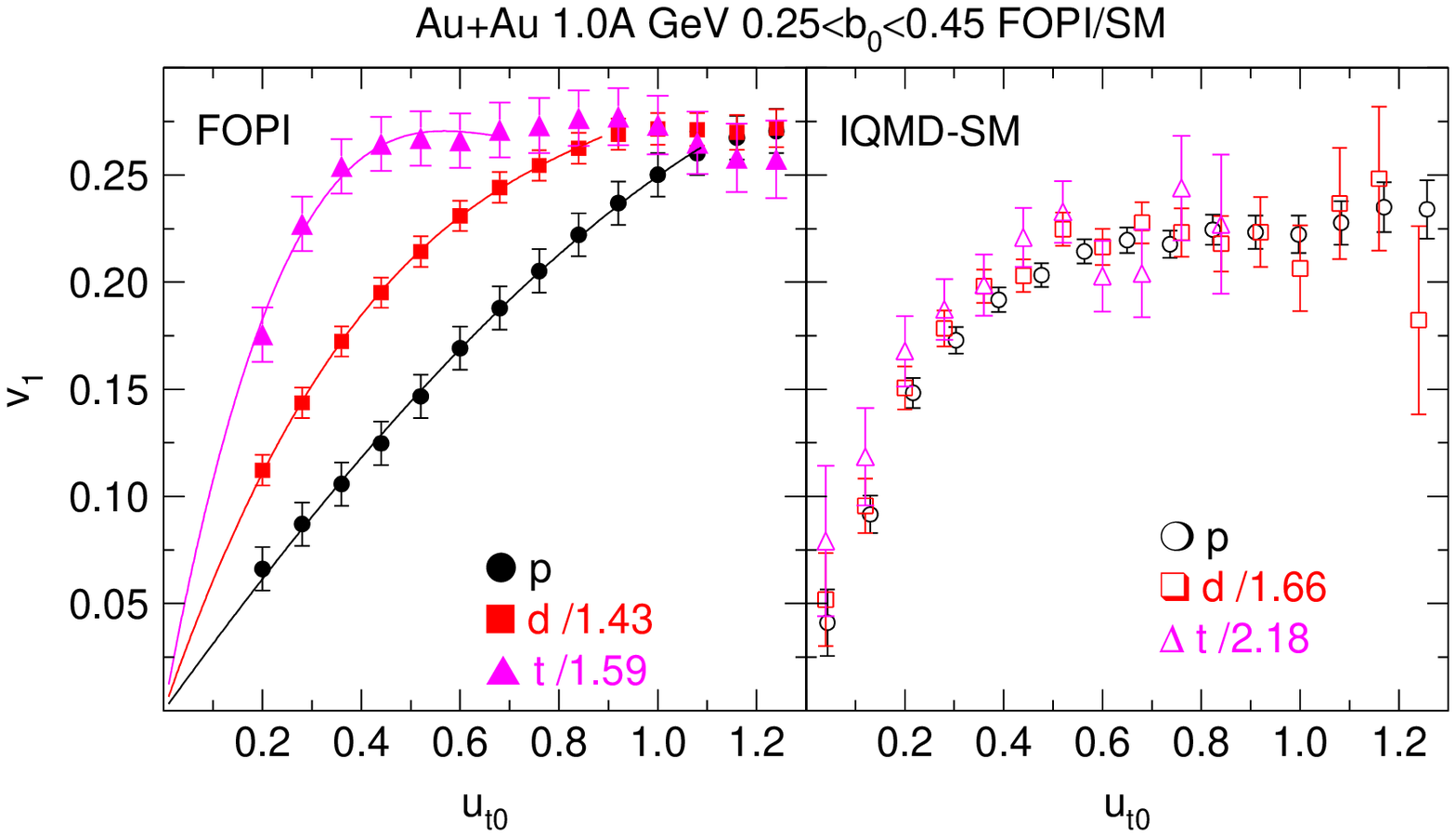,width=150mm}
\hspace{\fill}
\caption{
Flow $v_1(u_{t0})$ of hydrogen isotopes in Au+Au collisions at $E=1.0A$ GeV
and centrality $0.25<b_0<0.45$. The rapidity is constrained to the
interval $0.3<|y_0|<0.7$.
The experimental data (left panel) are compared to simulations using
IQMD-SM (right panel).
The ordinates of the deuteron and triton data are rescaled as indicated
in the panels.
Note the rescaling is different in the simulation.
Smooth fits to FOPI data are used to extrapolate to $u_{t0}=0$. 
}
\label{v1ut-au1000c2pdt-2}
\end{figure}

We proceed at an energy of $1A$ GeV in a similar way to what we did at 
$0.4A$ GeV showing flow data for protons, Fig. \ref{v1-au1000c2p-4}, and
deuterons, Fig. \ref{v1-au1000c2d-4} and also the remarkable disagreement with
the simulation concerning the mass dependence of $v_1(u_{t0})$,
Fig. \ref{v1ut-au1000c2pdt-2}.
The same cuts in scaled units as we used at $0.4A$ GeV are applied.

As before, the SM option gives a better description of the data,
especially of the momentum integrated rapidity dependences.
For the deuteron data, due to the higher sensitivity of clusters,
the preference for SM over HM is even more convincing.
In the upper left panels of Figs. \ref{v1-au1000c2p-4} and \ref{v1-au1000c2d-4}
we have also plotted the rapidity distributions of the yields,
 applying here also the
$u_{t0}>0.4$ constraint both for the $v_1(y_0)$ and $v_2(y_0)$ data.
Comparing to the simulation, we see again the now well known 
overestimation of single nucleon emissions accompanied for deuterons by the
underestimation of deuteron yields.
The more interesting point, made by us earlier \cite{reisdorf04b,reisdorf10},
is that when the system is stiffer (HM, dashed curves) this effect is even
more pronounced: there are less clusters.
We have associated this with the fact that stiff systems achieving less
maximal compression during the collision, undergo later a less efficient
cooling in the expansion period.

In contrast to the rapidity dependences, the momentum dependences of
proton flow are not so well reproduced. 
In Fig.~\ref{v1-au1000c2p-4} for example it is seen that for high momenta
($u_{t0}>1$) the preference for a given EoS is no longer clear.
In Fig. \ref{v1ut-au1000c2pdt-2}, where we show $v_1(u_{t0})$ for the three 
hydrogen isotopes in the rapidity interval $0.3<|y_0|<0.7$, we are
stressing again the qualitative discrepancy between experiment (left panel)
and simulation (right panel) by rescaling the $v_1$ of deuterons and
tritons so that they agree with the proton value at high scaled momenta.
(The factors are indicated in the figures).
While the IQMD-SM predictons are then superimposable for the three isotopes,
this is clearly not the case for the experimental data.
The latter show gradually steeper slopes at low $u_{t0}$ when shifting from
protons to deuterons and then to tritons.

\subsection{Comparisons at $1.5A$ GeV}

\begin{figure}
\hspace{\fill}
\epsfig{file=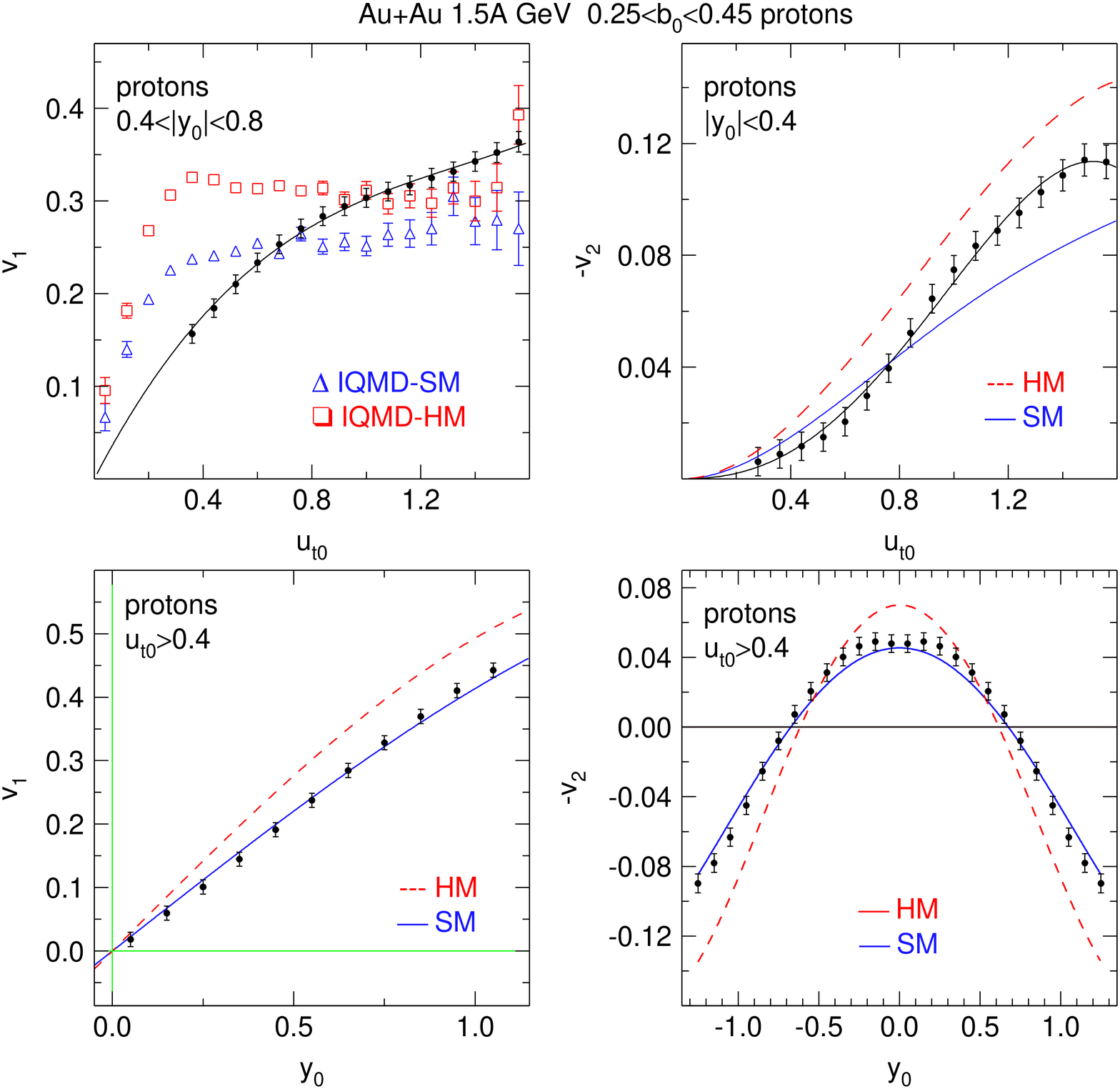,width=150mm}
\hspace{\fill}
\caption{
%
 Proton flow for  centrality $0.25<b_0<0.45$ 
 in collisions of Au+Au at
1.5A GeV.
The left panels show directed flow, elliptic flow is plotted in the
right panels. The lower panels show the rapidty dependence, while
the upper panels illustrate the momentum dependence.
Both the simulations (SM and HM) are shown together with the data (black dots).
For further details see text.
}
\label{v1-au1500c2p-4}
\end{figure}

\begin{figure}
\hspace{\fill}
\epsfig{file=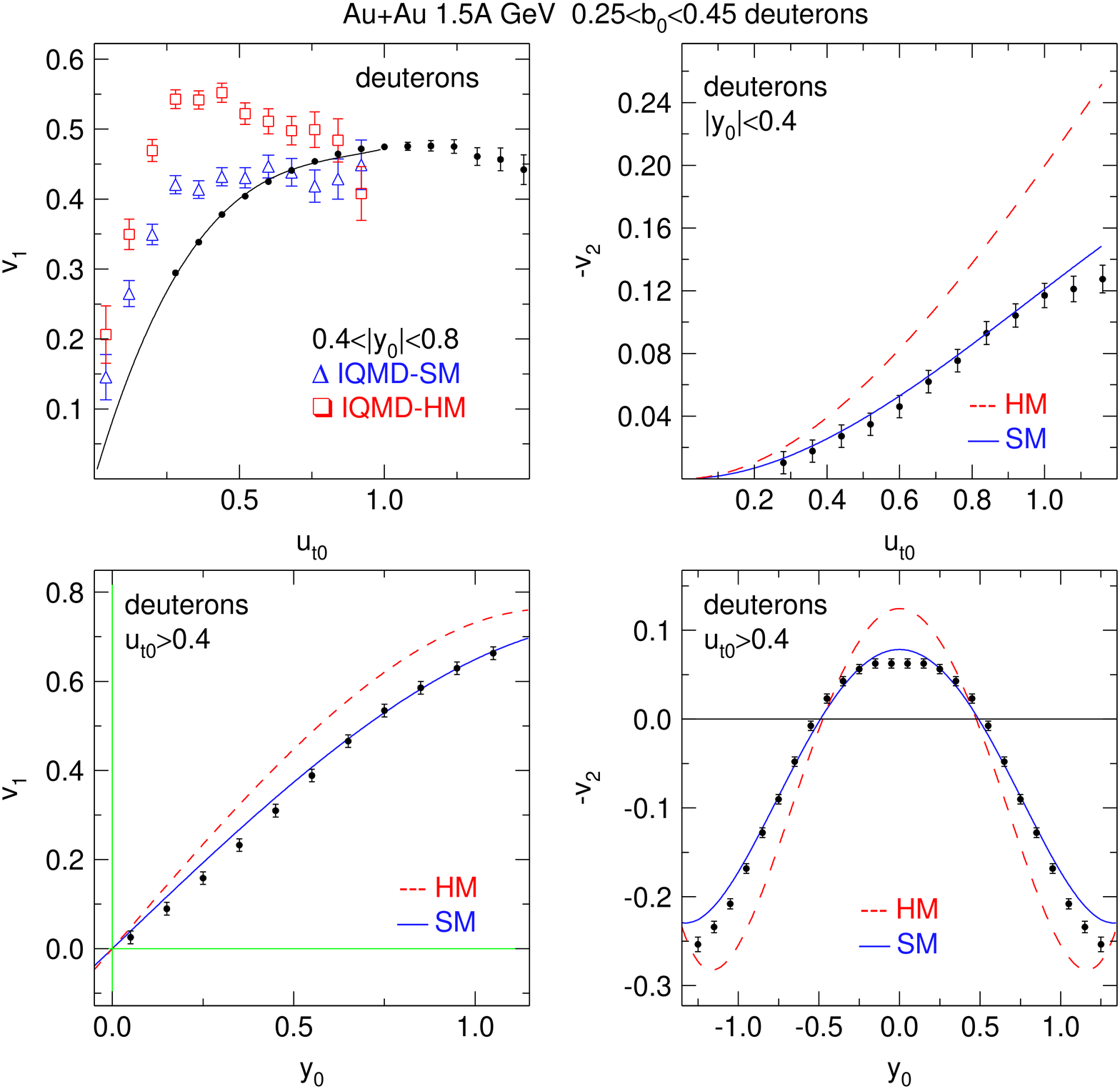,width=150mm}
\hspace{\fill}
\caption{
%
 Deuteron flow for  centrality $0.25<b_0<0.45$ 
 in collisions of Au+Au at
1.5A GeV.
The left panels show directed flow, elliptic flow is plotted in the
right panels. The lower panels show the rapidty dependence, while
the upper panels illustrate the momentum dependence.
Both the simulations (SM and HM) are shown together with the data (black dots).
For further details see text.
}
\label{v1-au1500c2Z1A2-4}
\end{figure}

We come finally to $1.5A$ GeV, the highest incident energy currently 
available at SIS for the heavy system Au+Au.
For the same centrality as before, we show the four flow projections
that we have studied before.
With the exception again of the $v_1(u_{t0})$ functions (around $y_0=0.5$)
and the high momentum part of $v_2(u_{t0})$ for protons, 
the description of the proton data, Fig. \ref{v1-au1500c2p-4}, and the
deuteron data, Fig. \ref{v1-au1500c2Z1A2-4}, is close to perfect if we use the
SM version.
Looking, specifically, at the elliptic flow $v_2(y_0)$ for deuterons,
we note that IQMD-SM is doing an excellent job from $0.15A$ GeV
(Fig. \ref{v2-au150c2p-pd}) to $1.5A$ GeV (Fig. \ref{v1-au1500c2Z1A2-4})
 Au on Au, i.e. over an order-of-magnitude change in energy, a remarkable
achievement of the transport code used.
We can say this despite the 'failures' concerning the yields of clusters 
and the mass hierarchy of the $v_1(u_{t0})$.
So more work is needed, but an encouraging start is done. 

\subsection{Excitation functions}
In Figs. \ref{v2-au150c2p-pd} to \ref{v1-au1500c2Z1A2-4} we have tried to reproduce with
simulations the full spectral information contained in the various
projections of directed and elliptic flow.
This is the most convincing way of checking the adequacy of the simulation.
To be able to represent such data in a readable form we had for a given plot to
fix many parameters, such as system size, incident energy, centrality,
particle type and finally the projection constraints.
To get a somewhat better overview of more general trends, such as excitation
functions, it is desirable to try to characterize flow with just one parameter.
For elliptic flow this is generally done by just showing its value at
mid-rapidity  as we have done in Fig. \ref{v20q-andro}.
For directed flow the mid-rapidity slope $dv_1(y0)/dy_0 \equiv v_{11}$
or its momentum-weighted alternative \cite{partlan95,gustafsson88a},
$u_{x01}$ in our scaled notation, is introduced.
Excitation functions for $v_{11}$ were shown in Fig. \ref{v11-auZ1A1-2}.
While this more compact form of displaying flow data is useful to display
general trends, one has to be aware that\\
a) information is lost (we showed the instructive shape changes of $v_2(y_0)$
associated with changes in the stiffness of the assumed EoS), and\\
b) a less 'robust' comparison is made: the midrapidity directed flow slopes 
depend on some technicalities such as the range of the fitted data and the
degree of the polynomials fitted, and clearly 'local' systematic errors
become more important when narrower data samples are taken.
 
\begin{figure}
\hspace{\fill}
\epsfig{file=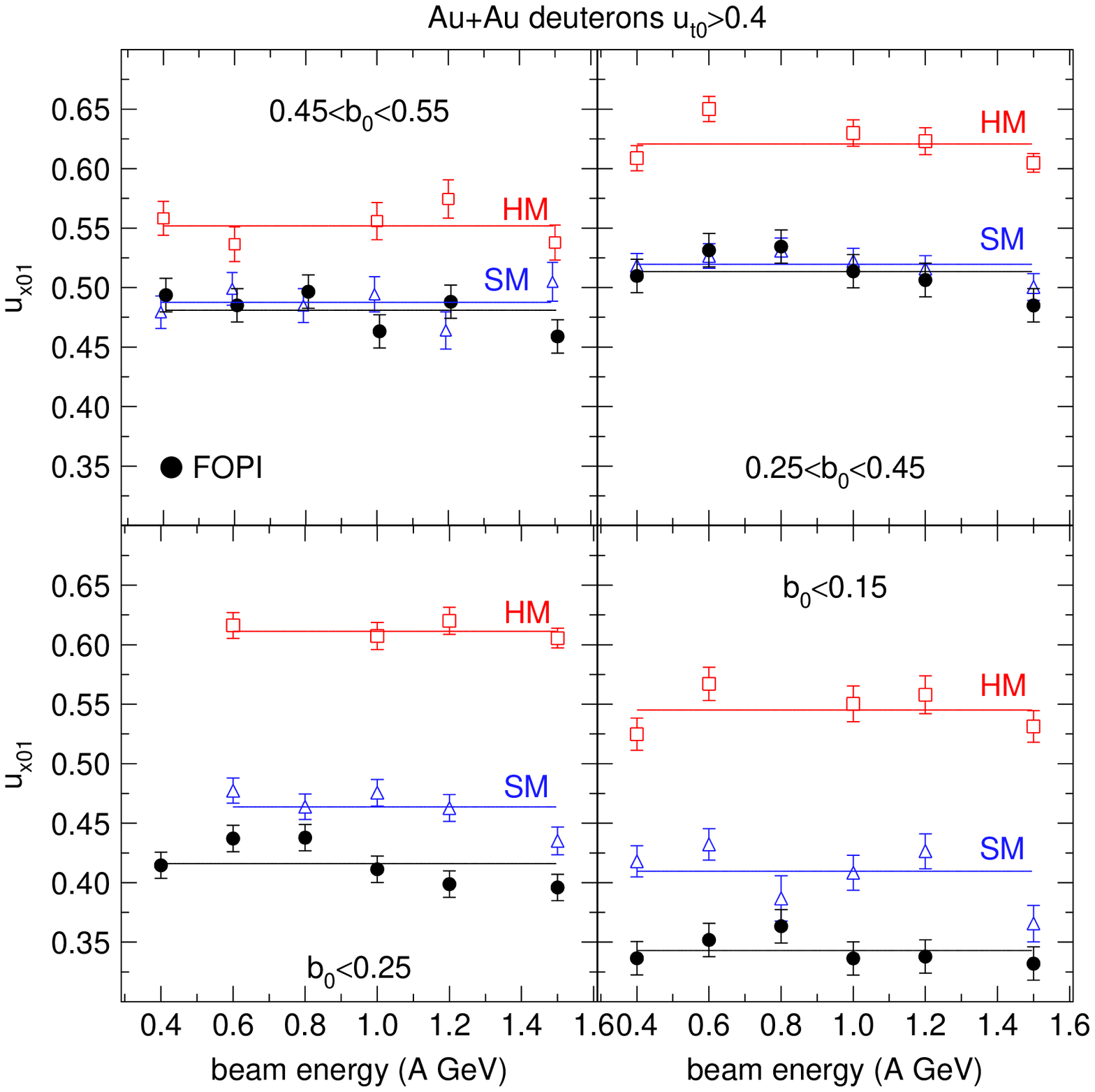,width=160mm}
\hspace{\fill}
\caption{
%
Excitation function for midrapidity slopes $u_{x01}$ of directed flow of
deuterons in 
 collisions of Au+Au with centrality $0.45<b_0<0.55$ (upper left panel),
$0.25<b_0<0.45$ (upper right panel),  $b_0<0.25$ (lower left panel)
and $b_0<0.15$ (lower right panel). 
A low-momentum cut-off was done ($u_{t0}>0.4$).
The FOPI data (black full circles) are compared to simulations with
IQMD-HM (red open squares) and IQMD-SM (blue open triangles).
The straight lines represent averages over the shown energy range.
}
\label{ux01}
\end{figure}

With this proviso in mind we present in Fig. \ref{ux01} a systematics of
 slopes $u_{x01}$ for deuterons as function of incident energy varying the
 centrality as indicated in the various panels.
We have used the same procedure throughout: fitting the interval 
$-1.1<y_0<1.1$ with a polynomial including a linear and a cubic term.
We choose deuterons rather than protons for two reasons:\\
1) as shown before, the flow signal is larger, and\\
2) in this energy range deuterons are less likely than protons to be resulting
in the last stages of the expansion from either late decays of larger clusters
or decay of baryonic resonances.\\
As shown earlier, the energy range we choose ($0.4-1.5A$ GeV) for this figure
is characterized by rather small changes of flow if scaled units are used.
We ignore in the figure therefore small variations with incident energy
and draw straight and constant lines through the data (and the HM and SM
calculations) that characterize the average behaviour.
Since in the previous figures we have shown comparisons almost exclusively for
one fixed centrality, $0.25<b_0<0.45$, we focus our interest here on varying
the centrality.
The upper right panel confirms the almost perfect agreement of the data
with the SM option.
While this agreement can be extended to a more peripheral interval,
$0.45<b_0<0.55$, in the upper left panel, the data shown in the lower panels
seem to suggest an even 'softer' behaviour than SM for the most
central collisions.
This strengthens the arguments tending to discard a stiff option like HM,
but also confirms that more work is necessary to still better understand the
'details'.
In particular it must be checked if stopping is described properly
(see later).
Also the missing Fermi motion in the simulation, discussed as possible reason
for the incorrect rendering of $v_1(u_{t0})$, especially of protons,
Figs. \ref{v1ut-au400c2pd-3} and \ref{v1ut-au1000c2pdt-2}, probably
 causes part of the overestimation of the $v_{11}$ slopes
even when the SM option is used. 
As noted before, interpreted classically, Fermi motion or
Brownian motion, has the effect of  higher apparent temperature lowering the
slopes $v_{11}$ in the experiment.

\begin{figure}
\hspace{\fill} 
\epsfig{file=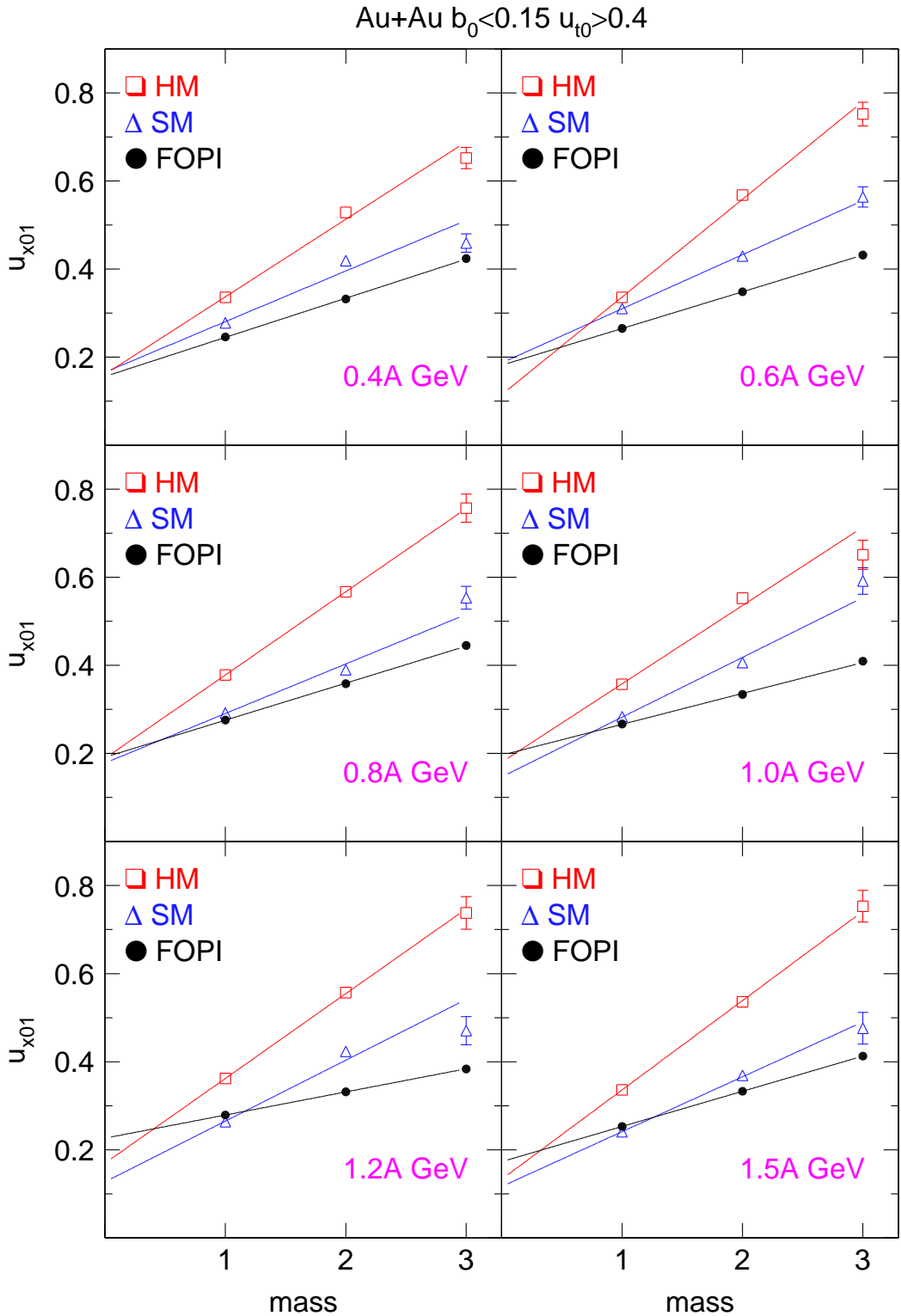,width=150mm}
\hspace{\fill}
\caption{
%
Midrapidity slopes $u_{x01}$ of directed flow of
hydrogen isotopes in 
 collisions of Au+Au with centrality $b_0<0.15$ at six different
indicated beam energies.
A low-momentum cut-off was done ($u_{t0}>0.4$).
The FOPI data (black dots) are compared to simulations with
IQMD-HM (red open squares) and IQMD-SM (blue open triangles).
The straight lines represent linear fits guiding the eye.
}
\label{ux01-auc1pdt}
\end{figure}
We have pointed out before the relatively small variation of directed flow 
with the centrality $b_0$ and associated this tentatively with  the
higher stopping in very central collisions compensating the loss of 
geometrical asymmetry.
Also we saw that heavier clusters showed more flow and this flow was more
sensitive to the EoS than flow of single nucleons.
Using the more compact form of flow description, we confirm these trends 
in Fig. \ref{ux01-auc1pdt} which presents a midrapidity slope systematics
($u_{x01}$) for the {\it most central} collisions ($b_0<0.15$, $2\%$
sharp-radius cross sections) as a function of ejectile mass for six
indicated incident energies.
For such pronounced centrality, using the {\it ERAT} observable,
 it is particularly
important that the simulations are done using the same selection methods as
in the experiment.
Looking first at the lowest right panel, which shows Au on Au data for
the highest energy, $1.5A$ GeV, we note that HM predictions for tritons are
almost $100\%$ (!) too high: this is a large signal {\it against} HM.
The preference for SM is true for all $E/u$, but at close inspection one
sees some fluctuations comparing the straight lines drawn through the
SM, HM and experimental data. 
These fluctuations seem to be of numerical origin and call for further efforts.
\subsection{The other observables: radial flow and stopping}
To raise the confidence in the physical adequacy of the simulation,
it is important not to limit the confrontation with experiment to directed
and elliptic flow.
Besides the chemical composition at freeze-out \cite{reisdorf10} there is
another related observable, the radial flow which is the main signal
of the expansion-cooling process.
Even more important is the so-called stopping.
A rather detailed presentation of these observables has been done earlier
\cite{reisdorf10}.
 
\begin{figure}
\hspace{\fill}
\epsfig{file=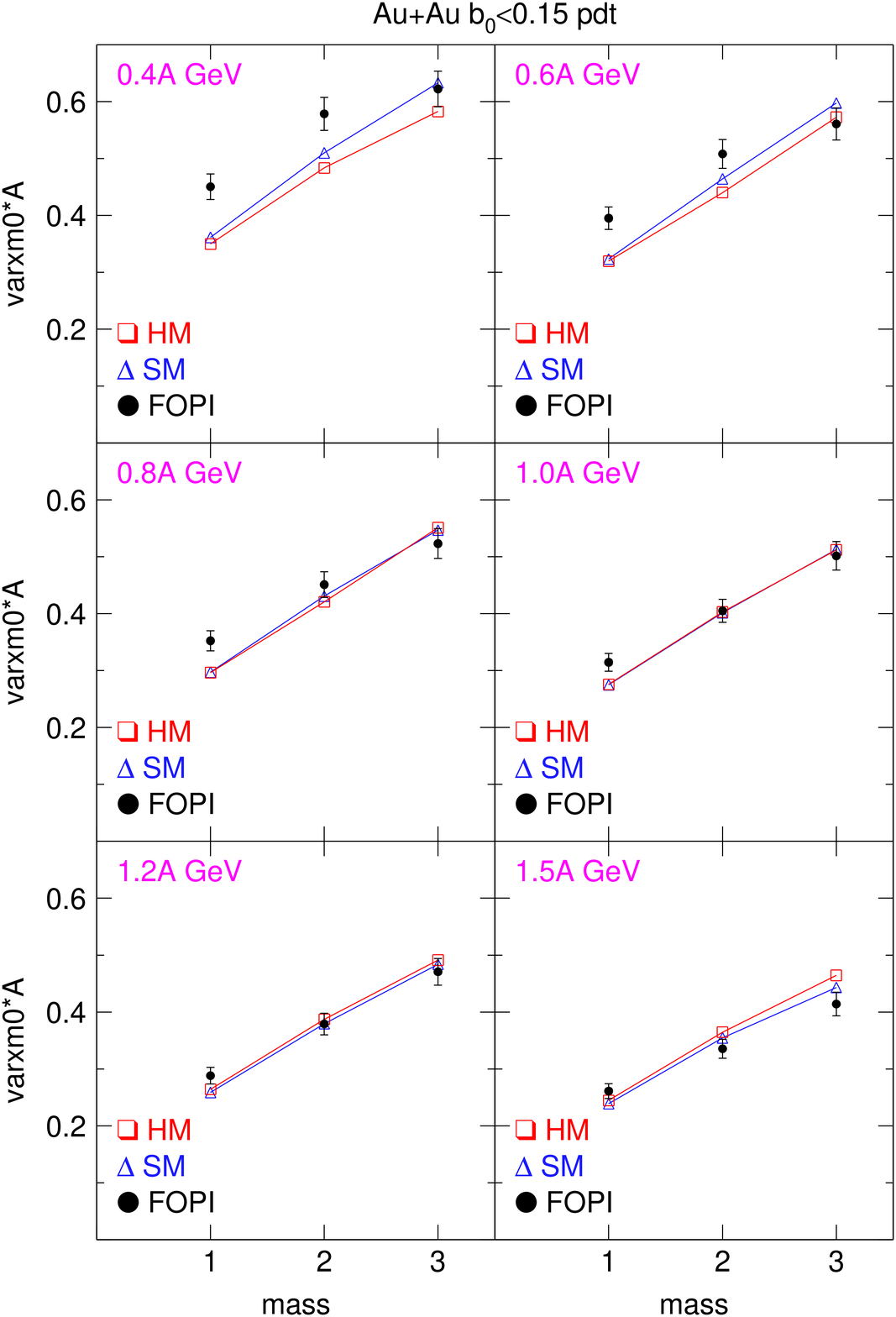,width=150mm}
\hspace{\fill}
\caption{
%
Mass-weighted scaled variances $varxm_{0}$ of constrained transverse
 mid-rapidity distributions for hydrogen isotopes in 
 collisions of Au+Au with centrality $b_0<0.15$ at six different
indicated beam energies.
The FOPI data (black dots) are compared to simulations with
IQMD-HM (red open squares) and IQMD-SM (blue open triangles).
The straight lines represent linear fits to the simulated data.
}
\label{varxm0-auc1pdt}
\end{figure}

\begin{figure}
\hspace{\fill}
\epsfig{file=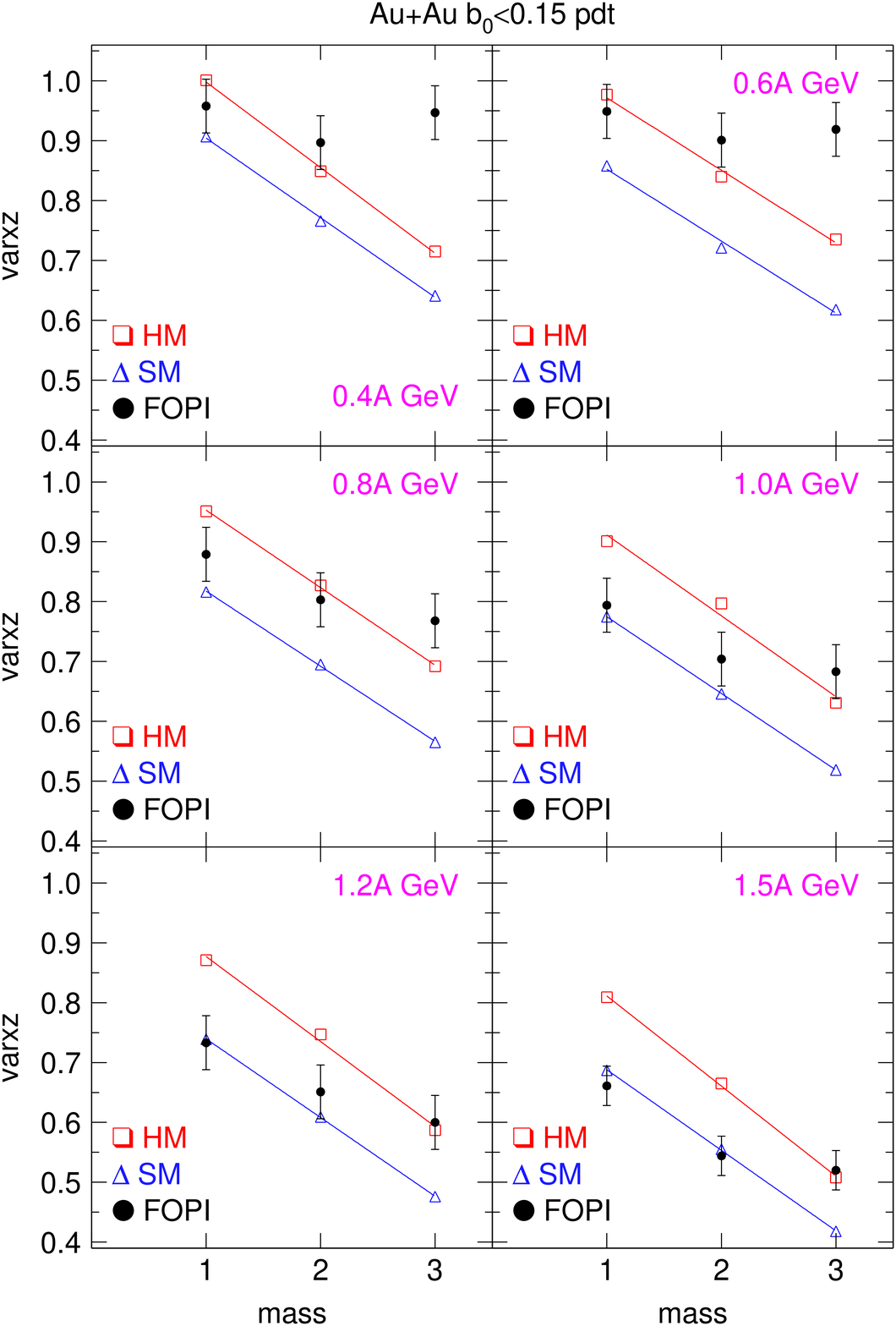,width=150mm}
\hspace{\fill}
\caption{
%
Stopping $varxz$  for hydrogen isotopes in 
 collisions of Au+Au with centrality $b_0<0.15$ at six different
indicated beam energies.
The FOPI data (black dots with error bars) are compared to simulations with
IQMD-HM (red open squares) and IQMD-SM (blue open triangles).
The straight lines represent linear fits to the simulated data guiding the eye.
}
\label{varxz-auc1pdt}
\end{figure}

In Fig. \ref{varxm0-auc1pdt} we present a systematics of radial flow for six 
incident energies together with transport code simulations.
What is plotted is the scaled variance,  $varxm_0$ of  constrained
(to $|y_0|<0.5$)  {\it transverse} rapidity distibutions \cite{reisdorf10}
as function of mass for the hydrogen isotopes.
Note that we are looking at the most central collisions here as for the
previous (and the next) figure.
In a globally thermal model (no flow, complete equilibration) these
variances should be (approximately, ignoring relativistic effects) 
proportional to $A^{-1}$, the inverse of the ejectile masses.
So then, when multiplied with the mass $A$, as done in the figure, they
should not depend on mass.
The fact that they grow with mass is interpreted as 'radial flow' in most
of the literature.
Such radial flow is also predicted by the IQMD calculation as can be seen in
the figure and, obviously, depends very little on the stiffness of the EoS.
This was known before, but is seen here to reproduce, {\it without any ad hoc 
adjustment}, the data very well at the highest energies (the two lower panels)
and overestimating the mass trend somewhat at the lowest energies (upper two
panels).
For hydrodynamically oriented people we add a word of caution:
due to the observed stopping hierarchy as function of mass 
(see Fig. \ref{varxz-auc1pdt})
the interpretation of slopes as seen in Fig. \ref{varxm0-auc1pdt} exclusively
in terms of a common 'radial-flow' velocity (profile) cannot be 
quantitatively  correct.

One can safely expect that maximal density is only achieved if full
stopping has occurred.
On the other hand incomplete stopping can be misinterpreted as 'ultra-soft'
EoS since it is associated with less static pressure.
Stopping must therefore be studied with great care.
Our registered events, when sorted according to {\it ERAT} are in fact
sorted according to the degree of stopping.
While {\it ERAT} is a global event qualifier affected partially by
apparatus effects (like multiplicity it is only used for selection),
we defined in our earlier work \cite{reisdorf10}
 an observable {\it varxz=varx/varz} where
{\it varx} is the variance of the transverse 1-dimensional rapidity
distribution (in an arbitrary direction $x$) and {\it varz} holds for the
rapidity distribution in the beam direction (the commonly used longitudinal
distribution).
{\it varxz} is determined individually for each particle type and
corrected for $4\pi$ acceptance.
It was observed \cite{reisdorf10} to follow a 'stopping hierarchy':
heavier clusters were associated with higher partial transparency (smaller
{\it varxz}).
Our simulations also predict this hierarchical trend for the hydrogen
isotopes as evident from Fig. \ref{varxz-auc1pdt}.
In contrast to radial flow, there is some sensitivity to the EoS
\cite{reisdorf10} as documented here also in Fig. \ref{varxz-auc1pdt}.
It is intuitively clear that a stiff repulsive mean field will favour
stopping.
We note in this context that our version of IQMD does not consider
in-medium modifications (besides Pauli blocking) of short-range
nucleon-nucleon interactions and cross sections.
The strong {\it axially symmetric} squeeze-out ($varxz>1$) predicted very early
\cite{scheid74} from ideal hydrodynamics is not observed:
it is hidden by (forward peaked !) surface transparency effects that
were not considered at the time: as one can see from the figure the
{\it varxz} values predicted by microscopic transport are well below one
in agreement with experiment (black dots in the figure).
On the other hand the hierarchy of stopping is somewhat less pronounced
in the experimental data and in particular the calculation misses
the absence of the hierarchy at $0.4-0.6A$ GeV where stopping was found
\cite{reisdorf04a} to be maximal.

\section{Conclusion}\label{conclusion}

The present work complements a systematic study of heavy ion collisions at
SIS energies ($0.09-1.5A$ GeV) with the aim of establishing a broad
data base for theoretical efforts to deduce constraints on the nuclear
EoS and other nuclear properties, such as viscosity.
In the frame of this systematics we have published \cite{reisdorf07}
extensive data on pion production,
 pion spectra and pion flow.
More recently, we have presented for central collisions an overview of the
chemical composition of the expanding fireballs, a systematics of stopping
and the evidence for radial flow in \cite{reisdorf10}.
Here we  have presented a  systematics of azimuthally
asymmetric fragment emissions in terms of the well established parameters
$v_1$, directed flow, and $v_2$, elliptic flow.
As already pointed out in some of our earlier work,
\cite{stoicea04,andronic01}, both $v_1$ and $v_2$ are complex multidimensional
functions: as a consequence we found it necessary to extract from our data
dependences on incident energy, system mass and charge, ejectile mass and
charge, centrality and, of course, longitudinal rapidity and transverse
momentum.

While the energy range spans more than an order of magnitude change,
the system sizes were varied from $^{40}$Ca+$^{40}$Ca to $^{197}$Au+$^{197}$Au
and the isospin of the system was separately varied using $^{96}$Ru+$^{96}$Ru
and $^{96}$Zr+$^{96}$Zr.
The centrality selection method was based primarily on binning the
distribution of the ratio of transverse and longitudinal energies, {\it ERAT},
i.e. on the degree of stopping achieved.
We stress that we use well defined {\it cross section} intervals,
corresponding to scaled impact parameter ranges, $b_0<0.15$, $b_0<0.25$,
$0.25<b_0<0.45$, and $0.45<b_0<0.55$.
We cover a large, sharply defined  two-dimensional, ($y_0$, $u_{t0}$), 
phase space both for $v_1$ and $v_2$.
All these features should ease precise duplication by transport model
simulations, a necessary condition to extract reliable nuclear matter
properties from the experimental data.
Of particular interest are the rapid rise of flow below $0.3A$ GeV,
followed by a relatively flat behavior above.
It will also be instructive to try to reproduce quantitatively the
correlations between system size dependent flow and stopping
\cite{reisdorf10,reisdorf04a}.

Using as transport code IQMD, the same code/version that we used in our
earlier studies on different observables, we have done an orientational study to
see how well we could reproduce the flow data and to which degree they were
sensitive to the assumed EoS.
To our surprise, although the used version of IQMD does not well reproduce
cluster yield data,  it reproduces well the flow of light, mass and charge
identified clusters, namely deuterons {\it if a soft equation of state is
 assumed}.
As a matter of fact, our confrontation with data virtually excludes a stiff
EoS.
This strengthens the case of a soft EoS proposed in earlier works
\cite{stoicea04,fuchs06,hartnack06} using different observables.
Another gratifying result is that we do not have any difficulty reproducing 
both directed and elliptic flow with the same assumptions.
At the highest energy studied here, we were also able to reproduce well the
radial flow and the stopping for the various hydrogen isotopes.
We note that our version of IQMD did not assume any in-medium modifications
of the microscopic cross sections other than Pauli blocking.
No {\it ad hoc} modification to achieve agreement with our data was done.

More specifically, for elliptic flow, $v_2(y_0)$, we were able to check the 
influence of the EoS on the shape of the function in a large rapidity range,
not just at mid-rapidity.
This allowed us to state that the differences of $v_2(y_0)$ between $^3$H and
$^3$He are consistent {\it over the complete rapidity gap}
 with a more repulsive force acting on the neutron richer cluster.
Our directed flow data on the mass three clusters are also consistent with
this conclusion.

Still, more work needs to be done.
On the experimental side a confirmation of the isospin effect will be helpful.
The future international facility FAIR at GSI, Darmstadt should allow to
access higher nuclear densities than has been possible here.
Theoretically it would be desirable to reproduce the cluster yields without
loosing the good performance on flow observables.
Also, using internally consistent theories such as \cite{gaitanos05}, 
a further fine tuning of our stopping data
is needed.
More subtle effects, to be further investigated,
are possible correlations of flow to clusterization visible in the
momentum dependences of single nucleon flow and, also, 
  the
experimental evidence of quantum effects on flow, such as Fermi motion
subsisting at freeze-out time.
Fermi motion tends to get 'lost' in a semiclassical transport calculation.
Such deficiency leads to overestimation of the predicted flow.
All of this  is an ambitious task beyond the scope of the present experimental
 work.
Hopefully there will be convergence of the various existing transport codes in
the future. 
An exciting prospect is a possible joining up of information from heavy ion
data with astrophysical observations on neutron stars \cite{oezel10,steiner10}.
\newpage
{\bf Appendix}

The correction factors \cite{ollitrault98} given in the table are defined by \\
$v_1=v'_1/d_i$   and  $v_2=v'_2/e_i$ \\
where the primed quantities are the uncorrected flow parameters and 
 the index $i$ distinguishes the various {\it ERAT} selected centralities,
 in order\\
$b_0<0.15$, $b_0<0.25$, $0.25<b_0<0.45$ and $0.45<b_0<0.55$.


\begin{tabular}{cccccccccc}
\hline
System & Energy(A GeV) & $d_1$ & $d_2$ & $d_3$ & $d_4$ 
                       & $e_1$ & $e_2$ & $e_3$ & $e_4$ \\
\hline
$^{197}$Au+$^{197}$Au & 0.09 & 0.372 & 0.475 & 0.607 & 0.598 
                            & 0.091 & 0.151 & 0.257 & 0.249\\  
$^{197}$Au+$^{197}$Au & 0.12 & 0.555 & 0.666 & 0.803 & 0.803 
                            & 0.211 & 0.316 & 0.496 & 0.496\\  
$^{197}$Au+$^{197}$Au & 0.15 & 0.701 & 0.789 & 0.879 & 0.857 
                            & 0.357 & 0.475 & 0.635 & 0.592\\  
$^{197}$Au+$^{197}$Au & 0.25 & 0.773 & 0.867 & 0.946 & 0.942 
                            & 0.451 & 0.611 & 0.805 & 0.795\\  
$^{197}$Au+$^{197}$Au & 0.40 & 0.854 & 0.916 & 0.965 & 0.956 
                            & 0.587 & 0.721 & 0.868 & 0.838\\  
$^{197}$Au+$^{197}$Au & 0.60 & 0.865 & 0.923 & 0.968 & 0.958 
                            & 0.608 & 0.741 & 0.877 & 0.845\\  
$^{197}$Au+$^{197}$Au & 0.80 & 0.864 & 0.919 & 0.966 & 0.960 
                            & 0.606 & 0.729 & 0.872 & 0.851\\  
$^{197}$Au+$^{197}$Au & 1.00 & 0.852 & 0.911 & 0.963 & 0.958 
                            & 0.583 & 0.708 & 0.860 & 0.845\\  
$^{197}$Au+$^{197}$Au & 1.20 & 0.840 & 0.902 & 0.958 & 0.954 
                            & 0.560 & 0.688 & 0.845 & 0.830\\  
$^{197}$Au+$^{197}$Au & 1.50 & 0.823 & 0.891 & 0.952 & 0.948 
                            & 0.530 & 0.662 & 0.824 & 0.813\\  
$^{129}$Xe+CsI & 0.15        & 0.578 & 0.684 & 0.811 & 0.790 
                            & 0.231 & 0.337 & 0.510 & 0.477\\  
$^{129}$Xe+CsI & 0.25        & 0.705 & 0.802 & 0.911 & 0.905 
                            & 0.361 & 0.495 & 0.708 & 0.694\\  
$^{96}$Ru+$^{96}$Ru & 0.40   & 0.729 & 0.817 & 0.912 & 0.911 
                            & 0.391 & 0.519 & 0.712 & 0.709\\  
$^{96}$Ru+$^{96}$Ru & 1.00   & 0.733 & 0.804 & 0.892 & 0.895 
                            & 0.397 & 0.499 & 0.663 & 0.671\\  
$^{96}$Ru+$^{96}$Ru & 1.50   & 0.704 & 0.781 & 0.889 & 0.892 
                            & 0.361 & 0.463 & 0.659 & 0.665\\  
$^{96}$Zr+$^{96}$Zr & 0.40   & 0.711 & 0.798 & 0.897 & 0.877 
                            & 0.368 & 0.489 & 0.676 & 0.631\\  
$^{96}$Zr+$^{96}$Zr & 1.50   & 0.702 & 0.771 & 0.871 & 0.837 
                            & 0.358 & 0.449 & 0.619 & 0.554\\  
$^{58}$Ni+$^{58}$Ni & 0.15   & 0.364 & 0.454 & 0.580 & 0.555 
                            & 0.087 & 0.138 & 0.232 & 0.211\\  
$^{58}$Ni+$^{58}$Ni & 0.25   & 0.505 & 0.604 & 0.752 & 0.756 
                            & 0.172 & 0.254 & 0.422 & 0.427\\  
$^{40}$Ca+$^{40}$Ca & 0.40   & 0.549 & 0.617 & 0.717 & 0.704 
                            & 0.206 & 0.267 & 0.376 & 0.361\\  
$^{40}$Ca+$^{40}$Ca & 0.60   & 0.575 & 0.646 & 0.745 & 0.731 
                            & 0.229 & 0.296 & 0.412 & 0.394\\  
$^{40}$Ca+$^{40}$Ca & 0.80   & 0.575 & 0.633 & 0.737 & 0.745 
                            & 0.229 & 0.283 & 0.402 & 0.412\\  
$^{40}$Ca+$^{40}$Ca & 1.00   & 0.587 & 0.641 & 0.738 & 0.739 
                            & 0.239 & 0.290 & 0.403 & 0.404\\  
$^{40}$Ca+$^{40}$Ca & 1.50   & 0.561 & 0.609 & 0.698 & 0.699 
                            & 0.216 & 0.259 & 0.353 & 0.354\\  
$^{40}$Ca+$^{40}$Ca & 1.93   & 0.511 & 0.552 & 0.643 & 0.663 
                            & 0.177 & 0.209 & 0.293 & 0.314\\  
\hline
\end{tabular}


\begin{ack}
This work has been supported by the German BMBF,
contract 06HD154, 
 by the DFG (Project 446-KOR-113/76) and DFG(Project 436POL 113/121/0-1),
by the Polish Ministry of Science and Higher Education under 
Grant No DFG /34/2007/,
the DAAD (PPP D /03/44611) and the IN2P3/GSI agreement 97/29.
This work was also supported by the National Research Foundation of Korea
 (NRF) under Grant (No. 2011-0003258)
\end{ack}




\begin{thebibliography}{30}

\bibitem{gustafsson84}H. A. Gustafsson, et al.,
Phys. Rev. Lett.  52 (1984) 1590.

\bibitem{renfordt84}R. E. Renfordt, et al.,
Phys. Rev. Lett.  53 (1984) 763.   

\bibitem{scheid74}W. Scheid, et al.,
Phys. Rev. Lett.  32 (1974) 741.

\bibitem{stoecker86}H. St\"{o}cker, W. Greiner,
Phys. Rep. 137 (1986) 277.

\bibitem{clare86}R. B. Clare, D. Strottman,
Phys. Rep. 141 (1986) 177.

\bibitem{hsong11}Huichao Song, Steffen A. Bass, U. Heinz, Tetsufumi Hirano,
Chun Shen,
Phys. Rev. Lett.  106 (2011) 192301.

\bibitem{niemi11}H. Niemi, G. S. Denisol, P. Huovinen, E. Moln\'{a}r,
D. Riscke,
Phys. Rev. Lett.  106 (2011) 212302.

\bibitem{bertsch88} G. F. Bertsch, S. Das Gupta,
Phys. Rep.  160 (1988) 189.

\bibitem{blaettel93} B. Blaettel, V. Koch, U. Mosel,
Rep. Prog. Phys. 56 (1993) 1. 

\bibitem{cassing00} W. Cassing, S. Juchem,
Nucl. Phys. A 665 (2000) 377;
Nucl. Phys. A 672 (2000) 417;
Nucl. Phys. A 677 (2000) 445.

\bibitem{aichelin91} J. Aichelin,
Phys. Rep.  202 (1991) 233.

\bibitem{hartnack98}C. Hartnack, et al.,
Eur. Phys. J. A  1 (1998) 151.

\bibitem{yzhang05} Yingxun Zhang,  Zhuxia Li,
Phys. Rev. C 71 (2005) 024604;
Phys. Rev. C 74 (2006) 014602.

\bibitem{bass98} S. A. Bass, et al.,
Prog. Part. Nucl. Phys. 41 (1998) 255.

\bibitem{qfli11} Qingfeng Li, Caiwan Shen, Chenchen Guo, Yongjia Wang,
Zhuxia Li, J. Lukasik, W. Trautmann, 
Phys. Rev. C 83 (2011) 044617. 

\bibitem{colonna04} M. Colonna, G. Fabbri, M. Di Toro, F. Matera, H. H. Wolter,
Nucl. Phys. A 742 (2004) 337.

\bibitem{santini05}E. Santini, T. Gaitanos, M. Colonna, M. Di Toro
Nucl. Phys.  A 756 (2005) 468.

\bibitem{baran05} V. Baran, M. Colonna, V. Greco, and M. Di Toro,
Phys. Rep. 410 (2005) 335.

\bibitem{gaitanos05} T. Gaitanos, C. Fuchs, H. H. Wolter,
Phys. Lett. B 609 (2005) 241.

\bibitem{lwchen04} L.-W. Chen, C. M. Ko, B.-A. Li,
Phys. Rev. C 69 (2004) 054606.

\bibitem{bali04} Bao-An Li, C. B. Das, S. Das Gupta and C. Gale,
Phys. Rev. C 69 (2004) 011603(R); Nucl. Phys. A 735 (2004) 563.

\bibitem{danielewicz92} P. Danielewicz, Q. Pan,
Phys. Rev. C 46 (1992) 2002.

\bibitem{danielewicz00} P. Danielewicz,
Nucl. Phys. A 673 (2000) 375.

\bibitem{buss11}O. Buss, T. Gaitanos, K. Gallmeister, H. van Hees, M. Kaskulov, 
             O. Lalakulich, A. B. Larionov, T. Leitner, J. Weil, U. Mosel,
  arXiv:1106.1344  (hep-ph). 

\bibitem{reisdorf97r}W. Reisdorf, H. G. Ritter,
Annu. Rev. Nucl. Part. Sci.  47 (1997) 663.

\bibitem{herrmann99}N. Herrmann, J. P. Wessels, T. Wienold,
Annu. Rev. Nucl. Part. Sci.  49 (1999) 581.

\bibitem{danielewicz02}P. Danielewicz, R. Lacey, W. G. Lynch,
Science 298 (2002) 1592.

\bibitem{partlan95} M. D. Partlan, et al. (EOS Collaboration),
Phys. Rev. Lett. 75 (1995) 2100.

\bibitem{andronic05} A. Andronic, et al. (FOPI Collaboration),
Phys. Lett. B 612 (2005) 173, arXiv: nucl-ex/0411024.

\bibitem{swang96} S. Wang, et al. (EOS Collaboration),
Phys. Rev. Lett. 76 (1996) 3911.

\bibitem{stoicea04}G. Stoicea, et al.,
Phys. Rev. Lett.  92 (2004) 072303, arXiv: nucl-ex/0401041.

\bibitem{peilert89} G. Peilert, et al.,                    
Phys. Rev. C 39 (1989) 1402.                 

\bibitem{sturm01} C. Sturm et al. (KaoS Collaboration),
Phys. Rev. Lett. 86 (2001) 39.

\bibitem{foerster07} A. F\"{o}rster, et al. (KaoS Collaboration),
Phys. Rev. C 75 (2007) 024906.

\bibitem{fuchs06}C. Fuchs,
Prog. Part. Nucl. Phys. 56 (2006) 1.

\bibitem{hartnack06}C. Hartnack, H.Oeschler and J. Aichelin,
Phys. Rev. Lett. 96 (2006) 012302.

\bibitem{gobbi93} A. Gobbi, et al.,
Nucl. Instr. Meth. A 324 (1993) 156.

\bibitem{ritman95}J. Ritman,
Nucl. Phys.  B 44 (1995) 708.

\bibitem{reisdorf07}W. Reisdorf, et al. (FOPI Collaboration),
Nucl. Phys. A 781 (2007) 459, arXiv: nucl-ex/0610025.      

\bibitem{reisdorf10}W. Reisdorf, et al. (FOPI Collaboration),
Nucl. Phys. A 848 (2010) 366, arXiv:1005.3418.      

\bibitem{RHIC05}
I. Arsene et al., Nucl. Phys.  A 757 (2005) 1;
B. B. Back et al., ibid.  A 757 (2005) 28;
J. Adams et al., ibid.  A 757 (2005) 102;
K. Adcox et al., ibid.  A 757 (2005) 184.

\bibitem{voloshin96} S. Voloshin, Y. Zhang,
Zeitsch. Phys. C (1996) 665.

\bibitem{poskanzer98} A. M. Poskanzer and S. A. Voloshin, 
Phys. Rev. C 58 (1998) 1671. 

\bibitem{bali01} Bao-An Li, W. Udo Schr\"{o}der (Editors), in
'Isospin Physics in Heavy-Ion Callisions at Intermediate Energies'
Nova Science Publishers, Inc., Huntington New York (2001).                    

\bibitem{andronic01} A. Andronic, et al. (FOPI Collaboration),
Nucl. Phys. A 679 (2001) 765, arXiv: nucl-ex/0008007.

\bibitem{andronic06} A. Andronic, J. Lukasik, W. Reisdorf, W. Trautmann,
Eur. Phys. J. A30 (2006) 31, arXiv: nucl-ex/0608015.

\bibitem{reisdorf97}W. Reisdorf, et al. (FOPI Collaboration),
Nucl. Phys. A 612 (1997) 493, arXiv:nucl-ex/9610009.      

\bibitem{reisdorf04b}W. Reisdorf, et al. (FOPI Collaboration),
Phys. Lett. B 595 (2004) 118, arXiv: nucl-ex/0405014. 

\bibitem{danielewicz85} P. Danielewicz, G. Odyniec,
Phys. Lett. B 157 (1985) 168.

\bibitem{gutbrod90f} H. H. Gutbrod, et al.,
Phys. Rev. C 42 (1990) 640.

\bibitem{barrette97} J. Barrette, et al. (E877 Collaboration), 
Phys. Rev. C 55 (1997) 1420, arXiv: nucl-ex/9610006.

\bibitem{ollitrault98} J.-Y. Ollitrault, 
arXiv: nucl-ex/9711003,
Nucl. Phys. A 638 (1998) 195c.

\bibitem{ramillien95} W. Ramillien, et al. (FOPI Collaboration),
Nucl. Phys. A 587 (1995) 802.      

\bibitem{bastid97} N. Bastid et al. (FOPI Collaboration),
Nucl. Phys. A 646 (1997) 753.

\bibitem{rami99} F. Rami, et al. (FOPI Collaboration),
Nucl. Phys. A 646 (1999) 367.

\bibitem{andronic01a} A. Andronic, et al. (FOPI Collaboration),
Phys. Rev. C 64 (2001) 041604(R), arXiv: nucl-ex/0108014.

\bibitem{andronic03} A. Andronic, et al. (FOPI Collaboration),
Phys. Rev. C 67 (2003) 034907, arXiv: nucl-ex/0301009.

\bibitem{bastid04} N. Bastid et al. (FOPI Collaboration),
Nucl. Phys. A 742 (2004) 29.

\bibitem{bastid05} N. Bastid et al. (FOPI Collaboration),
Phys. Rev. C 72 (2005) 011901, arXiv: nucl-ex/0504002.

\bibitem{doss86}K.G.R. Doss, et al.,
Phys. Rev. Lett. 57 (1986) 302.

\bibitem{lehaut10} G. Lehaut et al. (INDRA and ALADIN Collaborations),
Phys. Rev. Lett. 104 (2010) 232701.                  

\bibitem{doss87}K.G.R. Doss, et al.,
Phys. Rev. Lett. 59 (1987) 2720.

\bibitem{barrette99} J. Barrette, et al. (E877 Collaboration), 
Phys. Rev. C 59 (1999) 884, arXiv: nucl-ex/9805006.

\bibitem{schmidt93} W. Schmidt, U. Katscher, B. Walhauser, J. A. Maruhn,
H. St\"{o}cker, W. Greiner,
Phys. Rev. C 47 (1993) 2782.

\bibitem{phenix07a} A. Adare, et al. (PHENIX Collaboration),
Phys. Rev. Lett. 98 (2007) 162301.

\bibitem{reisdorf04a}W. Reisdorf, et al. (FOPI Collaboration),
Phys. Rev. Lett.  92 (2004) 232301, arXiv: nucl-ex/0404037.

\bibitem{gutbrod89r} H. H. Gutbrod, A. M. Poskanzer, H. G. Ritter,
Rep. Prog. Phys. 52 (1989) 1267.

\bibitem{leifels93} Y. Leifels et al.,
Phys. Rev. Lett. 71 (1993) 963.                  

\bibitem{russotto10} P. Russotto et al.,
Phys. Lett. B 697 (2010) 471.

\bibitem{gutbrod89} H. H. Gutbrod, et al.,
Phys. Lett. B 216 (1989) 267.

\bibitem{stoecker82}H. St\"{o}cker, et al.,
Phys. Rev. C 25 (1982) 1873.

\bibitem{pinkenburg99} C. Pinkenburg, et al. (E895 Collaboration),
Phys. Rev. Lett. 83 (1999) 1295.

\bibitem{giordano10} V. Giordano, M. Colonna, M. Di Toro, V. Greco, J. Rizzo,
Phys. Rev. 81 (2010) 044611.

\bibitem{brill96}D. Brill, et al.,
Z. Phys. A 355 (1996) 61. 

\bibitem{vandalen07} E. N. E. van Dalen, C. Fuchs, A. Faessler,
Eur. Phys. J. A 31 (2007) 29.

\bibitem{lattimer01} J. M. Lattimer and M. Prakash,
Astrophys. J. 550 (2001) 426.

\bibitem{demorest10} P. B. Demorest, T. Pennuci, S. M. Ransom,
  M. S. E. Roberts, W. T. Hessels,
Nature 467 (2010) 1081.

\bibitem{akmal98} A. Akmal, V. R. Pandharipande, D. G. Ravendall,
Phys. Rev. C 58 (1998) 1804.

\bibitem{gustafsson88a}H. A. Gustafsson, et al.,
Mod. Phys. Lett.  A3 (1988) 1323.


\bibitem{yzhang08} Yingxun Zhang, P. Danielewicz, M. Famiano, Zhuxia Li,
W. G. Lynch, M. B. Tsang,
Phys. Lett. B 664 (2008) 145.

\bibitem{oezel10} F. \"{O}zel, G. Baym, T. G\"{u}ver,
Phys. Rev. D 82 (2010) 101301(R).

\bibitem{steiner10} A. W. Steiner, J. M. Lattimer, E. F. Brown,
Astroph. J. 722 (2010) 33.

\end{thebibliography}
\end{document}